\documentclass[ecta,nameyear,draft]{econsocart-arxiv}

\RequirePackage[colorlinks,citecolor=blue,linkcolor=blue,urlcolor=blue,pagebackref]{hyperref}

\startlocaldefs

\theoremstyle{plain}
\newtheorem{assumption}{Assumption}
\newtheorem{theorem}{Theorem}
\newtheorem{lemma}{Lemma}
\newtheorem{corr}{Corollary}

\newtheorem{prop}{Proposition}
\newtheorem{con}{Condition}

\theoremstyle{definition}
\newtheorem{definition}{Definition}
\newtheorem{example}{Example}
\newtheorem{remark}{Remark}[section]

\endlocaldefs

\usepackage{graphicx}   % For including images
\usepackage{subcaption} % For subfigures and captions
\usepackage{pdflscape}  % For rotating the page to landscape

\begin{document}

\begin{frontmatter}

% \title{On the Asymptotic Properties of \\Debiased Machine Learning Estimators}
\title{Debiased Machine Learning with \\Many Cross-Fitting Folds} 
% \thankstext{T1}{This paper was previously circulated under the title ``On the Asymptotic Properties of Debiased Machine Learning Estimators.''}
% \runtitle{On the Asymptotic Properties of DML Estimators}

\begin{aug}
% use \particle for den|der|de|van|von (only lc!)
% [id=?,addressref=?,corref]{\fnms{}~\snm{}\ead[label=e?]{}\thanksref{}}
%
%% e-mail is mandatory for each author
%
%%% initials in fnms (if any) with spaces
%
\author[add1]{\fnms{Amilcar}~\snm{Velez}\ead[label=e1]{amilcare@cornell.edu}}
% \author[id=au3,addressref={add3}]
%%%%%%%%%%%%%%%%%%%%%%%%%%%%%%%%%%%%%%%%%%%%%%
%% Addresses                                %%
%%%%%%%%%%%%%%%%%%%%%%%%%%%%%%%%%%%%%%%%%%%%%%

\address[add1]{%
\orgdiv{Department of Economics},
\orgname{Cornell University}}
\end{aug}

\vspace{-1em}

%% Put support info here.  Reminder: do not thank the handling coeditor anonymously or by name
\begin{funding}
This paper was previously circulated under the title ``On the Asymptotic Properties of Debiased Machine Learning Estimators.'' This draft: August 2, 2026. I am deeply grateful to Ivan Canay, Federico Bugni, and Joel Horowitz for their guidance and support and for the extensive discussions that have helped shape the paper.
I also want to acknowledge helpful conversations with Eric Auerbach, Federico Crippa, Jacob Dorn, Bruno Fava, Igal Hendel, Diego Huerta, Danil Fedchenko, Chuck Manski, Pepe Montiel-Olea, Whitney Newey,  Giorgio Primiceri, Sebastian Sardon, Chris Walker, Thomas Wiemann, and Zhen Xie as well as helpful feedback from seminar participants at Duke, Rice, FGV-EPGE, UPenn, Rutgers, Michigan, Emory QTM, USC, Cornell, ESWC 2025, Cornell/PennState Conference on Econometrics and IO, Microeconometrics Class of 24/25, Georgetown U, XLIII Encuentro de Economistas del BCRP (Peru), LMU of Munich, University of Mannheim, University of Bonn, Harvard/MIT, Boston U, PennState, and NYU. Financial support from the Robert Eisner Memorial Fellowship and the Dissertation Year Fellowship at Northwestern University is gratefully acknowledged. Any and all errors are my own. 
\end{funding}
%% Put support info here.  Reminder: do not thank the handling coeditor anonymously or by name
% \support{Preliminary version. This draft: July 9, 2026.}
%
\begin{abstract}
This paper studies debiased machine learning (DML) when the number of cross-fitting folds, $K_n$, may grow with the sample size $n$. Existing fixed-$K$ asymptotic theory implies that DML1 and DML2, the two main DML variants, are asymptotically equivalent, providing no guidance on which variant to use or how to choose $K_n$. We show that this equivalence can break down when $K_n$ grows proportionally to $\sqrt{n}$: DML1 can exhibit asymptotic bias, in which case standard inference based on DML1 fails—as can occur, for instance, for the local average treatment effect (LATE)—whereas inference based on DML2 remains valid. Moreover, we show that, under an \emph{algorithmic-stability} condition, estimation and inference based on DML2 are valid for any $2\le K_n\le n$, including the leave-one-out case, $K_n=n$. Finally, for scalar DML2 estimators whose first-step estimators admit a stochastic linear expansion, we derive a second-order approximation showing that larger values of $K_n$ reduce the second-order asymptotic bias and mean-squared error, although the marginal improvements diminish.
\end{abstract}


\bigskip

\end{frontmatter} 
% \tableofcontents
%%%%%%%%%%%%%%%%%%%%%%%%%%%%%%%%%%%%%%%%%%%%%%%%%%%%%%%%%%%%%%%%%%%%%%%%%
%%%% Main text entry area:
%%%%%%%%%%%%%%%%%%%%%%%%%%%%%%%%%%%%%%%%%%%%%%%%%%%%%%%%%%%%%%%%%%%%%%%%%

\section{Introduction}

Debiased machine learning (DML) is a two-step estimation method for econometric settings in which the parameter of interest depends on unknown nuisance functions \citep{chernozhukov2018double}. By combining Neyman orthogonality with $K$-fold cross-fitting, DML attains standard asymptotic properties under weaker conditions than classical semiparametric methods, thereby allowing the use of machine-learning methods in the first step. In practice, however, implementing DML requires two choices: whether to use DML1 or DML2---two main variants that differ in how they aggregate fold-specific information---and how many folds $K$ to use for cross-fitting. Existing fixed-$K$ asymptotic theory implies that DML1 and DML2 are asymptotically equivalent, providing no guidance on either choice. Nevertheless, simulation evidence suggests that DML2 can outperform DML1 in some models \citep{chernozhukov2018double}, and that increasing $K$ can reduce the bias and mean-squared error (MSE) of DML2 \citep{ahrens2025introduction}. These findings raise two questions that fixed-$K$ asymptotic theory does not resolve: when and why should DML2 be preferred to DML1, and, if this preference has a theoretical basis, what formal guidance can be given for the choice of $K$ when implementing DML2. 

This paper studies these questions under an asymptotic framework in which the number of cross-fitting folds, $K_n$, may grow with the sample size $n$, with $2 \le K_n \le n$. This approach follows a tradition in econometrics of using refined asymptotic frameworks to study questions that standard (fixed-$K$) asymptotic theory cannot answer. Here, large values of $K_n$ are of interest because they increase the number of observations used by each first-step estimator, improving accuracy at the cost of greater dependence across estimators; see Remark~\ref{rem:cross-fitting}. Whether larger $K_n$ values provide a theoretical advantage for either DML1 or DML2 is unknown.
 
The econometric setting we consider is broad. The parameter of interest is finite dimensional and is identified by a moment condition involving unknown nuisance functions. The moment function is linear in the parameter of interest, and the nuisance function is a vector of known transformations of conditional expectations. This framework includes many examples studied in the literature, among them the average treatment effect (ATE), the average treatment effect on the treated  (ATT)  in difference-in-differences designs, the local average treatment effect, the partially linear model, and the partially linear IV model.  %Section~\ref{sec:setup} presents the formal setup and definitions.
 
This paper makes two main contributions. First, the growing-$K_n$ framework distinguishes DML1 from DML2 at first order. The distinction is characterized by a model-dependent quantity $\Lambda$; when $K_n$ grows proportionally to $\sqrt{n}$ and $\Lambda \neq 0$, the limiting distribution of DML1 exhibits asymptotic bias, in which case standard inference based on DML1 fails, whereas that of DML2 remains centered. Moreover, under an \emph{algorithmic-stability} condition on the first-step estimators, the standard first-order asymptotic theory for DML2 remains valid for any $2 \le K_n \le n$, including the leave-one-out case ($K_n = n$). Second, under stronger conditions on the first-step estimators, we derive a second-order asymptotic approximation for scalar DML2 estimators; in this regime, the performance of DML2, measured by second-order asymptotic bias and MSE, improves with more folds but with diminishing returns.
 
We first explain in Section~\ref{sec:result1} when and why DML2 should be preferred to DML1. Existing fixed-$K$ asymptotic theory implies that the two estimators have the same limiting distribution, which leaves the simulation evidence that motivated the literature's preference for DML2 unexplained. In contrast, under the growing-$K_n$ framework, the difference between DML1 and DML2 becomes first order. When $K_n$ grows proportionally to $\sqrt{n}$, the limiting distribution of DML1 may be shifted away from zero, and the size of this shift is proportional to $\Lambda$. Therefore, when $\Lambda\neq 0$, standard inference based on DML1 fails, whereas inference based on DML2 remains valid. We explain in Section \ref{sec:DML1-sensitive} that the sensitivity of DML1 in the growing-$K_n$ framework depends not on the first-step estimators, but on how DML1 aggregates information across folds. The case $\Lambda \neq 0$ arises in models used routinely in empirical work. For the ATE and the ATT in difference-in-differences designs, $\Lambda = 0$ and the two variants remain equivalent; for the local average treatment effect (LATE) and the partially linear IV model, $\Lambda$ is typically nonzero. In our LATE simulation design, the coverage probability of a nominal 95\% confidence interval based on DML1 with 30 folds falls below 87\%, while DML2 remains at the nominal level (Section~\ref{sec:simulations}). Whether $\Lambda$ is zero can be diagnosed on a model-by-model basis; see Remark \ref{rem:diagnosing-lambda}.

% Moreover, because $\Lambda$ is a known functional of the same objects that DML already estimates, it can be computed model by model, so the condition $\Lambda \neq 0$ is diagnosable in applications.

Moreover, we show that estimation and inference based on DML2 is valid for any $2\le K_n\le n$ under an additional algorithmic-stability condition on the first-step estimators. This condition controls the sensitivity of the first-step estimator to replacing a single observation with an independent copy, allowing us to handle the dependence induced by large $K_n$; see Remark \ref{rem:cross-fitting}. A consequence is that, under the maintained conditions, first-order asymptotics for DML2 offer no guidance on the choice of $K_n$, motivating the second-order analysis developed in Section~\ref{sec:result2}.

Finally, we provide in Section~\ref{sec:result2} theoretical guidance on the choice of $K_n$ for scalar DML2 estimators under stronger conditions on the first-step estimators. We focus on first-step estimators that admit a stochastic linear expansion, a class that includes local-polynomial estimators under suitable conditions. Within this framework, we derive a second-order asymptotic approximation for DML2 and show that larger values of $K_n$ reduce the leading terms of the second-order asymptotic bias and MSE, although these marginal improvements diminish. This yields two conclusions, both within this class of first-step estimators. First, the second-order analysis identifies DML2 with $n$ folds (the leave-one-out estimator) as a benchmark. Second, the number of folds should be set as large as computation allows.

\textbf{Related literature:} This paper contributes to the growing literature on DML and
orthogonal/debiased estimation, including  \cite{chernozhukov2018double}, \cite{chernozhukov2022locally}, \cite{chernozhukov2022automatic,chernozhukov2022debiased}, \cite{semenova2021debiased}, \cite{semenova2023adaptive,semenova2023debiased}, \cite{escanciano2023machine}, \cite{rafi2023efficient}, \cite{fava2024predicting}, among many others. Most of the papers use DML2, with exceptions such as \cite{chernozhukov2017double}, \cite{ji2023model}, \cite{noack2021flexible}, and \cite{cheng2023weight}, which use DML1. The existing literature develops first-order fixed-$K$ asymptotic theory. In contrast, we study DML when $K_n$ may increase with $n$, and use this framework both to distinguish DML1 from DML2 and to study the choice of $K_n$.

This paper also relates to the literature on double-robust and semiparametric estimators,
including \cite{robins1994estimation}, \cite{robins1995semiparametric}, \cite{scharfstein1999adjusting}, \cite{farrell2015robust}, \cite{sant2020doubly}, \cite{chang2020double}, \cite{callaway2021difference}, \cite{rothe2019properties}, and \cite{singh2024double}. Most closely related is \citet{rothe2019properties}, who study higher-order
properties of a leave-one-out double-robust estimator in a missing-data setting. We complement that work by studying how the number of folds affects DML2 through a second-order asymptotic approximation, finding that the choice $K_n=n$, associated with the leave-one-out estimator, is asymptotically optimal for DML2. More broadly, our paper is related to the semiparametric literature on higher-order asymptotics and refined approximations, including \citet{bickel1982adaptive}, \citet{robinson1988root}, \citet{newey1990efficient}, \citet{andrews1994asymptotics}, \citet{newey1994large}, \citet{newey1994asymptotic}, \citet{linton1995second},  \citet{bickel2003nonparametric}, \citet{newey2004higher}, \citet{cattaneo2018kernel}, \citet{bugni2021testing}, and \citet{cai2022linear}. The existing semiparametric literature provides conditions to study the asymptotic properties of plug-in and leave-one-out estimators (DML2 with $K_n=n$). We instead consider conditions for studying how to select $K_n$ for DML2 through second-order asymptotic approximations.

The remainder of the paper is organized as follows. Section~\ref{sec:setup} presents the setup and reviews the existing fixed-$K$ asymptotic theory. Section~\ref{sec:result1} develops first-order asymptotic theory for DML1 and DML2 under the growing-$K_n$ framework, explaining why inference based on DML1 can fail for large values of $K_n$ and when DML2 remains valid for any $K_n$. Section~\ref{sec:result2} derives a second-order asymptotic approximation for DML2 and uses it to study the choice of $K_n$. Section~\ref{sec:recommendations} summarizes the implications for practice. Section~\ref{sec:simulations} presents simulations. Section~\ref{sec:concluding-remarks} concludes. Appendices~\ref{sec:appendix_main_proofs} and~\ref{appendix:aux-results} contain proofs of the main results and present auxiliary results, respectively; Appendices~\ref{sec:appendix_examples}--\ref{appendix:local-polynomials} contain additional material.

\textit{Notation:}  $[L] = \{1,\ldots,L\}$, $\| \cdot \|$ denotes the $L_2$-operator norm for matrices and the $L_2$-norm for vectors, $\| \cdot \|_{\infty}$ denotes the element-wise supremum norm for matrices and vectors, $\partial_\eta m$ denotes the matrix of partial derivatives of $m$ with respect to $\eta$, and $\mathbb{I}_d$ is the $d \times d$ identity matrix. $a_n \asymp b_n $ denotes that $a_n$ is of the same order as $b_n$.

\section{Setup and Previous Results}\label{sec:setup}
 
The parameter of interest is $\theta_0 \in \Theta \subseteq \mathbf{R}^d$ and satisfies the following moment condition:
\begin{equation}\label{eq:moment_for_theta}
    E[m(W,\theta_0,\eta_0(X))] = 0_{d \times 1}~,
\end{equation}
where  $m: \mathcal{W}  \times \Theta  \times \mathcal{E} \to \mathbf{R}^d$ is a known moment function, $W \in \mathcal{W}   \subseteq \mathbf{R}^{d_w}$ is a random vector with distribution $F_0$, $X \in \mathcal{X} \subseteq \mathbf{R}^{d_x}$ is a sub-vector of $W$, and $\eta_0(X)$ take values in $\mathcal{E} \subseteq \mathbf{R}^p$, which is a bounded and convex set. The nuisance function $\eta_0: \mathcal{X}  \to \mathcal{E} $ is an unknown function of the covariates $X$ based on transformations of conditional expectations.

This paper considers moment functions $m$ that are linear in the parameter of interest: 
\begin{equation}\label{eq:moment_funct}
    m(W,\theta,\eta) = \psi^b(W,\eta) - \psi^a(W,\eta) \theta~,
\end{equation} 
where the moment function $m$ and function $\psi^a$ satisfy conditions specified in Assumption~\ref{asm:EM} in Section \ref{sec:result1}, which includes the identification condition, $E[\psi^a(W,\eta_0(X))] \in \mathbf{R}^{d \times d}$ is invertible, and guarantees a Neyman orthogonality condition,
\begin{equation*}
    E\left[  \partial_\eta m(W,\theta_0,\eta_0(X))    \mid X  \right] = 0~, \quad a.e.
\end{equation*} 
Throughout, $\partial_\eta m(W,\theta_0,\eta_0(X))$ denotes the derivative of $m(W,\theta_0,\eta)$ with respect to $\eta$, evaluated at $\eta = \eta_0(X)$.

A wide range of parameters of interest can be identified through moment conditions such as \eqref{eq:moment_for_theta} using a moment function like \eqref{eq:moment_funct}.  
Examples of $\theta_0$ include the average treatment effect (Example \ref{example_ate}), the average treatment effect on the treated in difference-in-differences designs (Example \ref{example_DID}), and the local average treatment effect (Example \ref{example_LATE}), among others. All these examples are in Appendix \ref{sec:appendix_examples} and additional examples appear in \cite{ahrens2025introduction}.

Consider the goal of estimating $\theta_0$ using a random sample $\{ W_i : 1 \le i \le n\}$ drawn from the distribution $F_0$. Since the parameter $\theta_0$  based on \eqref{eq:moment_for_theta} and \eqref{eq:moment_funct} can be identified as follows,
\begin{equation}\label{eq:theta0}
    \theta_0 = E[\psi^a(W,\eta_0(X))]^{-1}\ E[\psi^b(W,\eta_0(X))] ~,
\end{equation}
%\
an ideal estimator for $\theta_0$ is defined by replacing the expected values in \eqref{eq:theta0} with sample analogs. That is, the oracle estimator
\begin{equation}\label{eq:theta_hat_oracle}
    \hat{\theta}_n^* = \left( n^{-1}\sum_{i=1}^n \psi^a(W_i,\eta_i)\right)^{-1} \left(n^{-1}\sum_{i=1}^n \psi^b(W_i,\eta_i)\right)~,
\end{equation}
where $\eta_i = \eta_0(X_i)$ is the value of the nuisance function $\eta_0$ for observation $i$. However, the values of the $\eta_i$'s are unknown. As a result, the oracle estimator $\hat{\theta}_n^*$ is infeasible. For this reason, it is common to calculate first estimates $\hat{\eta}_i$ of $\eta_i$ that can be used later to obtain an estimator for $\theta_0$. Remark \ref{rem:plug-in} discusses the \textit{plug-in} estimator used in the semiparametric literature. In what follows, we explain how DML estimates first $\eta_0$ and then $\theta_0$. 

DML calculates the estimates $ \hat{\eta}_i$ of $\eta_i$ using a \emph{cross-fitting} procedure, which is a form of sample-splitting. The procedure has two steps and we assume that $n$ can be divided by $K$:\footnote{When $n$ is not divisible by $K$, with $ 2\le K \le n$, the number of observations in some folds will be $\lfloor n/K \rfloor$ while in others $\lfloor n/K \rfloor + 1$, where $\lfloor n/K  \rfloor$ is the greatest integer less than or equal to $n/K$.}
\begin{enumerate}
    \item[(i)] \textit{Sample splitting}: Randomly split the indices into $K$ equal-sized folds $\mathcal{I}_k$, i.e., $\cup_{k=1}^K \mathcal{I}_k = [n]$. The number of observations in fold $\mathcal{I}_k$ is denoted by $n_k = n/K$.

    \item[(ii)] \textit{Nuisance Function Estimates}: For each $i \in \mathcal{I}_k$, the estimates $ \hat{\eta}_i $ of $\eta_i$ are defined by $  \hat{\eta}_i = \hat{\eta}_k(X_i)$, where $\hat{\eta}_k(\cdot)$ is an estimator of $\eta_0(\cdot)$ using $\{ W_i : i \notin \mathcal{I}_k\}$, which is all the data except the ones with indices on $\mathcal{I}_k$. We then repeat the process for all $k \in [K]$. 
\end{enumerate}

Both DML estimators use the same estimates $\hat{\eta}_i$, but they differ in how they combine information across the different folds defined above. We explain this next.  

\begin{definition}[DML1]
    The DML1 estimator first calculates preliminary estimators $\tilde{\theta}_k$ by solving the moment condition \eqref{eq:moment_for_theta} within each fold $\mathcal{I}_k$ using the estimates $\hat{\eta}_i$, 
$$ \tilde{\theta}_k \quad \text{solves} \quad n_k^{-1} \sum_{i \in \mathcal{I}_k} m(W_i, \theta, \hat{\eta}_i) = 0 ~,$$
it then combines the information across the folds by averaging the $\tilde{\theta}_k$'s to obtain the proposed estimator for  $\theta_0$,
\begin{equation}\label{eq:theta_DML1}
    \hat{\theta}_{n,K}^{(1)} = K^{-1} \sum_{k=1}^K \tilde{\theta}_{k}~.
\end{equation} 
\end{definition}

Explicit expressions for $\tilde{\theta}_{k}$ is obtained by using \eqref{eq:moment_funct},
\begin{equation*}
     \tilde{\theta}_{k} = \left( n_k^{-1} \sum_{i \in \mathcal{I}_k} \psi^a(W_i,\hat{\eta}_i) \right)^{-1} \left( n_k^{-1}  \sum_{i \in \mathcal{I}_k} \psi^b(W_i,\hat{\eta}_i)\right)~, \quad \forall~ k \in [K]~.
\end{equation*} 
Note that $\tilde{\theta}_{k}$ is similar to \eqref{eq:theta_hat_oracle} but using the observations in $\mathcal{I}_k$ and estimates $\hat{\eta}_i$ instead of $\eta_i$. 
 
\begin{definition}[DML2]
    The DML2 estimator first combines the information across the folds $\mathcal{I}_k$ by averaging the sample analog of moment conditions like \eqref{eq:moment_for_theta} using the estimates $\hat{\eta}_i$, and then estimates $\theta_0$ by solving the average of moment conditions,
$$ \hat{\theta}_{n,K}^{(2)}  \quad \text{solves} \quad K^{-1} \sum_{k=1}^K \left( n_k^{-1} \sum_{i \in \mathcal{I}_k} m(W_i, \theta, \hat{\eta}_i) \right) = 0 ~.$$
\end{definition} 
An explicit expression for $\hat{\theta}_{n,K}^{(2)}$ is obtained by using \eqref{eq:moment_funct},%
\begin{equation}\label{eq:theta_DML2}
    \hat{\theta}_{n,K}^{(2)} = \left( n^{-1}\sum_{i=1}^n \psi^a(W_i,\hat{\eta}_i) \right)^{-1} \left( n^{-1}\sum_{i=1}^n \psi^b(W_i,\hat{\eta}_i)\right)~.
\end{equation}
Note that $\hat{\theta}_{n,K}^{(2)}$ is similar to \eqref{eq:theta_hat_oracle} but using the estimates $\hat{\eta}_i$ instead of $\eta_i$.

\begin{remark}\label{rem:dml1_dml2} 
DML1 and DML2 are numerically identical  when $\psi^a(W_i,\hat{\eta}_i)$ is a constant function. This is the case for the ATE (Example \ref{example_ate}). In contrast, if $\psi^a(W_i,\hat{\eta}_i)$ is not a constant, then DML1 and DML2 provide different estimates. This is the case for the ATT-DID (Example \ref{example_DID}) and LATE (Example \ref{example_LATE}), among other econometric models.
\end{remark}

\begin{remark}\label{rem:DML2>DML1}
\citet[Remark 3.1]{chernozhukov2018double} recommend using DML2 over DML1 based on simulation evidence. They note that for some econometric models the bias and MSE of DML1 and DML2 are similar, whereas for others the bias and MSE of DML2 are lower than those of DML1. The literature does not provide a theoretical explanation for these findings. Section~\ref{sec:result1} offers such an explanation, and shows that the difference between the two variants concerns the validity of inference, not only bias and MSE.  
\end{remark}

\begin{remark}\label{rem:cross-fitting}
Simulation evidence shows that increasing $K$ reduces DML2's finite-sample bias and MSE \citep{ahrens2024ddml,ahrens2024model}. An intuitive explanation is that increasing $K$ improves the accuracy of the first-step estimators by allowing them to use more data (e.g., the first-step estimators use $50\%$, $80\%$, and $90\%$ of the data when $K$ is $2$, $5$, and $10$, respectively). However, increasing $K$ also raises the dependence among the first-step estimators by increasing the number of observations they have in common (e.g., any two first-step estimators share $0\%$, $60\%$, and $80\%$ of the data when $K$ is $2$, $5$, and $10$, respectively). Therefore, it is not immediate whether large values of $K$ improve the estimation accuracy of $\theta_0$. Section~\ref{sec:result2} provides theoretical guidance on the choice of $K_n$ for DML2. 
\end{remark}

\begin{remark}\label{rem:plug-in}
Plug-in estimators of $\theta_0$ have been studied in the semiparametric literature, but their analysis typically requires strong conditions on the first-step estimators, such as Donsker-type restrictions, to derive standard properties, including asymptotic normality and parametric convergence rate. These conditions control the empirical-process terms that arise when the same observations are used for nuisance estimation and moment evaluation. Here, by a plug-in estimator, we mean the estimator of $\theta_0$ defined in \eqref{eq:theta_hat_oracle} with $\eta_i$ replaced by an estimate  $\hat{\eta}_i$, where $\hat{\eta}_i = \hat{\eta}(X_i)$ and $\hat{\eta}$ is obtained using the full sample. In contrast, DML relies on weaker first-step conditions, such as suitable $L_2$ convergence rates faster than $n^{-1/4}$, to obtain standard first-order properties. By combining Neyman orthogonality with cross-fitting, DML avoids the dependence effect that arises in plug-in estimators, which is known as the \textit{own-observation bias}; see \citet{newey2018cross}.
\end{remark}

\subsection{Previous Results}
In the fixed-$K$ asymptotic framework, \citet{chernozhukov2018double} showed under regularity conditions that DML1 and DML2 have the same limiting distribution,
\begin{equation}\label{eq:asymptotic_distr_fixed_K}
    \sqrt{n} \left( \hat{\theta}_{n,K}^{(j)} - \theta_0 \right) \overset{d}{\to} N(0,\Sigma)~,\quad \text{for} \quad j= 1, 2,
\end{equation}
where the variance of the limiting distribution is given by
\begin{equation}\label{eq:asymptotic-variance}
    \Sigma =  G^{-1}E \left[m(W,\theta_0,\eta_0(X))m(W,\theta_0,\eta_0(X))^\intercal \right] \left(G ^{-1}\right)^\intercal  ~,
\end{equation}
where $G = E\left[\psi^a(W,\eta_0(X))\right] \in \mathbf{R}^{d \times d}$. Note that $\Sigma$  depends only on the moment function $m$, the nuisance function $\eta_0$, and the data distribution $F_0$. Therefore, the fixed-$K$ asymptotic theory does not provide guidance for choosing between DML1 and DML2, or for selecting~$K$. 

The core idea of DML is that estimating  $\theta_0$ with DML1 or DML2 is as accurate as if the true $\eta_0$ had been used. Formally, both DML1 and DML2 are asymptotically equivalent to the oracle estimator $\hat{\theta}_n^*$, 
     \begin{equation}\label{eq:first-order-eq-fixed-K}
         \sqrt{n} \left( \hat{\theta}_{n,K}^{(j)} - \hat{\theta}_{n}^* \right) \overset{p}{\to} 0~,\quad j=1,2~.
     \end{equation}

To investigate the simulation findings of Remarks \ref{rem:DML2>DML1} and \ref{rem:cross-fitting}, we propose a framework in which the number of folds is allowed to depend on $n$; we denote it $K_n$ henceforth and refer to this setup as the growing-$K_n$ framework. Sections~\ref{sec:result1} and~\ref{sec:result2} develop first-order and second-order asymptotic theory for DML under this framework, respectively.

% In Section~\ref{sec:result1}, we show that the bias, MSE, and inference properties of DML1 are sensitive to large $K_n$, whereas those of DML2 are not, establishing in particular that inference based on DML1 can be invalid for large $K_n$, whereas inference based on DML2 remains valid. Under an algorithmic-stability condition on the first-step estimators, we extend the asymptotic validity of DML2 to any $2 \le K_n \le n$. In Section~\ref{sec:result2}, we study how the choice of $K_n$ affects scalar DML2 estimators. Under stronger conditions on the first-step estimators, we show that  the performance of DML2, measured by second-order bias and MSE, improves with more folds but with diminishing returns.
 
\section{First-Order Asymptotic Theory for DML1 and DML2 when $K$ increases}\label{sec:result1} 

We first explain when and why standard inference based on DML1 can fail while inference based on DML2 remains valid under the growing-$K_n$ framework (Theorem~\ref{theorem:sqrt-Kn}). We then explain why DML1 is sensitive to large $K_n$ in Section~\ref{sec:DML1-sensitive} and present, in Section~\ref{sec:DML2-valid}, the algorithmic-stability condition under which DML2 remains valid for any $K_n$.

Let $\Omega=E[m(W,\theta_0,\eta_0(X)) ~m(W,\theta_0,\eta_0(X))^\intercal ] \in \mathbf{R}^{d \times d}$. Recall $G =E[\psi^a(W,\eta_0(X))]$ and $\eta_i = \eta_0(X_i)$. We write $\psi^a(W,\eta) = [ \psi^a_{t,s}(W,\eta)]_{t,s}$ and $m(W,\theta,\eta) = [ m_t(W,\theta,\eta)]_{t}$. Let $C_G$, $C_1$, and $C_2$ be positive constants. The next assumption restricts the class of econometric models through moment, smoothness, identification, and orthogonality conditions on the moment function $m$ and function $\psi^a$ . 

\begin{assumption}[Econometric models]\label{asm:EM} 
\hspace{0.5cm}  
\begin{itemize}
    \item[(i)] $G$ and $\Omega$  are nonsingular and $\|G^{-1}\| < C_G$.
    
    \item[(ii)] $E[|m_{t}(W_i,\theta_0,\eta_i)|^4]<C_1$ and $E[|\psi^a_{t,s}(W_i,\eta_i)|^4] < C_1$ for all $t,s \in [d]$.   
    
    \item[(iii)] $m(W,\theta,\eta)$  is twice continuously differentiable with respect to $\eta \in \mathcal{E} $ and satisfies  $E[\partial_\eta m_t(W_i,\theta_0,\eta_i) | X_i] = 0$, $ \| E[ (\partial_\eta m_{t}(W_i,\theta_0,\eta_i))(\partial_\eta m_{t}(W_i,\theta_0,\eta_i))^\intercal | X_i] \|_\infty < C_2$, $\sup_{\eta \in \mathcal{E}} \| \partial_\eta^2 m_t (W_i,\theta_0,\eta)\|_\infty \le C_2$,  and $\| E[ (\partial_\eta \psi^a_{t,s}(W_i,\eta_i))(\partial_\eta \psi^a_{t,s}(W_i,\eta_i))^\intercal \mid X_i] \|_\infty$ $< C_2$ a.e. for all $t, s \in [d]$.
      
    \item[(iv)] $\psi^a(W,\eta)$ is twice continuously differentiable with respect to $\eta \in \mathcal{E} $ and satisfies  $E[\partial_\eta \psi^a_{t,s}(W_i,\eta_i) \mid X_i] =  0$   and  $\sup_{\eta \in \mathcal{E}} \| \partial_\eta^2 \psi^a_{t,s} (W_i,\eta)\|_\infty \le C_2$ a.e. for all $t,s \in [d]$. 
\end{itemize} 
\end{assumption}

Parts (i) and (ii) of Assumption \ref{asm:EM} are standard conditions that guarantee identification of the parameter of interest and stochastic expansions for the oracle estimator, similar to \citet[Assumptions 2 and 3]{newey2004higher}. Part (iii) presents a Neyman-orthogonality condition, $E[\partial_\eta m_t(W_i,\theta_0,\eta_i) | X_i] =  0$, involving standard partial derivatives as in \cite{belloni2017program} and \cite{farrell2025deep} rather than functional derivatives as in \cite{chernozhukov2018double}. In general, conditions of this type are necessary for feasible estimators to be asymptotically equivalent to the infeasible estimators that use the true nuisance functions; see, for instance, \citet[Eq.~(2.12)]{andrews1994asymptotics}, where an asymptotic orthogonality condition of this form is shown to be necessary and sufficient under stochastic equicontinuity. Under certain conditions, it is possible to transform a moment function into one that satisfies a Neyman-orthogonality condition; see Remark \ref{rem:orthogonal-moment} for existing methods. Part (iii)  also includes standard conditions ensuring that nonlinear effects of first-step estimation error are negligible when they are sufficiently accurate (e.g., when their $L_2$-convergence rates are faster than $n^{-1/4}$). Finally, part (iv) of Assumption \ref{asm:EM} implies that we can construct a DML estimator for each component of $G = E[\psi^a(W_i,\eta_i)]$. It holds automatically when $\psi^a$ does not depend on $\eta$. We use part (iv) for the first-order asymptotic analysis of DML1, and for the second-order asymptotic analysis of DML2.

Assumption \ref{asm:EM} holds in many common econometric models studied in the literature, including the examples in Appendix \ref{sec:appendix_examples} and \citet[Appendix C]{ahrens2025introduction}. However, it excludes settings in which the moment function is not differentiable in $\eta$; see Remark \ref{rem:non-smooth-m} for examples of such nonsmooth models where DML estimators have been proposed.

 Let  $N = (1-\tfrac{1}{K_n})n$ be the number of observations in the sample $\{W_i : i \notin \mathcal{I}_k\}$ used by $\hat{\eta}_k$ to estimate $\eta_0$. Note that $N \asymp n$ for any $2 \le K_n \le n$.   Let $\tau_N$ be a sequence of positive numbers converging to zero.  We now present conditions on the first-step estimators that hold for any $k \in [K_n]$, and for $X$ independent of $\{W_i:1\leq i \leq n\}$:

\begin{assumption}[$L_2$-convergence rate]\label{asm:L2rate}
 $E\left[\| \hat{\eta}_k(X)-\eta_0(X)\|^2 \right]^{1/2} \le N^{-1/4} \tau_{N}$~. 
\end{assumption}

Assumption~\ref{asm:L2rate} is an out-of-sample $L_2$ prediction-error rate condition. It holds for several first-step estimators considered in the DML literature. For instance, it holds for deep neural networks as in \citet{Schmidt-Hieber2020} and \citet{farrell2021deep}, and, under a sparsity condition, for LASSO and related penalized estimators of linear models \citep{tibshirani1996regression,van2008high,belloni2011square,montiel2026distributionally}. Local-polynomial and series estimators \citep{newey1997convergence, belloni2015some, chen2007large} satisfy it after appropriate trimming to ensure well-conditioned Gram matrices. It also holds for an honest version of random forests \citep[Theorem 9.4.3]{chernozhukov2024applied}; see \citet{chi2022asymptotic} for convergence rates of high-dimensional random forests.
 
  As is common in the DML literature, a Neyman orthogonality condition on the moment function (Assumption \ref{asm:EM}) and sufficiently accurate first-step estimators (Assumption \ref{asm:L2rate}) are enough to derive the limiting distribution of DML estimators. 
 The next theorem presents the limiting distributions of DML1 and DML2 under the growing-$K_n$ framework. 

\begin{theorem}\label{theorem:sqrt-Kn}
    Let Assumptions \ref{asm:EM} and \ref{asm:L2rate} hold and let $K_n$ be such that $2 \le K_n \le n$ and $K_n/\sqrt{n} \to c \in [0,\infty)$ as $n \to \infty$. Then, 
    $$ \sqrt{n} \left( \hat{\theta}_{n,K_n}^{(1)} - \theta_0 \right) \overset{d}{\to} N(c \Lambda ,\Sigma)$$
    and
    $$ \sqrt{n} \left( \hat{\theta}_{n,K_n}^{(2)} - \theta_0 \right) \overset{d}{\to} N(0,\Sigma)~,$$
    where $ \hat{\theta}_{n,K_n}^{(1)}$, $ \hat{\theta}_{n,K_n}^{(2)}$, and $\Sigma$ are defined in  \eqref{eq:theta_DML1}, \eqref{eq:theta_DML2}, and \eqref{eq:asymptotic-variance}, respectively, and
    \begin{equation}\label{eq:Lambda}
        \Lambda = -G^{-1} E\left[ \psi^a(W,\eta_0(X)) ~G^{-1}~ m(W,\theta_0,\eta_0(X)) \right]~.
    \end{equation}
\end{theorem}

Theorem \ref{theorem:sqrt-Kn} provides an asymptotic result that explains the discrepancy found in simulations between DML1 and DML2, which motivated the literature’s recommendation of DML2 (Remark~\ref{rem:DML2>DML1}). This theorem now provides the theoretical explanation. It shows that DML2 is asymptotically better than DML1 in terms of bias and MSE, and in terms of the asymptotic coverage of the associated confidence intervals, when $c >0$ and $\Lambda \neq 0$; otherwise, both share the same limiting distribution. Our explanation through $\Lambda$ emerges under the growing-$K_n$ framework, providing insights not captured by the existing fixed-$K$ asymptotic theory or simulation-based evidence.

Within the growing-$K_n$ framework, the distinction between DML1 and DML2 relies only on the model-dependent quantity $\Lambda$ rather than on the first-step estimator $\hat{\eta}$. Remark~\ref{rem:diagnosing-lambda} discusses how $\Lambda$ can be evaluated model by model.

When $\Lambda \neq 0$, DML1 becomes increasingly sensitive to large values of $K_n$ regarding bias, MSE, and coverage probability of its associated confidence interval. In contrast, the limiting distribution of DML2 is unaffected by the choice of $K_n$. Motivated by the theorem, when we replace $c$ by $K_n/\sqrt{n}$, the finite-sample distribution of DML1, $\sqrt{n} \left( \hat{\theta}_{n,K_n}^{(1)} - \theta_0 \right)$, can be approximated by $ N(  (K_n/\sqrt{n})  \Lambda ,\Sigma)$,  which is not centered at the origin and whose centering error increases with $K_n$ when $\Lambda \neq 0$. This implies that the standard recommended DML1 confidence interval \citep{chernozhukov2018double} is not valid and its coverage deteriorates as $K_n$ increases, explaining  the simulation evidence in Section~\ref{sec:simulations}, where the coverage probability of DML1 with $30$ folds for the LATE is below $87\%$ for a nominal $95\%$ level.    
 
\begin{remark}[Diagnosing $\Lambda$]\label{rem:diagnosing-lambda} 
Whether $\Lambda$ is zero can be diagnosed on a model-by-model basis. In particular, the form of $\Lambda$ is determined by $(\psi^a,m,\eta_0,F_0)$ and can often be evaluated analytically. For the examples in Appendix~\ref{sec:appendix_examples} and \citet[Appendix C]{ahrens2025introduction}, this calculation yields $\Lambda=0$ for the ATE, partially linear model, and ATT-DID, whereas $\Lambda$ is typically nonzero for the LATE and the partially linear IV model. Thus, a researcher using DML1 can assess whether large values of $K_n$ may invalidate inference in the model under consideration. Using DML2 makes this diagnostic unnecessary; see Section~\ref{sec:recommendations}.
\end{remark}

\begin{remark}\label{rem:orthogonal-moment}
We can obtain moment functions satisfying Neyman orthogonality by adding adjustment terms to the original moment functions. The adjustment terms are constructed using first-step influence functions, as developed in \citet{newey1994asymptotic}, \citet{hahn2013asymptotic}, \citet{ichimura2022influence}, and \citet{farrell2025deep}, among others.
Since the analytical construction can be tedious, the DML literature has focused on automatic construction of orthogonal moments: procedures that take the original moment function as input and automatically return a DML2 estimate. Examples include \citet{chernozhukov2022locally}, \citet{escanciano2023automatic}, and \citet{arganaraz2025automatic}.
\end{remark}

% \begin{remark}\label{rem:n-o-c}  
% Neyman orthogonality of the moment function is necessary for a feasible
% estimator to be first-order equivalent to its oracle version, as in \eqref{eq:first-order-eq-fixed-K}. To see this, suppose $\psi^a(W,\eta) = 1$, let $\psi^b(W,\eta)$ be a linear function of $\eta$ and let $\eta_0$ be an unknown finite-dimensional nuisance parameter. Define the feasible estimator $\hat{\theta}_n = n^{-1} \sum_{i=1}^n \psi^b(W_i,\hat{\eta})$ and its oracle version $\hat{\theta}^*_n = n^{-1} \sum_{i=1}^n \psi^b(W_i,{\eta}_0)$, where $\hat{\eta}$ estimates $\eta_0$ and satisfies $n^{1/2}(\hat{\eta} - \eta_0) \overset{d}{\to} N(0,V_\eta)$ with $V_\eta$ invertible. A first-order expansion shows
% %
% $$ n^{1/2}(\hat{\theta}_n - \hat{\theta}_n^*) =  n^{1/2} (\hat{\eta} - \eta_0)^\intercal E[ \partial_\eta m(W_i, \theta_0, \eta_0)] + o_p(1)~.$$
% %
% Therefore, the right-hand side is $o_p(1)$ if and only if
% $E \left[\partial_\eta m(W_i,\theta_0,\eta_0)\right]=0$. Hence
% first-order asymptotic equivalence holds in this example exactly when the Neyman
% orthogonality condition of Assumption~\ref{asm:EM}(iii) holds. See also \citet[Eq.~(2.12)]{andrews1994asymptotics}.
% \end{remark} 

\begin{remark}\label{rem:non-smooth-m}
Beyond smooth moment conditions, DML methods have been successfully applied to nonsmooth econometric models. \citet{chernozhukov2022locally} develop DML estimators for quantile regression coefficients, while \citet{semenova2023debiased} propose methods for support functions in set-identified models. Related approaches have been used to study algorithmic fairness \citep{liu2024inference,liu:moli:vel2026} and to conduct inference on welfare under optimal treatment rules \citep{park2024debiased,whitehouse2025inference}. See also \citet{kitagawa2025leave} for the use of DML ideas in policy targeting. 
\end{remark}  

\subsection{Why DML1 is Sensitive to Large $K_n$}\label{sec:DML1-sensitive}

We now provide a high-level explanation for DML1's sensitivity to large $K_n$. The main reason is that an oracle version of DML1 is already sensitive to large $K_n$ when the model-dependent quantity $\Lambda \neq 0$. Therefore, the discussion that we provided for smooth moment functions continues to apply to nonsmooth econometric models  whenever DML1 is asymptotically equivalent to its oracle version, i.e., when \eqref{eq:DML1-oracle-equivalence} below holds.

The oracle version of DML1 is defined as DML1 but using  $\eta_i$'s instead of $\hat{\eta}_i$, that is, assuming perfect knowledge of $\eta_0$,
\begin{equation}\label{eq:theta_DML1-oracle}
    \hat{\theta}_{n,K}^{*,(1)} = K^{-1} \sum_{k=1}^K \tilde{\theta}_{k}^*~,
\end{equation} 
where $\tilde{\theta}_{k}^*$'s are preliminary estimators solving the moment condition \eqref{eq:moment_for_theta} within each fold $\mathcal{I}_k$ and using the true values of $\eta_i$,  that is $\tilde{\theta}_k^*$ solves $n_k^{-1} \sum_{i \in \mathcal{I}_k} m(W_i, \theta, {\eta}_i) = 0$.  

Standard arguments \citep{newey2004higher} show that the second-order asymptotic bias of $\tilde{\theta}_k^*$ is $\Lambda K_n/n$ because $\tilde{\theta}_k^*$ is based on $n_k = n/K_n$ observations. The same bias carries over to the oracle DML1 estimator because it is the average of these preliminary estimators. Hence, the second-order asymptotic bias of $\sqrt{n}\left( \hat{\theta}_{n,K_n}^{*,(1)} - \theta_0\right)$ is $\Lambda K_n/\sqrt{n}$. Consequently, when $K_n$ grows proportionally to $\sqrt{n}$ and $\Lambda \neq 0$, this bias becomes first-order, preventing the limiting distribution of the oracle DML1 estimator from being correctly centered. The following theorem formalizes this result.

\begin{theorem}\label{theorem:DML1-oracle}
    Let Assumption \ref{asm:EM}~(i)--(ii) hold and let $K_n$ be such that $2 \le K_n \le n$ and $K_n/\sqrt{n} \to c \in [0,\infty)$ as $n \to \infty$. Then,
    $$ \sqrt{n} \left( \hat{\theta}_{n,K_n}^{*,(1)} - \theta_0 \right) \overset{d}{\to} N(c \Lambda ,\Sigma)~,$$
    where $\Sigma$ and $\Lambda$ are defined in  \eqref{eq:asymptotic-variance} and \eqref{eq:Lambda}, respectively. 
\end{theorem}

Theorem~\ref{theorem:DML1-oracle} holds under mild regularity conditions that do not require the moment function $m$ to be smooth in $\eta$, and therefore covers a large class of econometric models, including nonsmooth ones. 

Theorem~\ref{theorem:DML1-oracle} shows that the oracle DML1 has the same limiting distribution as DML1, which we derive in Theorem~\ref{theorem:sqrt-Kn}. This is because the proof of Theorem~\ref{theorem:sqrt-Kn} establishes that DML1 and the oracle DML1 are asymptotically equivalent under Assumptions~\ref{asm:EM} and~\ref{asm:L2rate}:
\begin{equation}\label{eq:DML1-oracle-equivalence}
    \sqrt{n} \left( \hat{\theta}_{n,K_n}^{(1)} -  \hat{\theta}_{n,K_n}^{*,(1)}  \right) \overset{p}{\to} 0 \quad \text{as} \quad n \to \infty~.
\end{equation}
Part~(iv) of Assumption~\ref{asm:EM} is what guarantees that \eqref{eq:DML1-oracle-equivalence} holds.
In general, the equivalence in \eqref{eq:DML1-oracle-equivalence} together with Assumption~\ref{asm:EM}(i)--(ii) is sufficient to conclude that DML1 is sensitive to large $K_n$ when $\Lambda \neq 0$. Part~(i) of Lemma~\ref{lemma:sqrt-Kn-DML1-DML2} shows that Assumptions~\ref{asm:EM} and~\ref{asm:L2rate} imply \eqref{eq:DML1-oracle-equivalence}.

\subsection{When DML2 remains valid for any $K_n$}\label{sec:DML2-valid}

This section introduces the algorithmic-stability condition (Assumption~\ref{asm:AS}) under which DML2  remains valid for any $2 \le K_n \leq n$. This condition controls the sensitivity of the first-step estimator to the replacement of a single observation by another i.i.d.\ draw and allows us to handle the dependence induced by large values of $K_n$; see Remark \ref{rem:cross-fitting}.

Theorem \ref{theorem:sqrt-Kn} shows that DML2 is valid when $K_n$ grows proportionally to $\sqrt{n}$. The next assumption allows us to extend this result to any $2 \le K_n \le n$. Let  $\hat{\eta}_k^{\ell}(\cdot)$ be identical to $\hat{\eta}_k(\cdot)$ except that it replaces $W_\ell$ with an independent copy $\widetilde{W}_\ell $ drawn from $F_0$, so that $\hat{\eta}_k^{\ell}-\hat{\eta}_k$ is the change in the first-step estimator from resampling one training observation. Recall $N = (1-\tfrac{1}{K_n})n$ is the sample size used by first-step estimators, $\tau_{N} = o(1)$, and $N \asymp n$.  For any $k\in[K_n]$, and for $X$ independent of $\{W_i:1\leq i\leq n\}$:

\begin{assumption}[Algorithmic Stability]\label{asm:AS}  
     $ \max_{\ell \notin \mathcal{I}_k} E[\| \hat{\eta}_k(X)-\hat{\eta}_k^{\ell}(X)\|^2]^{1/2} \le  N^{-1/2} \tau_{N}$.
\end{assumption} 
 
Assumption~\ref{asm:AS} measures the stability of the first-step estimators under replacement of a single observation, in out-of-sample $L_2$-norm. It is a sufficient condition that we use to control the dependence among the prediction errors of the moment function due to first-step estimation errors, $\{m(W_i,\theta_0,\hat{\eta}_i)-m(W_i,\theta_0,\eta_i): 1 \le i \le n \}$. A stronger version of this assumption appears in  \citet[Corollary 4]{chen2022debiased}, which is satisfied by bagged estimators; see also \citet[Assumption 4]{Chernozhukov17022026}. Conditions controlling the sensitivity of learning algorithms to the replacement of a single training observation have received considerable attention in the machine-learning literature on algorithmic stability and generalization; see \cite{bousquet2002stability} and \cite{hardt2016train}. Local-polynomial and series estimators satisfy this condition after an appropriate trimming, and LASSO satisfies it under strong sparsity. We do not know whether it holds for deep neural networks; characterizing when it holds remains an open question.

The next theorem shows that DML2  is valid for any $2 \le K_n \le n$ when the first-step estimators satisfy the algorithmic-stability condition. 

\begin{theorem}\label{theorem:DML2-Kn}
    Let  Assumptions~\ref{asm:EM}(i)--(iii), \ref{asm:L2rate}, and \ref{asm:AS} hold and let $K_n$ be  such that $2 \le K_n \le n$. Then,
    $$ \sqrt{n} \left( \hat{\theta}_{n,K_n}^{(2)} - \theta_0 \right) \overset{d}{\to} N(0 ,\Sigma)~,$$ 
    where $\hat{\theta}_{n,K_n}^{(2)}$ and $\Sigma$ are as in \eqref{eq:theta_DML2} and \eqref{eq:asymptotic-variance}, respectively.
\end{theorem}

The proof of Theorem \ref{theorem:DML2-Kn} relies on the asymptotic equivalence between DML2 and its oracle version, that is
\begin{equation}\label{eq:DML2-oracle-equivalence}
    \sqrt{n} \left( \hat{\theta}_{n,K_n}^{(2)} -  \hat{\theta}_{n,K_n}^{*,(2)}  \right) \overset{p}{\to} 0 \quad \text{as} \quad n \to \infty~,
\end{equation}
where the oracle version of DML2 is defined by
\begin{equation}\label{eq:theta_DML2-oracle}
        \hat{\theta}_{n,K_n}^{*,(2)} = \left( \frac{1}{n}\sum_{i=1}^n \psi^a(W_i,{\eta}_i) \right)^{-1} \left( \frac{1}{n}\sum_{i=1}^n \psi^b(W_i,{\eta}_i)\right)~.
\end{equation}
Note that the oracle version of DML2 is the same as the ideal estimator defined in \eqref{eq:theta_hat_oracle}. Therefore, the oracle version of DML2 does not depend on the choice of $K_n$ and its limiting distribution is exactly the same as in fixed-$K$ asymptotics. 
We show in part (ii) of Lemma \ref{lemma:sqrt-Kn-DML1-DML2} that Assumptions~\ref{asm:EM}(i)--(iii) and \ref{asm:L2rate} are sufficient to verify the asymptotic equivalence in \eqref{eq:DML2-oracle-equivalence} when $K_n $ grows proportionally to $\sqrt{n}$. To guarantee that \eqref{eq:DML2-oracle-equivalence} continues to hold for any $2 \le K_n \le n$, we additionally use Assumption \ref{asm:AS}. We formalize the result in Lemma \ref{lemma:DML2_Kn}. 

Theorem~\ref{theorem:DML2-Kn} shows that the standard asymptotic theory for DML2 extends well beyond the fixed-$K$ setting. In particular, DML2 remains valid with $K_n=n$, in which case it coincides with the leave-one-out estimator commonly used in semiparametric models; see, for example, \citet{robinson1988root}, \citet{linton1995second}, and \citet{rothe2019properties}. By contrast, DML1 cannot generally be used with $K_n=n$, because it may be inconsistent unless DML1 and DML2 coincide; see Remark~\ref{rem:dml1_dml2}.
 
The next corollary turns to inference, showing that the standard confidence intervals for DML2 presented in \citet[Corollary 3.1]{chernozhukov2018double} remain valid in the growing-$K_n$ framework. Let $z_{1-\alpha/2}$ denote the $(1-\alpha/2)$ quantile of the standard normal distribution and define $\hat{G}_n = n^{-1}\sum_{i=1}^n \psi^a(W_i,\hat{\eta}_i)$, $\hat{\Omega}_n = n^{-1}\sum_{i=1}^n \hat{m}_i \hat{m}_i^\intercal$ with $\hat{m}_i = m(W_i,\hat{\theta}^{(2)}_{n,K_n},\hat{\eta}_i)$, and $\hat{\Sigma}_n = \hat{G}_n^{-1}\hat{\Omega}_n(\hat{G}_n^{-1})^\intercal$.

\begin{corr}\label{corr:CI-DML2}
Suppose that either (i) Assumptions~\ref{asm:EM} and \ref{asm:L2rate} hold and $K_n/\sqrt{n} \to c \in [0,\infty)$, or (ii) Assumptions~\ref{asm:EM}, \ref{asm:L2rate}, and \ref{asm:AS} hold and $2 \le K_n \le n$. Then, $\hat{\Sigma}_n \overset{p}{\to} \Sigma$ and, for any $v \in \mathbf{R}^d$ with $v \neq 0$ and any $\alpha \in (0,1)$,
$$ P\left( v^\intercal\theta_0 \in \left[\, v^\intercal\hat{\theta}^{(2)}_{n,K_n} \pm z_{1-\alpha/2}\, \big(v^\intercal\hat{\Sigma}_n v / n\big)^{1/2} \right]\right) \to 1-\alpha~. $$
\end{corr}

Two features of Corollary~\ref{corr:CI-DML2} are worth noting. First, the consistency of $\hat{\Sigma}_n$ holds for any $2 \le K_n \le n$ under Assumptions~\ref{asm:EM} and \ref{asm:L2rate} alone; Assumption~\ref{asm:AS} enters only through the limiting distribution of $\hat{\theta}^{(2)}_{n,K_n}$. Second, in combination with Theorem~\ref{theorem:sqrt-Kn}, the corollary implies that the same interval centered at DML1 has asymptotic coverage below $1-\alpha$ whenever $c\Lambda \neq 0$, since $\hat{\Sigma}_n$ remains consistent while the center is asymptotically biased.

% \begin{remark}
%     One of the main benefits of using DML2 with  $ n$ folds is that it ensures replicability. DML2 with  $n$ folds is uniquely determined by the data and therefore eliminates the random-split variability that exists when  $2 \le K_n < n$, where different random splits yield different DML2 estimates. However, its implementation in practice may appear challenging due to the computational burden of estimating $n$ first-step estimators.  
% \end{remark} 

This section has shown that first-order asymptotic theory for DML2 under the growing-$K_n$ framework does not provide formal guidance for the choice of $K_n$: under the maintained conditions, all choices of $2 \le K_n \le n$ deliver asymptotically equivalent estimators. The next section complements this analysis with a second-order approximation that ranks these first-order equivalent choices, in the tradition of using higher-order approximations to compare asymptotically equivalent estimators \citep{rothenberg1984approximating, linton1995second, donald2001choosing, newey2004higher}.

\section{Second-Order Asymptotic Approximation for DML2 when $K_n$ increases }\label{sec:result2}

Section~\ref{sec:result1} shows that the literature's preference for DML2 has a theoretical basis. It also shows that, under the conditions of Theorem~\ref{theorem:sqrt-Kn} or~\ref{theorem:DML2-Kn}, DML2 estimators based on different values of $K_n$ share the same first-order asymptotic distribution. Any asymptotic guidance on the number of folds must therefore come from a higher-order analysis, which requires additional structure on the first-step estimators. This section accordingly imposes stronger conditions than Section~\ref{sec:result1}---and assumes $d=1$, so $\theta_0$ is scalar---by considering first-step estimators that admit a stochastic linear expansion (Assumption~\ref{asm:SLE}), such as local-polynomial estimators with MSE-rate-optimal bandwidths. We then derive a second-order asymptotic approximation and use it to analyze how $K_n$ affects the performance of DML2.

Our results show that  the performance of DML2, in terms of second-order asymptotic bias and MSE, improves with more folds but with diminishing returns. See Appendix \ref{sec:relative-efficiency-loss} for the cost of using a fixed number of folds instead of the benchmark $K_n = n$.

The next assumption introduces additional information on the class of first-step estimators through their estimation error, which is a stronger requirement than those imposed in Section~\ref{sec:result1}. 
Recall that $\hat{\eta}_k(x)$ is an estimator for $\eta_0(x)$ using  $\{W_j: j \notin \mathcal{I}_k\}$ with sample size $N=(1-\tfrac{1}{K_n})n$, and  $\tau_{N} = o(1)$. For simplicity we assume that $n$ can be divided by $K_n$ throughout this section. Let $c_\delta$, $c_b$,  $C_3$, and $C_4$ be positive constants.

\begin{assumption}[Stochastic Linear Expansion] \label{asm:SLE} 
    There are functions $\delta_{N}: \mathcal{W} \times \mathcal{X} \to \mathbf{R}^p$ and $b_{N}: \mathcal{X} \times \mathcal{X} \to \mathbf{R}^p$, and $\varphi \in (1/4,1/2)$, such that for $X$ independent of $\{W_j: j \notin \mathcal{I}_k\}$:
    \begin{enumerate} 
      \item[(i)] $E[\|\hat{R}_k(X)\|^2]^{1/2} \le C_3 N^{-2\varphi}$, where 
      \begin{align*}
          \hat{R}_k(x) &= \hat{\eta}_{k}(x) - \eta_0(x)  - \Delta_{k}(x)~,\\
          \Delta_{k}(x) &= N^{-1/2 } \sum_{j \notin \mathcal{I}_k} N^{-\varphi}\delta_{N}(W_j,x)  + N^{-1} \sum_{j \notin \mathcal{I}_k} N^{-\varphi} b_{N}(X_{j},x)~.
      \end{align*}
      
       \item[(ii)] $E[\delta_{N}(W_{j},x) \mid X_{j}] = 0$, $E \left[ \|\delta_{N}(W_{j},x)\|^{2} \right] \in (c_\delta,C_4)$, and $E\left[ \|\delta_N(W_j,x)\|^4\right] \le C_4 N^{1-2\varphi}$ for all $x \in \mathcal{X}$.

       \item[(iii)] $\|E[b_N(X_j,x)]\| \in (c_b,C_3)$ and $E[ \| b_N(X_j,x)\|^{2s}] \le C_{4} N^{(2s-1)(1-2\varphi)}$ for all $x \in \mathcal{X}$ and $s \in \{1,2\}$. 
       
        \item[(iv)] $\| N^{1/2-\varphi}E[ m(W_j,\theta_0,\eta_j) \delta_N(W_j,x) ] \|  \le C_4 $, $ \| E[m(W_j,\theta_0,\eta_j) b_N(X_j,x) ]\|  \le C_4$, and $ E[ \| m(W_j,\theta_0,\eta_j) \delta_N(W_j,x)\|^2 ] \le C_3$ for all $x \in \mathcal{X}$.    
    \end{enumerate}
\end{assumption} 

Assumption~\ref{asm:SLE} is a high-level condition requiring the first-step estimators to admit a stochastic linear expansion. Part~(i) decomposes the estimation error $\hat{\eta}_k - \eta_0$ into a leading linear term $\Delta_k(x)$ and a higher-order remainder $\hat{R}_k(x)$, and requires the out-of-sample $L_2$-norm rate of the remainder to be sufficiently fast. The linear term $\Delta_k(x)$ splits into a mean-zero stochastic component based on $\delta_N$, which determines the variance of the estimator, and a deterministic component based on $b_N$, which captures the bias, mirroring the standard bias--variance structure of nonparametric estimators \citep{masry1996multivariate,newey1997convergence}. Parts~(ii) and~(iii) impose moment and rate conditions on these two components: the bound on $\delta_N$ in part~(ii) implies that the variance of $\hat{\eta}_k$ is of order $N^{-2\varphi}$, while the bounds on $b_N$ in part~(iii) imply that its squared bias is of the same order. Together with the negligible remainder from part~(i), this gives $E[\|\hat{\eta}_k(X) - \eta_0(X)\|^2]^{1/2} = O(N^{-\varphi})$, implying the out-of-sample $L_2$ convergence rate of Assumption~\ref{asm:L2rate} since $\varphi \in (1/4,1/2)$. Part~(iv) bounds the interaction of these components with the moment function $m$.  
 
The second-order approximation we develop below combines the stochastic linear expansion of the first-step estimation error  with a Taylor expansion of the moment function. Stating it requires some additional notation. Define 
%$\mathcal{T}_{n}^*  =  n^{-1/2} \sum_{i=1}^n G^{-1}m_i~,$
%
% $ \mathcal{T}_{n,K_n}^{l}  =   n^{-1/2 } \sum_{k = 1}^{K_n} \sum_{i \in \mathcal{I}_k} \Delta_{k,i}^\intercal   \left(G^{-1} \partial_\eta m_i \right)~,  \mathcal{T}_{n,K_n}^{q}  =  \tfrac{1}{2}  n^{-1/2 } \sum_{k = 1}^{K_n} \sum_{i \in \mathcal{I}_k} \Delta_{k,i}^\intercal   \left(G^{-1} \partial_\eta^2 m_i \right)  \Delta_{k,i}~,$
%
% and $\mathcal{T}_{n}^{*,(2)}  =  - n^{-1/2} \left( n^{-1/2} \sum_{i=1}^n G^{-1}m_i \right) \left(   n^{-1/2} \sum_{i=1}^n G^{-1}(\psi^a_i-G)\right),$
\begin{align*} 
    \mathcal{T}_{n}^*  &=  n^{-1/2} \sum_{i=1}^n G^{-1}m_i~,\\
    \mathcal{T}_{n,K_n}^{l} &=   n^{-1/2 } \sum_{k = 1}^{K_n} \sum_{i \in \mathcal{I}_k} \Delta_{k,i}^\intercal   \left(G^{-1} \partial_\eta m_i \right)~,  \\ 
    \mathcal{T}_{n,K_n}^{q} &=  \tfrac{1}{2}  n^{-1/2 } \sum_{k = 1}^{K_n} \sum_{i \in \mathcal{I}_k} \Delta_{k,i}^\intercal   \left(G^{-1} \partial_\eta^2 m_i \right)  \Delta_{k,i}~,  \\
    \mathcal{T}_{n}^{*,(2)} &=  - n^{-1/2} \left( n^{-1/2} \sum_{i=1}^n G^{-1}m_i \right) \left(   n^{-1/2} \sum_{i=1}^n G^{-1}(\psi^a_i-G)\right)~,
\end{align*}
where $m_i = m(W_i,\theta_0,\eta_i)$, $\partial_\eta m_i = \partial_\eta m(W_i,\theta_0,\eta_i)$, $\partial_\eta^2 m_i = \partial_\eta^2m(W_i,\theta_0,\eta_i)$, $\psi^a_i = \psi^a(W_i,\eta_i)$, $\Delta_{k,i} = \Delta_{k}(X_i)$,  and $G=E[\psi^a_i]$.

With this notation, the scaled DML2 estimation error admits the second-order expansion
\begin{equation}\label{eq:so-expansion}
 n^{1/2}\left(\hat{\theta}_{n,K_n}^{(2)} - \theta_0 \right) = \mathcal{T}_{n,K_n} + \mathcal{R}_{n,K_n},
\end{equation}
where our second-order approximation is defined as 
$$ \mathcal{T}_{n,K_n} = \mathcal{T}_{n}^* + \mathcal{T}_{n,K_n}^{\ell} + \mathcal{T}_{n,K_n}^{q} + \mathcal{T}_{n}^{*,(2)}~,$$ 
and $\mathcal{R}_{n,K_n}$ is a remainder term. The leading term $\mathcal{T}_n^*$ is asymptotically normal and does not depend on $K_n$. The terms $\mathcal{T}_{n,K_n}^{\ell}$ and $\mathcal{T}_{n,K_n}^{q}$ are second-order linear and quadratic corrections arising from the first-step estimation error and the Taylor approximation, and $\mathcal{T}_n^{*,(2)}$ is the correction from estimating $G$. Because $\mathcal{T}_n^*$ and
$\mathcal{T}_n^{*,(2)}$ are invariant to $K_n$, the dependence of the
second-order approximation on $K_n$ arises entirely through
$\mathcal{T}_{n,K_n}^{\ell}$ and $\mathcal{T}_{n,K_n}^{q}$.

We require two additional assumptions to ensure that $\mathcal{T}_{n,K_n}$ in~\eqref{eq:so-expansion} is a valid second-order approximation for any $2 \le K_n \le n$. The first (Assumption~\ref{asm:leading-terms}) makes the second-order analysis nondegenerate, ensuring that $\mathcal{T}_{n,K_n}^{\ell}$ and $\mathcal{T}_{n,K_n}^{q}$ carry information beyond $\mathcal{T}_n^*$. The second (Assumption~\ref{asm:AS-SO}) guarantees that the remainder $\mathcal{R}_{n,K_n}$ is of smaller order than $\mathcal{T}_{n,K_n}$ for any $2 \le K_n \le n$, so that the performance of DML2 can be evaluated  using  $\mathcal{T}_{n,K_n}$ in the growing-$K_n$ framework. We present these conditions next.

\begin{assumption}\label{asm:leading-terms}
(i) $m$ is three-times continuously differentiable on $\eta \in \mathcal{E} \subseteq \textbf{R}^p$ and  $\sup_{\eta \in \mathcal{E}} \| \partial_\eta^3 m(W_i, \theta_0, \eta) \|_{\infty} \le c_2$, (ii) for any $i \neq j$, the next limits exist and are finite,  
    \begin{align}
        \mathcal{V}^{(\ell)} &= \lim_{N \to \infty}    E\left[ \left( \delta_{N}(W_j,X_i)^\intercal (G^{-1} \partial_\eta m_i) +\delta_{N}(W_i,X_j)^\intercal (G^{-1} \partial_\eta m_j)  \right)^2 \right] ~, \label{eq:var-linear}\\
        \mathcal{B} &=  \lim_{N \to \infty}    E\left[ ( \delta_{N} (W_j,X_i) +  \tilde{b}_{N}(X_i))^\intercal \left(  G^{-1} \partial_\eta^2 m_i \right) ( \delta_{N} (W_j,X_i) +  \tilde{b}_{N}(X_i)) \right]~, \label{eq:bias_nl} 
    \end{align}
    where $ \tilde{b}_{N}(x) = E[b_N(X_j,x)]$, 
    (iii) $\mathcal{V}^{(\ell)} >0$ and $\mathcal{B} \neq 0$, and (iv) $Var[\mathcal{T}_{n,K_n}^{q}] \le C_4 n^{-2\varphi}$. 
\end{assumption}

Parts (i), (ii), and (iii) are mild regularity conditions. Part~(i) holds for several econometric models studied in the semiparametric literature, where the moment function is a quadratic polynomial in $\eta$; see examples in Appendix \ref{sec:appendix_examples}. Part~(ii) holds with subsequences under Assumptions~\ref{asm:EM} and~\ref{asm:SLE}. Part~(iii) guarantees the second-order approximation is nondegenerate, which we need for our analysis and holds as long as the bias term $\tilde{b}_{N}(x)$ is nonzero and the moment function is nonlinear on $\eta$. Part~(iv) is a high-level condition that simplifies the exposition. Appendix \ref{appendix:local-polynomials} verifies these conditions for first-step estimators based on local-polynomial regressions and the econometric models in Appendix~\ref{sec:appendix_examples}.

The next assumption is an algorithmic-stability condition for the higher-order remainder term $\hat{R}_{k}(x)$ that appears in Assumption~\ref{asm:SLE}. Let $\hat{R}_{k}^{\ell}(x)$ be the corresponding remainder for $\hat{\eta}_k^{\ell}(x)$, i.e. the remainder obtained when $W_\ell$ is replaced by an independent copy $\widetilde{W}_\ell$ drawn from $F_0$. Recall $N$ is the sample size used by first-step estimators, $C_3>0$, and $N \asymp n$.

\begin{assumption}\label{asm:AS-SO}
      $ \max_{\ell \notin \mathcal{I}_k} E[\| \hat{R}_k(X)-\hat{R}_k^{\ell}(X)\|^2]^{1/2} \le  C_3 N^{-1/2-2\varphi}$~.
\end{assumption}

We now state the formal result establishing the validity of the second-order approximation  $\mathcal{T}_{n,K_n}$  to the scaled DML2 estimation error, $n^{1/2}(\hat{\theta}_n^{(2)} - \theta_0)$.

\begin{theorem}\label{thm:stochastic-expansion-DML2}
Let Assumptions \ref{asm:EM}, \ref{asm:SLE}, \ref{asm:leading-terms}, and \ref{asm:AS-SO} hold and let $K_n$ be such that $2 \le K_n \le n$. Then, $\mathcal{R}_{n,K_n} = O_p(n^{1/2-3\varphi})$. Furthermore, $\mathcal{T}_{n,K_n}-\mathcal{T}_n^* = O_p(n^{-\varphi})+O(n^{1/2-2\varphi})$. 
\end{theorem} 

Theorem~\ref{thm:stochastic-expansion-DML2} shows that $\mathcal{T}_{n,K_n}$ is a more accurate approximation than the first-order term $\mathcal{T}_n^*$ alone. The remainder $\mathcal{R}_{n,K_n}$ in \eqref{eq:so-expansion} is of order $O_p(n^{1/2-3\varphi})$, which is smaller than the second-order correction $\mathcal{T}_{n,K_n} - \mathcal{T}_n^* = O_p(n^{-\varphi}) + O(n^{1/2-2\varphi})$. The improvement comes from the linear and quadratic correction terms $\mathcal{T}_{n,K_n}^{\ell}$ and $\mathcal{T}_{n,K_n}^{q}$ that contribute the stochastic component of order $O_p(n^{-\varphi})$, while $\mathcal{T}_{n,K_n}^{q}$ also carries a deterministic bias of order $O(n^{1/2-2\varphi})$. The correction from estimating $G$, $\mathcal{T}_n^{*,(2)}$, is of smaller order $O_p(n^{-1/2})$ and does not contribute to the leading second-order behavior since $\varphi < 1/2$.

\begin{remark}\label{rem:DML2-second-order}
Assumption~\ref{asm:AS-SO} guarantees the remainder $\mathcal{R}_{n,K_n}$ is $ O_p(n^{1/2-3\varphi})$ for any $2 \le K_n \le n$. Without it,  the same bound can be established when $K_n$ grows proportionally to $\sqrt{n}$ under Assumptions~\ref{asm:EM}, \ref{asm:SLE}, and \ref{asm:leading-terms}. Therefore, Assumption~\ref{asm:AS-SO} mirrors the role of Assumption~\ref{asm:AS} in Section~\ref{sec:DML2-valid} by extending the second-order approximation $n^{1/2}(\hat{\theta}_{n,K_n}^{(2)} - \theta_0) = \mathcal{T}_{n,K_n} + O_p(n^{1/2-3\varphi})$ from the regime $K_n\asymp\sqrt n$ to any $2\le K_n\le n$.
\end{remark}

\begin{remark}\label{rem:so-dml2-sample-split}
The second-order asymptotic approximation $\mathcal{T}_{n,K_n}$ depends on the sample-splitting partition through $\mathcal{T}_{n,K_n}^{\ell}$ and $\mathcal{T}_{n,K_n}^{q}$; in contrast, the first-order asymptotic approximation $\mathcal{T}_{n}^*$ does not. Therefore, $\mathcal{T}_{n,K_n}$ can be used to study the effects of sample splitting in DML2, which is an interesting research direction left for future work. 
\end{remark} 
 
\subsection{Second-order Asymptotic bias and MSE for DML2}\label{sec:so-mse-dml2}

We define the second-order asymptotic bias and MSE of DML2 as the mean and second moment of $\mathcal{T}_{n,K_n}$, respectively. These quantities allow us to compare how different choices of the number of folds affect DML2 in higher-order asymptotics. Theorem~\ref{theorem:so-bias-mse-DML2} gives explicit leading-order expressions for both quantities and shows how they depend on \(K_n\).

The analysis of the second-order asymptotic MSE relies on an additional high-level assumption. In Appendix \ref{appendix:local-polynomials} we verify this assumption for first-step estimators based on local-polynomial regressions and several econometric models studied in the semiparametric literature. Let $S(x) = G^{-2} E\left[ m_i ~\partial_\eta m_i \mid X_i = x \right]$, $H(x) = G^{-1} E\left[  \partial_\eta^2 m_i \mid X_i = x \right]$, and $\tilde{\delta}_{N}(x) = N^{1/2-\varphi} G^{-1} E\left[ m_j \delta_N(W_j,X_i)  \mid X_i = x\right]$ for $j \neq i$ and $x \in \mathcal{X}$. 
 
\begin{assumption}\label{asm:information}
    There is a function $\tilde{\delta}_{\infty}:\mathcal{X} \to \mathbf{R}^p$ such that  
    \begin{enumerate}
        \item[(i)] $E [~ \| \tilde{\delta}_{N}(X) - \tilde{\delta}_{\infty}(X)\|^2  ~]^{1/2} = O(N^{-\varphi})$.
        \item[(ii)] $S(x) + H(x)\tilde{\delta}_{\infty}(x) = 0 $ for all $x \in \mathcal{X}$.
    \end{enumerate} 
\end{assumption}

The next result shows that the performance of DML2, in terms of second-order bias and MSE, improves with more folds but with diminishing returns.

\begin{theorem}\label{theorem:so-bias-mse-DML2}
    Let Assumptions \ref{asm:EM}, \ref{asm:SLE}, and  \ref{asm:leading-terms}  hold and let $K_n$ be such that $2 \le K_n \le n$. Then,
    \begin{equation}
        E[\mathcal{T}_{n,K_n}]  = \tfrac{1}{2} \mathcal{B} \left( 1 + \tfrac{1}{K_n-1} \right)^{2\varphi} n^{1/2-2\varphi} + o(n^{1/2-2\varphi})~, \label{eq:bias-DML2}
    \end{equation}
    where $\mathcal{B}$ is defined in \eqref{eq:bias_nl}. Furthermore, if either $\varphi \in (1/4,~1/3)$ or Assumption~\ref{asm:information} holds, then
    \begin{equation}
        E[\mathcal{T}_{n,K_n}^2]  =  \Sigma + \tfrac{1}{4}\mathcal{B}^2 \left( 1 + \tfrac{1}{K_n-1} \right)^{4\varphi} n^{1-4\varphi} + o(n^{1-4\varphi})~.\label{eq:mse-DML2}
    \end{equation}  
\end{theorem}

Assumption \ref{asm:leading-terms}(iii) is key to guarantee that our analysis is nondegenerate, i.e., $\mathcal{B} \neq 0$. 
Theorem \ref{theorem:so-bias-mse-DML2} is consistent with existing simulation findings \citep{ahrens2024ddml,ahrens2024model}, and our simulation results in Section \ref{sec:result_simulations}, in the sense that it ranks the performance of DML2 decreasing on the number of folds.\footnote{Since our result provides an asymptotically theoretical comparison, it is possible that for some simulation designs the predicted monotonicity does not hold in small sample sizes even if the assumptions hold.} In particular, for a sample size n sufficiently large, and among popular choices, DML2 with 10 folds asymptotically dominates DML2 with 5 folds in terms of second-order asymptotic bias and MSE ($|E[\mathcal{T}_{n,5}]| > |E[\mathcal{T}_{n,10}]| $ and $  E[\mathcal{T}_{n,5}^2] > E[\mathcal{T}_{n,10}^2]$), and DML2 with $n$ folds (the leave-one-out estimator) asymptotically dominates DML2 with any fixed number of folds $K$ ($|E[\mathcal{T}_{n,5}]| > |E[\mathcal{T}_{n,10}]|$ and $E[\mathcal{T}_{n,5}^2] > E[\mathcal{T}_{n,10}^2]$). The leave-one-out case thus serves as a theoretical benchmark; its practical implementation can be challenging due to the computational burden of estimating $n$ first-step estimators. 

\begin{remark}\label{rem:higher-order-fixed-K}
When $K_n$ is fixed as $n \to \infty$, the second-order bias and
MSE for DML1 have the same leading terms as those in \eqref{eq:bias-DML2}
and \eqref{eq:mse-DML2}. A second-order analysis therefore cannot
distinguish DML1 from DML2 under fixed-$K$ asymptotics. In
Section~\ref{sec:result1}, by contrast, we distinguished the two using an
asymptotic framework with $K_n \asymp \sqrt{n}$, which scales up terms that
fixed-$K$ asymptotics treat as negligible, making the distinction visible.
\end{remark}

\section{Recommendations for Practitioners}\label{sec:recommendations}
 
This section summarizes the practical implications of our theoretical results. Our recommendations complement recent practitioner-oriented discussions of DML implementation, such as \citet{ahrens2025introduction}.

\subsection{Prefer DML2 over DML1}

Our main recommendation is to prefer DML2 over DML1 because it is more robust to the choice of the number of folds. Under the growing-$K_n$ framework studied in this paper, DML1 can become invalid for inference purposes at large values of $K_n$, since its limiting distribution can be incorrectly centered  whenever $\Lambda \neq 0$. By contrast, DML2 is correctly centered and supports valid inference under the same conditions.

The distinction between DML2 and DML1 matters for empirical practice when researchers want to increase the number of folds to improve accuracy of first-step estimators. DML2 is  the safer default since it avoids the large-$K_n$ bias that may arise for DML1 when $\Lambda \neq 0$, and it does not require the practitioner to determine in advance whether $\Lambda = 0$ in the model at hand (although Remark~\ref{rem:diagnosing-lambda} explains how this can be done). Unlike the guidance on the number of folds in Section~\ref{sec:choosing-k}, this recommendation does not rely on the stronger conditions of Section~\ref{sec:result2}: it requires no restriction on the first-step estimators beyond those already imposed in Section~\ref{sec:result1}, and when $\Lambda = 0$ the two variants are asymptotically equivalent, so following it entails no cost. Moreover, under the algorithmic-stability condition, the standard asymptotic theory for DML2 remains valid for any $2 \le K_n \le n$.

\subsection{Choosing $K_n$ for DML2}\label{sec:choosing-k}

For researchers who want theory-based guidance on the number of folds, the second-order analysis of Section~\ref{sec:result2} provides it for scalar DML2 estimators whose first-step estimators admit a stochastic linear expansion: within this class, the performance of DML2 improves with more folds but with diminishing returns. In particular, among popular choices, we recommend using 10 folds rather than 5.
 
This recommendation should be read as theory-based guidance within the class of estimators studied in Section~\ref{sec:result2}, rather than as a universal prescription for all DML implementations. Moreover, it relies on asymptotic analysis and holds for sufficiently large sample sizes.

\section{Simulations}\label{sec:simulations}
 
This section examines how well the asymptotic approximations in Sections \ref{sec:result1} and \ref{sec:result2} capture finite-sample behavior in two models: (i) ATT-DID \citep{sant2020doubly} and (ii) LATE \citep{hong2010semiparametric}. We calculate the bias, MSE, and coverage probability of confidence intervals for DML1 and DML2 for several values of $K$. We use confidence intervals based on \citet[Corollary 3.1]{chernozhukov2018double}.  
 
\subsection{Design: ATT-DID and LATE }
 
\paragraph*{ATT-DID:} Our first design implements Example \ref{example_DID}  of Appendix \ref{sec:appendix_examples}, building on \citet{sant2020doubly}. The observed outcome in the pre-treatment period and the potential outcomes in the post-treatment period are defined by
\begin{align*}
    Y_{0,i} &= f_{\text{reg}}(X_i) + v(X_i,A_i) + \varepsilon_{0,i} \\
     Y_{1,i} (a) &= 2 f_{\text{reg}}(X_i) + v(X_i,A_i) + \varepsilon_{1,i}(a)~,\quad a = 0, 1
\end{align*}
where $f_{\text{reg}}(X) = 210 + 6.85 X_1^2 + 3.425(X_2 + X_3)$ and $v(X_i,A_i) = A_i f_{\text{reg}}(X_i) + \varepsilon_{v,i}$, and $(\varepsilon_{0,i}, \varepsilon_{1,i}(0), \varepsilon_{1,i}(1), \varepsilon_{v,i})$ is distributed as $N(0,\mathbb{I}_4)$, where $\mathbb{I}_4$ is the $4 \times 4$ identity matrix. 
The treatment assignment is defined by $A_i \sim \text{Bernoulli}( p(X_i) )$, where
\begin{align*}
     p(X_i) &= \frac{ \exp( f_{\text{ps}}(X_i) ) }{1+ \exp( f_{\text{ps}}(X_i) )} \\
     f_{\text{ps}}(X)  &= 0.25 ( -X_1 + 0.5 X_2 - 0.25 X_3 )~.
\end{align*}
Finally, the vector of covariates is $X_i = (X_{1,i}, X_{2,i}, X_{3,i}) \in [0,1]^3$ and all its coordinates are independent uniform random variables (i.e., $X_{1,i} \sim \text{Uniform}[0,1]$).  

\paragraph*{LATE:} Our second design implements Example \ref{example_LATE}  of Appendix \ref{sec:appendix_examples}, building on \citet{hong2010semiparametric}. The potential treatment decisions are defined as
\begin{align*}
    D_i(1) &= I\{   X_i + 0.5  \ge V_i \}~,\\
    D_i(0) &= I\{   X_i -0.5   \ge V_i \}~,
\end{align*}
where $X_i \sim \text{Uniform}[0,1]$ and $V_i \sim N(0,1)$ are independent. The potential outcomes are defined by
\begin{align*}
    Y_i(1) &= \sigma_\lambda \left( \xi_{1,i} + \xi_{3,i} I\{D_i(1) = 1, D_i(0) = 1\} + \xi_{4,i} I\{D_i(1) = 0, D_i(0) = 0\} \right)~,\\
    Y_i(0) &= \sigma_\lambda \left( \xi_{2,i} + \xi_{3,i} I\{D_i(1) = 1, D_i(0) = 1\} + \xi_{4,i} I\{D_i(1) = 0, D_i(0) = 0\} \right)~,
\end{align*}
where $\xi_{1,i} \sim \text{Poisson}(\exp(X_i/2)) $, $\xi_{2,i} \sim \text{Poisson}(\exp(X_i/2)) $, $\xi_{3,i} \sim \text{Poisson}(2) $, and $\xi_{4,i} \sim \text{Poisson}(1) $. The terms involving $\xi_{3,i}$ and $\xi_{4,i}$ allow outcome heterogeneity across compliance types. The random variables $\xi_{1,i},\ldots,\xi_{4,i}$ are independent conditional on $X_i$. The treatment assignment is defined by $Z_i \sim \text{Bernoulli}(\Phi(X_i-0.5)) $. We calibrate $\sigma_\lambda$ so that the asymptotic variance $\Sigma$ in \eqref{eq:asymptotic-variance} equals $10$, which yields $\sigma_\lambda \approx 0.283$.

\subsection{Results: DML1 is sensitive to large $K$ for the LATE design, but not for the ATT-DID}\label{sec:result_simulations}

This section provides simulation evidence---in bias, MSE, and coverage---that DML2 outperforms DML1 for LATE, while the two perform similarly for ATT-DID. This is consistent with Section~\ref{sec:result1} since the model-dependent quantity $\Lambda$ that characterizes the performance difference between DML1 and DML2 is nonzero for LATE and zero for ATT-DID.

We calculate DML1 and DML2 for the ATT-DID and LATE for different values of $K \in \{2,5,10,15,20,25,30\}$. The nuisance function $\eta_0$ for the ATT-DID and LATE are presented in Examples \ref{example_DID} and \ref{example_LATE} in Appendix~\ref{sec:appendix_examples}, respectively. 
We estimate each component of $\eta_0$ by local-linear regressions with cross-fitting (see Section \ref{sec:setup}), so that each first-step estimator uses sample size $N = (1 - 1/K)n$. We use the Epanechnikov kernel, with bandwidth $h_j = c_j N^{-1/7}$ for the ATT-DID and $h_j = c_j N^{-1/5}$ for the LATE; the constant  $c_j$ is chosen by cross-validation for each component of $\eta_0$.\footnote{Simulations based on Nadaraya-Watson estimators appear in a previous version of this paper; see \citet[Section 5 and Appendix D]{velez2024asymptotic}. The results show that the sensitivity of DML1 is due to its oracle version rather than to the first-step estimators.} 
 
\begin{figure}[t]
     \centering
    \begin{subfigure}{0.45\textwidth}
        \centering
        \includegraphics[width= \textwidth]{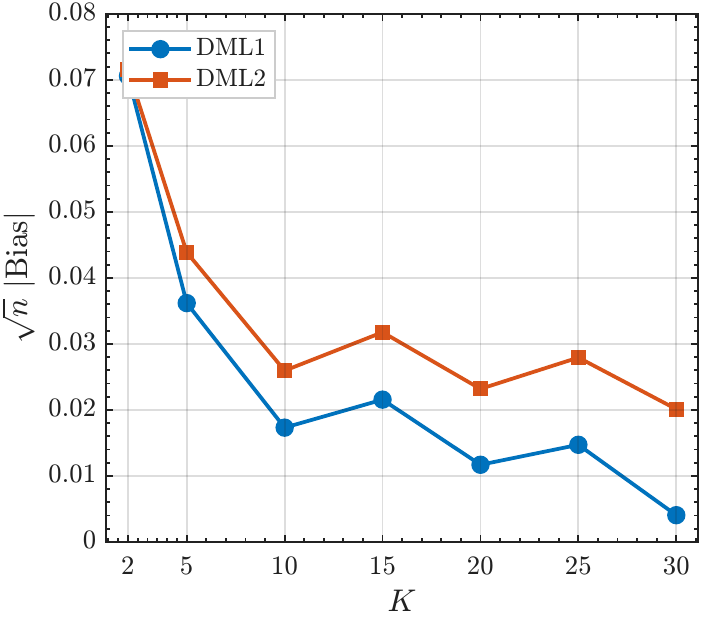}  
        \caption{ATT-DID ($\Lambda = 0$)}
    \end{subfigure}%\hspace{0.02\textwidth}
     \hfill
    \begin{subfigure}{0.45\textwidth}
        \centering
        \includegraphics[width= \textwidth]{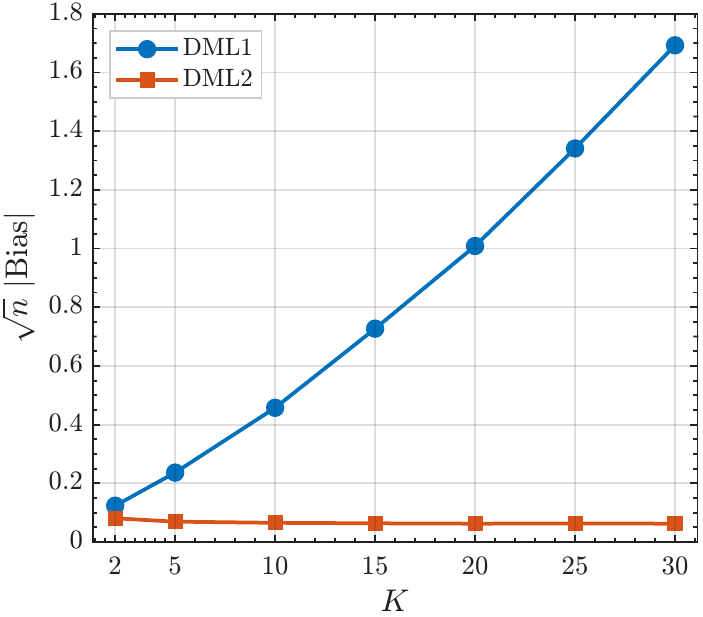} 
        \caption{LATE ($\Lambda \neq 0$)}
    \end{subfigure} 
    \caption{Bias of DML1 and DML2 for ATT-DID and LATE. Sample size $n = 3,000$; 2,000 simulations.}\label{fig_bias}
\end{figure}
\begin{figure}[h!]
     \centering
    \begin{subfigure}{0.45\textwidth}
        \centering
        \includegraphics[width=1\textwidth]{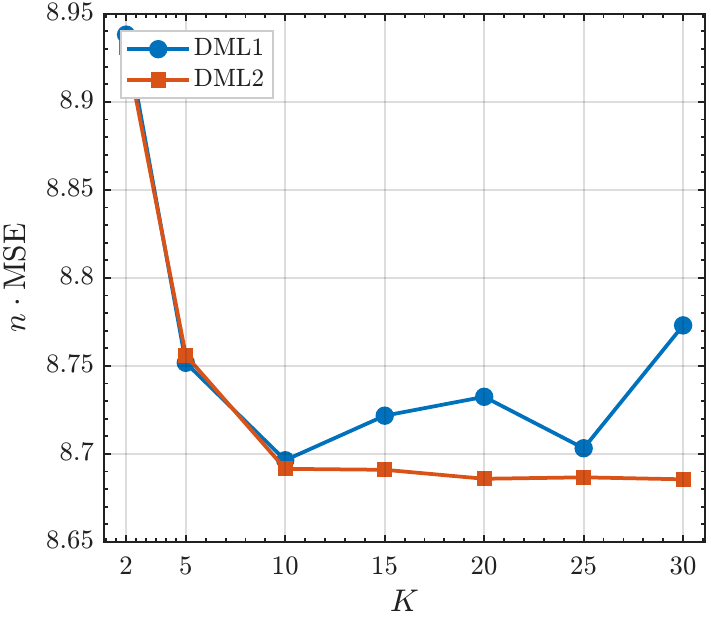}  
        \caption{ATT-DID ($\Lambda = 0$)}
    \end{subfigure}
    \hfill
    \begin{subfigure}{0.45\textwidth}
        \centering
        \includegraphics[width=0.97\textwidth]{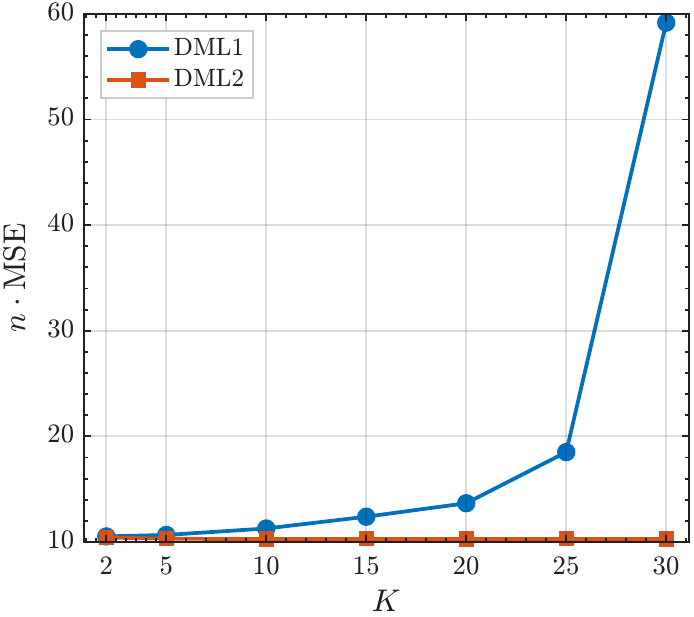}  
        \caption{LATE ($\Lambda \neq 0$)}
    \end{subfigure} 
    \caption{MSE of DML1 and DML2 for ATT-DID and LATE. Sample size $n = 3,000$; 2,000 simulations.}\label{fig_mse}
\end{figure}
% \subsubsection{Bias} 

\paragraph*{Bias:}
Figure~\ref{fig_bias} presents the bias of DML1 and DML2 for several values of $K$ in two models: ATT-DID in panel~(a) and LATE in panel~(b). Panel~(a) shows that DML1 and DML2 perform similarly in terms of bias, whereas panel~(b) shows that the bias of DML1 increases almost linearly with $K$. These findings are consistent with Theorem~\ref{theorem:sqrt-Kn}, since $\Lambda=0$ in panel~(a), whereas $\Lambda\neq0$ in panel~(b). Furthermore, the bias of DML2 decreases from $K=2$ to $K=10$ and remains approximately constant thereafter, consistent with Theorem~\ref{theorem:so-bias-mse-DML2} and the factor $(1+1/(K-1))^{2\varphi}$ in \eqref{eq:bias-DML2} that changes little once $K\ge10$.

\begin{figure}[h!]
     \centering
    \begin{subfigure}{0.45\textwidth}
        \centering
        \includegraphics[width=1\textwidth]{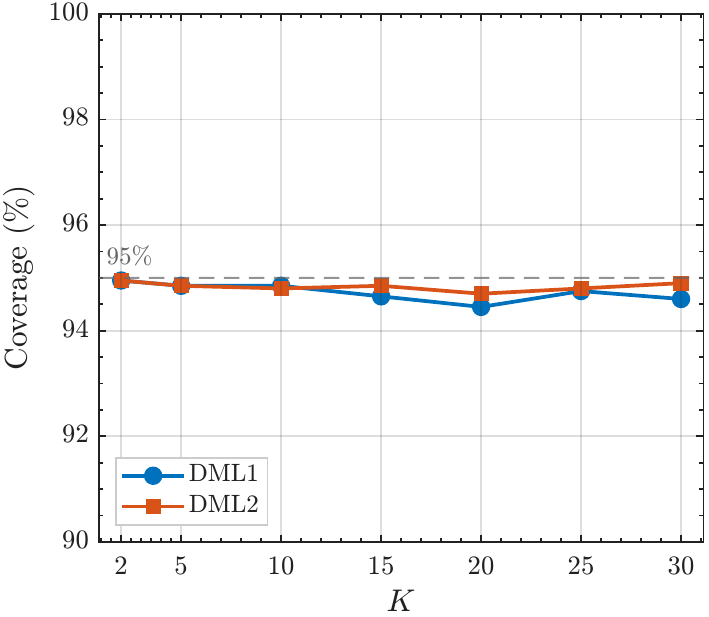}  
        \caption{ATT-DID ($\Lambda = 0$)}
    \end{subfigure}
     \hfill
    \begin{subfigure}{0.45\textwidth}
        \centering
        \includegraphics[width=1\textwidth]{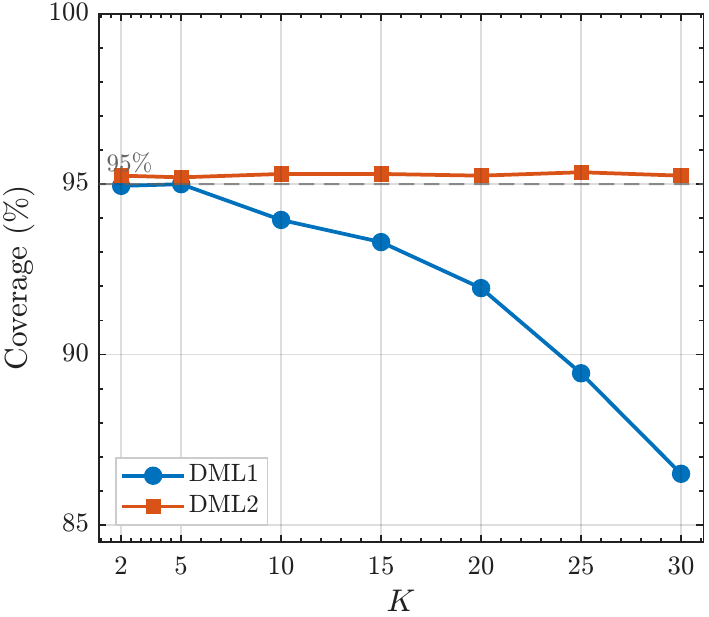}  
        \caption{LATE ($\Lambda \neq 0$)}
    \end{subfigure} 
    \caption{Coverage probability of 95\%-confidence intervals based on DML1 and DML2 for ATT-DID and LATE. Sample size $n = 3,000$ and 2,000 simulations.}\label{fig_cp}
\end{figure} 

\paragraph*{MSE:}
Figure~\ref{fig_mse} presents the MSE of DML1 and DML2 for several values of $K$ in two models: ATT-DID in panel~(a) and LATE in panel~(b). In panel~(a), the MSEs of DML1 and DML2 are nearly identical across all values of $K$, consistent with Theorem~\ref{theorem:sqrt-Kn} since $\Lambda=0$. In contrast, panel~(b) shows that the MSE of DML1 increases rapidly with $K$, reflecting the first-order bias that arises when $\Lambda\neq0$.\footnote{Additional simulation results in the previous version of this paper show that DML1  and the oracle DML1 exhibit similar MSE values; see  \citet[Figure 10 in Appendix D]{velez2024asymptotic}.}  By contrast, the MSE of DML2 decreases from $K=2$ to $K=10$ and remains essentially unchanged thereafter. This behavior is consistent with Theorem~\ref{theorem:so-bias-mse-DML2}, which predicts that increasing the number of folds improves the second-order MSE of DML2, although with diminishing gains.

\paragraph*{Coverage probability:}
Figure~\ref{fig_cp} presents the coverage probabilities of 95\% confidence intervals based on DML1 and DML2 for several values of $K$ in two models: ATT-DID in panel~(a) and LATE in panel~(b). Panel~(a) corresponds to the case $\Lambda=0$ and shows that both DML1 and DML2 achieve coverage close to the nominal 95\% level across all values of $K$. In contrast, panel~(b) corresponds to the case $\Lambda\neq0$ and shows that the coverage of DML1 deteriorates steadily as $K$ increases, whereas DML2 maintains coverage close to the nominal level. At $K=30$, the coverage of the DML1 confidence interval is below 87\% for a nominal 95\% level; the deterioration is already visible at $K=10$. These findings are consistent with Theorem~\ref{theorem:sqrt-Kn}, which implies that DML1 may fail to support valid inference for large values of $K$, while DML2 remains valid under the same conditions.

\section{Concluding Remarks}\label{sec:concluding-remarks}

This paper studies debiased machine learning under an asymptotic framework in which the number of cross-fitting folds, $K_n$, may grow with the sample size. This framework reveals distinctions that are not visible under the existing fixed-$K$ asymptotic theory. In particular, although DML1 and DML2 are first-order asymptotically equivalent when $K$ is fixed, they can behave differently when $K_n$ grows. The difference is governed by a model-dependent quantity $\Lambda$: when $\Lambda \neq 0$ and $K_n $ grows proportionally to $ \sqrt{n}$, DML1 has asymptotic bias, in which case standard inference based on DML1 fails---as occurs, for instance, for the LATE---whereas DML2 remains correctly centered.  

Our analysis also clarifies why this difference arises. The sensitivity of DML1 is already present in its oracle version and comes from the way DML1 aggregates fold-specific estimators. By contrast, DML2 aggregates the moment conditions before solving for the parameter. When an algorithmic-stability condition on the first-step estimators holds, this structure allows the standard first-order asymptotic theory for DML2 to remain valid for any $2 \le K_n \leq n$, including the leave-one-out case (DML2 with $n$ folds).

We then study the choice of $K_n$ for scalar DML2 estimators under stronger conditions on the first-step estimators. We rely on a second-order asymptotic approximation that shows that the performance of DML2, in terms of second-order bias and MSE, improves with more folds
but with diminishing returns. This result provides theory-based guidance on the choice of the number of folds for DML2 within the class of estimators covered by Section~\ref{sec:result2}. In particular, DML2 with $n$ folds (the leave-one-out case) defines a natural benchmark.

Finally, our results support two default recommendations. First, DML2 should be preferred to DML1; this recommendation requires no conditions on the first-step estimators beyond those DML already imposes. Second, within the class of first-step estimators covered by Section~\ref{sec:result2},  the number of folds for DML2 should be set as large as computation allows. Our analysis also leaves several directions for future work, including the study of the algorithmic-stability condition for modern machine-learning methods, second-order asymptotic comparisons beyond the class of first-step estimators considered here, and the consequences of sample-splitting variability using our second-order approximation.

\bibliographystyle{ecta-fullname} % Style BST file
\bibliography{references}  % Bibliography file (usually '*.bib')

\begin{appendix}

\section{Proof of Main Results}\label{sec:appendix_main_proofs}

\subsection{Proofs for Section \ref{sec:result1}}\label{appendix:main-results} 

\begin{lemma}\label{lemma:sqrt-Kn-DML1-DML2}
    Let $K_n$ be such that $K_n/\sqrt{n} \to c \in [0,\infty)$ and $ 2 \le K_n \le n$. In addition,
    \begin{itemize}
        \item[(i)] Let Assumptions~\ref{asm:EM} and \ref{asm:L2rate} hold. Then, equation \eqref{eq:DML1-oracle-equivalence} holds.
        \item[(ii)] Let Assumptions~\ref{asm:EM}~(i)--(iii) and \ref{asm:L2rate} hold. Then, equation \eqref{eq:DML2-oracle-equivalence} holds.
    \end{itemize}
    
\end{lemma}

\begin{proof}
We use notation  and auxiliary results presented in  Appendix \ref{appendix:aux-results}.

    \noindent \textbf{\textit{Part (i)}:} Let $\hat{a}_k$, $\hat{b}_k $, ${a}_k $, and ${b}_k $ denote the quantities defined in Appendix \ref{appendix:aux-results}.   We write
    $ \sqrt{n} \left( \hat{\theta}_{n,K_n}^{(1)} -  \hat{\theta}_{n,K_n}^{*,(1)}\right) = A + B$,
    where
    \begin{align*}
        A &= K_n^{-1/2} \sum_{k=1}^{K_n} \left( \mathbb{I}_d + n_k^{-1/2} \hat{b}_k \right)^{-1} \left( \hat{a}_k - a_k \right) \\
        B &= K_n^{-1/2} \sum_{k=1}^{K_n} \left \{  \left( \mathbb{I}_d + n_k^{-1/2} \hat{b}_k \right)^{-1} - \left( \mathbb{I}_d + n_k^{-1/2} {b}_k \right)^{-1} \right \} a_k
    \end{align*} 
    The identity \eqref{eq:identity-appendix} and triangle inequality imply 
        \begin{equation*}
            \| A \| \le \|I_1\| + \max_{1\le k \le K_n}  \left | \left| (\mathbb{I}_d + n_k^{-1/2} \hat{b}_k)^{-1}  \right| \right| \times ( I_{2} + I_3) \\
        \end{equation*}
where $ I_2 = n^{-1/2} \sum_{k=1}^{K_n}  \left | \left| \hat{b}_k - b_k  \right| \right| \times \left | \left|   \hat{a}_k - a_k   \right| \right| $, $I_3 = n^{-1/2} \sum_{k=1}^{K_n}  \left | \left| {b}_k  \right| \right| \times \left | \left|   \hat{a}_k - a_k   \right| \right|$, and
\begin{align}
        I_1 &=   K_n^{-1/2} \sum_{k=1}^{K_n}    \hat{a}_k - a_k ~. \label{eq:I_1}   
\end{align}
 The  inequality \eqref{eq:inequality-appendix} and triangle inequality imply 
        \begin{equation*}    
            \| B \| \le \max_{1\le k \le K_n}  \left | \left| (\mathbb{I}_d + n_k^{-1/2} {b}_k)^{-1}  \right| \right| \times \max_{1\le k \le K_n}  \left | \left| (\mathbb{I}_d + n_k^{-1/2} \hat{b}_k)^{-1}  \right| \right| \times I_4~,
        \end{equation*}
    where $  I_4 = n^{-1/2} \sum_{k=1}^{K_n} \left | \left|    \hat{b}_k - b_k    \right| \right| \times \left | \left|     a_k  \right| \right| $. 

    We next show that $I_j = o_p(1)$ for $j=1,2,3,4$, which is sufficient to complete the proof of part (i) since Lemma \ref{lemma:sqrt-Kn} guarantees that both $ \max_{1\le k \le K_n}  \left | \left| (\mathbb{I}_d + n_k^{-1/2} {b}_k)^{-1}  \right| \right| $ and $ \max_{1\le k \le K_n}  \left | \left| (\mathbb{I}_d + n_k^{-1/2} \hat{b}_k)^{-1}  \right| \right| $ are $O_p(1)$ when $K_n = O(n^{1/2})$.

    \noindent \textbf{Claim 1:} $I_1 = o_p(1)$. We use Taylor expansion with Lagrange remainder term for each coordinate and notation defined in Appendix \ref{appendix:aux-results} to write $I_1 = I_{1,1} + I_{1,2}$, where
    \begin{align}
        I_{1,1} &= n^{-1/2} \sum_{k=1}^{K_n}  \sum_{i \in \mathcal{I}_k}  D_\eta m_i[\hat{\eta}_i - \eta_i] \label{e:I_{1,1}}\\
        I_{1,2} &=  n^{-1/2} \sum_{k=1}^{K_n} \sum_{i \in \mathcal{I}_k} \frac{1}{2}  D_\eta^2 \tilde{m}_i[\hat{\eta}_i - \eta_i, \hat{\eta}_i - \eta_i]~. \label{e:I_{1,2}}
    \end{align}
    By the Law of Iterated Expectations and part (iii) of Assumption \ref{asm:EM}, $E[I_{1,1}] = 0$. Let $e_j$ be the $j$-th column of the identity matrix $\mathbb{I}_d$. To conclude that $I_{1,1} = o_p(1)$, it is sufficient to show $E[(e_j^\intercal I_{1,1})^2] \to 0$. To see this, consider the following derivations,
    \begin{align*}
        E[(e_j^\intercal I_{1,1})^2] &\overset{(1)}{\le}  n^{-1} K_n \sum_{k=1}^{K_n}  E\left[ \left( \sum_{i \in \mathcal{I}_k}  e_j^\intercal \left( D_\eta m_i[\hat{\eta}_i - \eta_i] \right) \right)^2  \right] \\
        &\overset{(2)}{=}  n^{-1} K_n \sum_{k=1}^{K_n}  E\left[  \sum_{i \in \mathcal{I}_k}  \left( e_j^\intercal \left( D_\eta m_i[\hat{\eta}_i - \eta_i] \right) \right)^2  \right] \\
        &\overset{(3)}{\le} C  (n^{-1/2} K_n) n^{-1/2} \sum_{k=1}^{K_n}  \sum_{i \in \mathcal{I}_k} E\left [\left | \left|    \hat{\eta}_k(X_i) - \eta_0(X_i)  \right| \right|^2 \right ]  \overset{(4)}{=}  O(1) \times o(1)~,
    \end{align*} 
    where (1) holds by Jensen's inequality, (2) holds because $\{ e_j^\intercal \left( D_\eta m_i[\hat{\eta}_i - \eta_i] \right) : i \in \mathcal{I}_k\}$ are uncorrelated random variables, (3) holds by Lemma \ref{lemma:bounds-L2CR}, and (4) holds since $K_n = O(n^{1/2})$ and by Assumption \ref{asm:L2rate}.

    The next derivation shows that $I_{1,2} = o_p(1)$, 
    \begin{align}
        E[\|I_{1,2}\|]  \overset{(1)}{\le} C n^{-1/2} \sum_{k=1}^{K_n}  \sum_{i \in \mathcal{I}_k} E\left [\left | \left|    \hat{\eta}_k(X_i) - \eta_0(X_i)  \right| \right|^2 \right ]    \overset{(2)}{=} o(1) \label{eq:I_{1,2}-derivations}
    \end{align}
    where (1) holds by the triangle inequality and Lemma \ref{lemma:bounds-L2CR}, and (2) holds by Assumption \ref{asm:L2rate}.
    
     \noindent \textbf{Claim 2:} $I_2 = o_p(1)$. We first use Taylor expansion with Lagrange remainder term for each coordinate and notation defined in Appendix \ref{appendix:aux-results} to write
     \begin{align*}
         \hat{a}_k - a_k &= n_k^{-1/2} \sum_{i \in \mathcal{I}_k} D_\eta m_i[\hat{\eta}_i-\eta_i] + \frac{1}{2} D_\eta^2\tilde{m}_i[\hat{\eta}_i-\eta_i,\hat{\eta}_i-\eta_i] \\
         \hat{b}_k - b_k &= n_k^{-1/2} \sum_{i \in \mathcal{I}_k} D_\eta \psi^a_i[\hat{\eta}_i-\eta_i] + \frac{1}{2} D_\eta^2\tilde{\psi^a}_i[\hat{\eta}_i-\eta_i,\hat{\eta}_i-\eta_i]~.
     \end{align*} 
     Let $\mathcal{D}m_n$ and $\mathcal{D}\psi^a_n$ be as in Appendix \ref{appendix:aux-results}. 
     Then,
     \begin{align*}
         I_2 &\overset{(1)}{\le} (n^{-1/2} K_n) \times \mathcal{D}m_n \times \mathcal{D}\psi^a_n  + C \left( \mathcal{D}m_n + \mathcal{D}\psi^a_n \right) \times n^{-1/2} \sum_{k=1}^{K_n} \left (    n_k^{-1/2} \sum_{i \in \mathcal{I}_k} \left | \left| \hat{\eta}_i-\eta_i \right| \right|^2   \right) \\
         &+ C^2 n^{-1/2} \sum_{k=1}^{K_n} \left (    n_k^{-1/2} \sum_{i \in \mathcal{I}_k} \left | \left| \hat{\eta}_i-\eta_i \right| \right|^2   \right)^2 \\
         &\overset{(2)}{\le} O(1) o_p(1)   +  \left( K_n^{1/2}  n^{-1/2} \right) \times o_p(1) + O(1) \times  \left( n^{-1/2} \sum_{i=1}^n  \left | \left| \hat{\eta}_i-\eta_i \right| \right|^2 \right)^2  \overset{(3)}{=} o_p(1)~,
     \end{align*}
     where (1) holds by the triangle inequality and Lemma \ref{lemma:bounds-L2CR}, (2) and (3) hold by Lemmas \ref{lemma:sqrt-Kn} and \ref{lemma:Kn} and since $K_n = O(n^{1/2})$. This completes proof of claim 2.

    \noindent \textbf{Claim 3:} $I_3 = o_p(1)$. As in the proof of Claim 2, we use the Taylor expansion and $\mathcal{D}m_n$ defined in Appendix \ref{appendix:aux-results} to obtain,
    \begin{align*}
        I_3 &\overset{(1)}{\le}  (\mathcal{D}m_n ) \times n^{-1/2} \sum_{k=1}^{K_n}  \left | \left| b_k \right| \right| + C n^{-1/2} \sum_{k=1}^{K_n}  \left | \left| b_k \right| \right| \times \left (    n_k^{-1/2} \sum_{i \in \mathcal{I}_k} \left | \left| \hat{\eta}_i-\eta_i \right| \right|^2   \right) \\
        &\overset{(2)}{\le} o_p(1)  \times \left(n^{-1/2} \sum_{k=1}^{K_n}  \left | \left| b_k \right| \right| \right) \\
        &~ + C n^{-1}  K_n^{1/2} \left( \sum_{k=1}^{K_n}  \left | \left| b_k \right| \right|^2 \right)^{1/2} \times \left( \sum_{k=1}^{K_n}  \left (    \sum_{i \in \mathcal{I}_k} \left | \left| \hat{\eta}_i-\eta_i \right| \right|^2   \right) ^2 \right)^{1/2} \\
        &\overset{(3)}{\le} o_p(1) \times O_p(1) + n^{-1} K_n \times O_p(1) \times \left( \sum_{k=1}^{K_n}       \sum_{i \in \mathcal{I}_k} \left | \left| \hat{\eta}_i-\eta_i \right| \right|^2   \right)  \overset{(4)}{=} o_p(1)~,
    \end{align*}
    where (1) holds by the triangle inequality and Lemma \ref{lemma:bounds-L2CR}, (2) holds by Lemma \ref{lemma:sqrt-Kn} and the Cauchy-Schwarz inequality, (3) holds by part (ii) of Assumption \ref{asm:EM}, using the definition of $b_k$, and since $K_n = O(n^{1/2})$, and (4) holds by Lemma \ref{lemma:Kn} and since $K_n = O(n^{1/2})$.
     
     \noindent \textbf{Claim 4:} $I_4 = o_p(1)$. The proof is similar to Claim 3 but using $\mathcal{D}\psi^a_n$ instead of $\mathcal{D}m_n$.

    \noindent \textbf{\textit{Part (ii)}:}  Let $\hat{a}_k$, $\hat{b}_k $, ${a}_k $, and ${b}_k $ denote the quantities defined in Appendix \ref{appendix:aux-results}.   We write
    $ \sqrt{n} \left( \hat{\theta}_{n,K_n}^{(2)} -  \hat{\theta}_{n,K_n}^{*,(2)}\right) = A + B~,$
    where
    \begin{align*}
        A &= \left(  \mathbb{I}_d +  n_k^{-1/2} K_n^{-1}\sum_{k=1}^{K_n}  \hat{b}_k \right)^{-1} \left( K_n^{-1/2} \sum_{k=1}^{K_n} \hat{a}_k - a_k \right) \\
        B&= \left \{ \left( \mathbb{I}_d  +  n_k^{-1/2} K_n^{-1} \sum_{k=1}^{K_n}  \hat{b}_k \right)^{-1} - \left(  \mathbb{I}_d + n_k^{-1/2} K_n^{-1} \sum_{k=1}^{K_n}   {b}_k \right)^{-1}\right \} \left( K_n^{-1/2} \sum_{k=1}^{K_n}  a_k \right) 
    \end{align*}
%   %
    and $\hat{a}_k$, $\hat{b}_k $, ${a}_k $, and ${b}_k $ are defined in Appendix \ref{appendix:aux-results}.

    $A$ is $o_p(1)$ due to two results. First, $\left(  \mathbb{I}_d +  n_k^{-1/2} K_n^{-1}\sum_{k=1}^{K_n}  \hat{b}_k \right)^{-1}$ is $O_p(1)$ by Lemma \ref{lemma:Kn}. Second, $I_1 = K_n^{-1/2} \sum_{k=1}^{K_n} \hat{a}_k - a_k$ is $o_p(1)$ by claim 1 in the proof of part (i). 

    To show that $B$ is $o_p(1)$, we consider the following derivations
    \begin{align*}
        \|B\| &\overset{(1)}{\le} \left | \left| \left( \mathbb{I}_d  +  n_k^{-1/2} K_n^{-1} \sum_{k=1}^{K_n}  \hat{b}_k \right)^{-1} \right| \right| \times \left | \left| \left( \mathbb{I}_d  +  n_k^{-1/2} K_n^{-1} \sum_{k=1}^{K_n}   {b}_k \right)^{-1} \right| \right| \\
        & ~~\times \left | \left|   n^{-1/2} K_n^{-1/2} \sum_{k=1}^{K_n}  \hat{b}_k-b_k  \right| \right| \times \left | \left|   K_n^{-1/2} \sum_{k=1}^{K_n}  a_k  \right| \right| \\
        &\overset{(2)}{\le} O_p(1) \times  n^{-1/2}  \left | \left|  K_n^{-1/2} \sum_{k=1}^{K_n}  \hat{b}_k-b_k  \right| \right|  \overset{(3)}{=} o_p(1)~,
    \end{align*}
    where (1) holds by the inequality \eqref{eq:inequality-appendix} in Appendix \ref{appendix:aux-results}, (2) holds by Lemma \ref{lemma:Kn} and the Central Limit Theorem, and  (3) holds by  Lemma \ref{lemma:Kn}. 
\end{proof}

\begin{proof}[Proof of Theorem \ref{theorem:sqrt-Kn}]
   First,  Lemma \ref{lemma:sqrt-Kn-DML1-DML2} implies that, for $j=1,2$, we have
    $$  \sqrt{n} \left( \hat{\theta}_{n,K_n}^{(j)} -  \hat{\theta}_{n,K_n}^{*,(j)}  \right) \overset{p}{\to} 0~ \quad \text{as} \quad n \to \infty~.$$
    Second, since $\hat{\theta}_{n,K_n}^{*,(2)}$ = $\hat{\theta}_n^*$, we conclude $\sqrt{n}(\hat{\theta}_{n,K_n}^{*,(2)}-\theta_0) \overset{d}{\to} N(0,\Sigma)$ by standard arguments. 
    Finally, Theorem \ref{theorem:DML1-oracle} in Section \ref{sec:DML1-sensitive} demonstrates that $\sqrt{n}(\hat{\theta}_{n,K_n}^{*,(1)}-\theta_0) \overset{d}{\to} N(c\Lambda,\Sigma)$. 
\end{proof}

\subsubsection{Proof for Section \ref{sec:DML1-sensitive}}

\begin{proof}[Proof of Theorem \ref{theorem:DML1-oracle}]
    We first use the definition of $\hat{\theta}_{n,K_n}^{*,(1)}$ to write 
    \begin{align*}
        \sqrt{n} \left( \hat{\theta}_{n,K_n}^{*,(1)} - \theta_0 \right) &= K_n^{-1/2} \sum_{k=1}^{K_n} \left( \mathbb{I}_d + n_k^{-1/2} b_k \right)^{-1} a_k ~,
    \end{align*}
    where $a_k$ and $b_k$ are defined in Appendix \ref{appendix:aux-results}. We then apply identity \eqref{eq:identity-appendix} from Appendix \ref{appendix:aux-results} twice to write
    $ \sqrt{n} \left( \hat{\theta}_{n,K_n}^{*,(1)} - \theta_0 \right) - (K_n/\sqrt{n}) \Lambda = I_1 - I_2 + I_3$,
    % \begin{align*}
    %     \sqrt{n} \left( \hat{\theta}_{n,K_n}^{*,(1)} - \theta_0 \right) - (K_n/\sqrt{n}) \Lambda %&= K^{-1/2} \sum_{k=1}^K_na_k -  n_k^{-1/2} b_k a_k +(\mathbb{I}_k + n_k^{-1/2} b_k)^{-1} b_k^2 a_k \\
    %     %
    %     &= I_1 - I_2 + I_3
    % \end{align*}
%
where $I_1 =  K_n^{-1/2} \sum_{k=1}^{K_n} a_k $, $I_2 =  K_n^{-1/2} \sum_{k=1}^{K_n}    n_k^{-1/2} b_k a_k +  (K_n/\sqrt{n}) \Lambda$, and $I_3 =  K_n^{-1/2} \sum_{k=1}^{K_n} n_k^{-1} (\mathbb{I}_d + n_k^{-1/2} b_k)^{-1} b_k^2 a_k$.
% \begin{align*}
%     I_1 &=  K_n^{-1/2} \sum_{k=1}^{K_n} a_k    \\
%         %
%     I_2 &=  K_n^{-1/2} \sum_{k=1}^{K_n}    n_k^{-1/2} b_k a_k +  (K_n/\sqrt{n}) \Lambda \\
%     %
%     I_3 &=  K_n^{-1/2} \sum_{k=1}^{K_n} n_k^{-1} (\mathbb{I}_d + n_k^{-1/2} b_k)^{-1} b_k^2 a_k 
% \end{align*}

Claims 1 and 2 below show that $I_2 = o_p(1)$ and $I_3 = o_p(1)$, which is sufficient to complete the proof of this lemma since $I_1 \overset{d}{\to} N(0,\Sigma)$ by the Central Limit Theorem.

\noindent \textbf{Claim 1:} $I_2 = o_p(1)$. To show this, we first note that $E[I_2] = 0$ since $E[b_k a_k ] = - \Lambda$. It is sufficient to show that $E[\|I_2\|^2] \to 0$. Algebra shows
\begin{align*}
    E[\|I_2\|^2] &\overset{(1)}{=} E\left[ \left | \left| n^{-1/2}  \sum_{k=1}^{K_n}  (b_k a_k - E[b_k a_k] ) \right| \right|^2\right] \\
    &\overset{(2)}{=} n^{-1}  \sum_{k=1}^{K_n} E\left[ \left | \left|   (b_k a_k - E[b_k a_k] )  \right| \right|^2\right] \\
    &\overset{(3)}{\le} n^{-1} K_n  E\left[ \left | \left|  b_k a_k  \right| \right|^2\right] \\
    &\overset{(4)}{\le} n^{-1} K_n   E\left[ \left | \left|  b_k \right| \right|^4\right]^{1/2}  E\left[ \left | \left|  a_k  \right| \right|^4\right]^{1/2} \overset{(5)}{=}  n^{-1} K_n \times O(1) \times O(1)~,
\end{align*}
 where (1) holds since $E[b_k a_k ] = - \Lambda$, (2) and (3) hold because $\{ (b_k a_k - E[b_k a_k]) : 1 \le k \le K_n\}$ are i.i.d. zero mean random vectors, (4) holds by Cauchy-Schwarz inequality, and (5) holds by part (ii) Assumption \ref{asm:EM} and using the definition of $a_k$ and $b_k$. Therefore,  $ E[\|I_2\|^2] = O(n^{-1/2})$ since $K_n = O(n^{1/2})$. 

\noindent \textbf{Claim 2:} $I_3 = o_p(1)$. To show this, first note that
$$ \| I_3 \| \le  \max_{k=1,\ldots,K_n}  \left | \left| (\mathbb{I}_d + n_k^{-1/2} b_k)^{-1}  \right| \right| \times K_n^{-1/2} \sum_{k=1}^{K_n} n_k^{-1} \| b_k^2 a_k\| ~, $$
where $ \max_{k=1,\ldots,K_n}  \left | \left| (\mathbb{I}_d + n_k^{-1/2} b_k)^{-1}  \right| \right|  = O_p(1)$ due to Lemma \ref{lemma:sqrt-Kn}. Then, it is sufficient to show that $  K_n^{-1/2} \sum_{k=1}^{K_n} n_k^{-1} \| b_k^2 a_k\|  = o_p(1)$, which holds by the next derivations
\begin{align*}
    E[K_n^{-1/2} \sum_{k=1}^{K_n} n_k^{-1} \| b_k^2 a_k\| ] &\overset{(1)}{\le} K_n^{3/2} n ^{-1}E[\| b_k\|^4]^{1/2} E[ \|a_k\|^2]^{1/2} \\
    &\overset{(2)}{\le}  K_n^{3/2} n ^{-1} \times O(1) \times O(1)  \overset{(3)}{=} O(n^{-1/4})~,
\end{align*}
where (1) holds because $\{ b_k^2 a_k : 1 \le k \le K_n\}$ are i.i.d. random vectors and Cauchy-Schwarz inequality, (2) holds by part (ii) of Assumption \ref{asm:EM} and using the definition of $a_k$ and $b_k$, and (3) holds since $K_n = O(n^{1/2})$. This completes the proof of claim 2. 
\end{proof}

\subsubsection{Proof for Section \ref{sec:DML2-valid}}

\begin{lemma} \label{lemma:DML2_Kn}
Let Assumptions \ref{asm:EM}~(i)--(iii), \ref{asm:L2rate}, and \ref{asm:AS} hold and let $K_n$ be such that~$2\le K_n\le n$. Then, equation \eqref{eq:DML2-oracle-equivalence} holds. 
\end{lemma}
\begin{proof}
    \noindent The proof of part (ii) in Lemma \ref{lemma:sqrt-Kn-DML1-DML2} relies on Lemma \ref{lemma:Kn} and $I_1 = o_p(1)$, where $I_1$ is defined in \eqref{eq:I_1}. Lemma \ref{lemma:Kn} holds for $K_n = O(n)$, but the proof of $I_1=o_p(1)$ relies on $K_n = O(n^{1/2})$. Therefore, the validity of the previous proof does not apply to the case $K_n = O(n)$. To adapt the proof of part (ii) in Lemma \ref{lemma:sqrt-Kn-DML1-DML2}, we show that $I_1=o_p(1)$ also holds when $K_n = O(n)$, provided we add Assumption \ref{asm:AS}.
    
    Recall that $I_1 = I_{1,1} + I_{1,2}$, where $I_{1,1}$ and $I_{1,2}$ are defined in \eqref{e:I_{1,1}} and \eqref{e:I_{1,2}}, respectively. Note that the proof of $I_{1,2} = o_p(1) $ also applies when $K_n = O(n)$; see derivations in \eqref{eq:I_{1,2}-derivations}. Therefore, it is sufficient to show that $I_{1,1} = o_p(1)$. Since $E[I_{1,1}] = 0$, it is sufficient to show that $E[(e_j^\intercal I_{1,1})^2] = o(1)$, where $e_j$ is the $j$-th column of $\mathbb{I}_d$.   
    Consider the next derivations,
    \begin{align*}
        E[(e_j^\intercal I_{1,1})^2] &= E\left[ \left( n^{-1/2} \sum_{k=1}^{K_n}  \sum_{i \in \mathcal{I}_k}  e_j^\intercal (D_\eta m_i[\hat{\eta}_i - \eta_i]) \right)^2 \right]  
        %
       % &= n^{-1} \sum_{k_1=1}^{K_n} \sum_{k_2=1}^{K_n} \sum_{i_1 \in \mathcal{I}_{k_1}} \sum_{i_2 \in \mathcal{I}_{k_2}}  E[e_j^\intercal (D_\eta m_{i_1}[\hat{\eta}_{i_1} - \eta_{i_1}])e_j^\intercal (D_\eta m_{i_2}[\hat{\eta}_{i_2} - \eta_{i_2}])] \\
        %
         = I_{1,1,1} + I_{1,1,2} + I_{1,1,3} 
    \end{align*}
    where 
    \begin{align*}
        I_{1,1,1} &= n^{-1} \sum_{k=1}^{K_n}  \sum_{i \in \mathcal{I}_k} E\left[    \left( e_j^\intercal (D_\eta m_i[\hat{\eta}_i - \eta_i]) \right)^2 \right] \\
        I_{1,1,2} &=  n^{-1} \sum_{k=1}^{K_n}  \sum_{i_1,i_2 \in \mathcal{I}_k} E[e_j^\intercal (D_\eta m_{i_1}[\hat{\eta}_{i_1} - \eta_{i_1}])e_j^\intercal (D_\eta m_{i_2}[\hat{\eta}_{i_2} - \eta_{i_2}])] I\{i_1 \neq i_2\} \\
        I_{1,1,3} &=  n^{-1} \sum_{k_1,k_2=1}^{K_n}  \sum_{i_1 \in \mathcal{I}_{k_1}} \sum_{i_2 \in \mathcal{I}_{k_2}}  E[e_j^\intercal (D_\eta m_{i_1}[\hat{\eta}_{i_1} - \eta_{i_1}])e_j^\intercal (D_\eta m_{i_2}[\hat{\eta}_{i_2} - \eta_{i_2}])]  I\{k_1 \neq k_2\}
    \end{align*} 

    Note that $I_{1,1,1} = o(1)$ and $I_{1,1,2} = 0$. The former holds by Lemma \ref{lemma:bounds-L2CR} and Assumption \ref{asm:L2rate}, while the latter by part (iii) of Assumption \ref{asm:EM} and the Law of Iterated Expectations.

    We now show that $I_{1,1,3} = o(1)$ using Assumption \ref{asm:AS}. We proceed in three steps. First, for $i_1 \in \mathcal{I}_{k_1}$, $i_2 \in \mathcal{I}_{k_2}$, and $k_1 \neq k_2$, let $\hat{\eta}_{i_1}^{i_2} = \hat{\eta}_{k_1}^{i_2}(X_{i_1})$ and $\hat{\eta}_{i_2}^{i_1} = \hat{\eta}_{k_2}^{i_1}(X_{i_2})$. We have
    $$ E[e_j^\intercal (D_\eta m_{i_1}[\hat{\eta}_{i_1}^{i_2} - \eta_{i_1}])e_j^\intercal (D_\eta m_{i_2}[\hat{\eta}_{i_2} - \eta_{i_2}])] = 0~,$$
    which holds by the Law of Iterated Expectations (conditional on $X_{i_2}$ and $\{ W_{i}: 1\le i \le n, i \neq  i_2\}$), part (iii) of Assumption \ref{asm:EM}, and the definition of $\hat{\eta}_{k_1}^{i_2}(X_{i_1})$. Second, we use that $D_\eta m_{i_1}$ is a linear operator (i.e., $D_\eta m_{i_1}[\hat{\eta}_{i_1} - {\eta}_{i_1}] = D_\eta m_{i_1}[\hat{\eta}_{i_1} - \hat{\eta}_{i_1}^{i_2}] + D_\eta m_{i_1}[\hat{\eta}_{i_1}^{i_2} - {\eta}_{i_1}]$) and the previous step to write
    $$ I_{1,1,3} =  n^{-1} \sum_{k_1,k_2=1}^{K_n}  \sum_{i_1 \in \mathcal{I}_{k_1}} \sum_{i_2 \in \mathcal{I}_{k_2}}  E[e_j^\intercal (D_\eta m_{i_1}[\hat{\eta}_{i_1} - \hat{\eta}_{i_1}^{i_2}])e_j^\intercal (D_\eta m_{i_2}[\hat{\eta}_{i_2} - \hat{\eta}_{i_2}^{i_1}])]  I\{k_1 \neq k_2\}~.$$
    Finally, we use the previous step, the Cauchy-Schwarz inequality, and Lemma \ref{lemma:bounds-L2CR} to obtain
    \begin{align*}
        I_{1,1,3} &\le  C n^{-1} \sum_{k_1,k_2=1}^{K_n}  \sum_{i_1 \in \mathcal{I}_{k_1}} \sum_{i_2 \in \mathcal{I}_{k_2}}  E\left[ \left | \left| \hat{\eta}_{i_1}-\hat{\eta}_{i_1}^{i_2} \right| \right|^2 \right]^{1/2} E\left[ \left | \left| \hat{\eta}_{i_2}-\hat{\eta}_{i_2}^{i_1} \right| \right|^2 \right]^{1/2}  \overset{(1)}{=} o(1)~,
    \end{align*}
    where (1) holds by Assumption \ref{asm:AS}. This completes the proof of $I_{1}=o_p(1)$.   
\end{proof} 

\begin{proof}[Proof of Theorem \ref{theorem:DML2-Kn}]
    By Lemma \ref{lemma:DML2_Kn}, $  \sqrt{n} \left( \hat{\theta}_{n,K_n}^{(2)} -  \hat{\theta}_{n,K_n}^{*,(2)}  \right) \overset{p}{\to} 0~,$ for $K_n = O(n)$, which is sufficient to conclude the theorem since $\sqrt{n}(\hat{\theta}_{n,K_n}^{*,(2)}-\theta_0) \overset{d}{\to} N(0,\Sigma)$.
\end{proof}
 
\begin{proof}[Proof of Corollary \ref{corr:CI-DML2}]
    Note that in both cases we have  $\hat{\theta}^{(2)}_{n,K_n} \overset{p}{\to} \theta_0$. Lemma~\ref{lemma:variance-consistency} with $\bar{\theta}_n = \hat{\theta}^{(2)}_{n,K_n}$ gives $\hat{G}_n \overset{p}{\to} G$ and $\hat{\Omega}_n \overset{p}{\to} \Omega$. Since $G$ is nonsingular (Assumption~\ref{asm:EM}(i)), $\hat{G}_n$ is nonsingular with probability approaching one, and $\hat{\Sigma}_n \overset{p}{\to} \Sigma$ by the Continuous Mapping Theorem. Since $\Omega$ is nonsingular, $v^\intercal \Sigma v > 0$ for any $v \neq 0$, and the conclusion follows from Slutsky's theorem.  
\end{proof}  

\subsection{Proofs for Section \ref{sec:result2}}\label{appendix:main-results2}

\begin{proof}[Proof of Theorem \ref{thm:stochastic-expansion-DML2}] Recall that $\varphi \in (1/4,1/2)$ by Assumption~\ref{asm:SLE}. By similar arguments as in \citet{newey2004higher}, we have $\sqrt{n} \left( \hat{\theta}_{n,K_n}^{*,(2)} - \theta_0 \right) = \mathcal{T}_n^* + \mathcal{T}_n^{*,(2)} + O_p(n^{-1})$. Therefore, it is sufficient to show  $\sqrt{n} \left( \hat{\theta}_{n,K_n}^{(2)} - \hat{\theta}_{n,K_n}^{*,(2)} \right) = \mathcal{T}_{n,K_n}^{\ell} + \mathcal{T}_{n,K_n}^{q} + O_p(n^{1/2-3\varphi})$ since $O_p(n^{-1}) = o_p(n^{1/2-3\varphi})$. To show this, we write
$ 
        \sqrt{n} \left( \hat{\theta}_{n,K_n}^{(2)} - \hat{\theta}_{n,K_n}^{*,(2)} \right) = A + B$, 
where
    \begin{align*}
        A &= \left( 1 +  n_k^{-1/2} K_n^{-1}\sum_{k=1}^{K_n}  \hat{b}_k \right)^{-1} \left( K_n^{-1/2} \sum_{k=1}^{K_n} \hat{a}_k - a_k \right) \\
        B&= \left \{ \left( 1 +  n_k^{-1/2} K_n^{-1} \sum_{k=1}^{K_n}  \hat{b}_k \right)^{-1} - \left( 1+ n_k^{-1/2} K_n^{-1} \sum_{k=1}^{K_n}   {b}_k \right)^{-1}\right \} \left( K_n^{-1/2} \sum_{k=1}^{K_n}  a_k \right)~. 
    \end{align*}
    Claims 1 and 2 below show that $A = \mathcal{T}_{n,K_n}^{\ell} + \mathcal{T}_{n,K_n}^{q} + O_p(n^{1/2-3\varphi})$  and $B = O_p(n^{-2\varphi})$, respectively, which is sufficient to complete the proof since $O_p(n^{-2\varphi}) = o_p(n^{1/2-3\varphi})$.

    \noindent \textbf{Claim 1:} $A = \mathcal{T}_{n,K_n}^{\ell} + \mathcal{T}_{n,K_n}^{q} + O_p(n^{1/2-3\varphi})$. To show this, we first use the definition of $\hat{a}_k$ and $a_k$ to write 
    $ K_n^{-1/2} \sum_{k=1}^{K_n} \hat{a}_k - a_k = n^{-1/2} \sum_{i=1}^n G^{-1}\left( \widehat{m}_i - m_i \right)$,
    and we then use Lemma~\ref{lemma:approximation-dml2}(i) and (iv) to write 
    $$ A = (1+ O_p(n^{-1/2})) \left( \mathcal{T}_{n,K_n}^{\ell} + \mathcal{T}_{n,K_n}^{q} + O_p(n^{1/2-3\varphi}) \right)~.$$
    It is sufficient to show that $O_p(n^{-1/2})\left(\mathcal{T}_{n,K_n}^{\ell} + \mathcal{T}_{n,K_n}^{q} \right) = O_p(n^{1/2-3\varphi})$ to conclude. This holds because  $\mathcal{T}_{n,K_n}^{\ell} = O_p(n^{-\varphi})$ by Lemma~\ref{lemma:moments-so-dml2}(i), $\mathcal{T}_{n,K_n}^{q}-E[\mathcal{T}_{n,K_n}^{q}] = O_p(n^{-\varphi})$ by Assumption~\ref{asm:leading-terms}(iv), and $E[\mathcal{T}_{n,K_n}^{q}] = O(n^{1/2-2\varphi})$ by Lemma \ref{lemma:moments-so-dml2}(i). Note that these results also imply that $\mathcal{T}_{n,K_n}-\mathcal{T}_n^* = O_p(n^{-\varphi})+O(n^{1/2-2\varphi})$.

    \noindent \textbf{Claim 2:} $B = O_p(n^{-2\varphi})$. To show this, we first use the definition of $\hat{b}_k$ and $b_k$ to write $ n_k^{-1/2} K_n^{-1} \sum_{k=1}^{K_n} (\hat{b}_k - b_k) = n^{-1} \sum_{i=1}^n G^{-1} \left( \hat{\psi}^a_i - {\psi}^a_i  \right)$, and we then  write
    \begin{align*}
        |B| &\overset{(1)}{\le} O_p(1) \times \left  | n^{-1} \sum_{i=1}^n G^{-1} \left( \hat{\psi}^a_i - {\psi}^a_i  \right)\right| \times \left| K_n^{-1/2} \sum_{k=1}^{K_n}  a_k  \right| \overset{(2)}{=}  O_p(1) \times  O_p(n^{-2\varphi}) \times   O_p(1) ~,
    \end{align*}
    where (1) holds by the inequality \eqref{eq:inequality-appendix}, Lemma~\ref{lemma:approximation-dml2}(iii) and (iv), and (2) holds by Lemma~\ref{lemma:approximation-dml2}(ii) and the definition of $a_k$ and the Central Limit Theorem. 
\end{proof}

\begin{proof}[Proof of Theorem \ref{theorem:so-bias-mse-DML2}]
    Recall that $\varphi \in (1/4,1/2)$ by Assumption~\ref{asm:SLE} and $N \asymp n$ since $N =(1-\tfrac{1}{K_n})n$. 
    Note that \eqref{eq:bias-DML2} follows by Proposition \ref{prop:so-bias-variance-dml2}, Assumption~\ref{asm:leading-terms} ($\mathcal{B}_N = \mathcal{B} + o(1)$), and since $n^{-1/2} = o(n^{1/2-2\varphi})$. To prove that \eqref{eq:mse-DML2} holds, it is sufficient to show that 
    \begin{equation}\label{eq:aux1_proof_theorem-so-bias-mse-DML2}
        Var[\mathcal{T}_{n,K_n}] = \Sigma + o(n^{1-4\varphi})
    \end{equation}
     since $E[\mathcal{T}_{n,K_n}^2] = E[\mathcal{T}_{n,K_n}]^2 + Var[\mathcal{T}_{n,K_n}]$ and \eqref{eq:bias-DML2} holds.

    When $\varphi \in (1/4,1/3)$ holds, Proposition \ref{prop:so-bias-variance-dml2} and $\mathcal{C}_N = O(1)$ imply \eqref{eq:aux1_proof_theorem-so-bias-mse-DML2}. 
    
    When Assumption~\ref{asm:information} holds, Proposition \ref{prop:so-bias-variance-dml2} and \eqref{eq:aux2_proof_theorem-so-bias-mse-DML2} imply \eqref{eq:aux1_proof_theorem-so-bias-mse-DML2} since $n^{-2\varphi} = o(n^{1-4\varphi})$.%, where
    \begin{equation}\label{eq:aux2_proof_theorem-so-bias-mse-DML2}
        \mathcal{C}_N = O(N^{-\varphi})~.
    \end{equation}
    To verify \eqref{eq:aux2_proof_theorem-so-bias-mse-DML2}, we consider the following derivations
    \begin{align*}
        | \mathcal{C}_N | % &= \tfrac{1}{2} E\left[   \tilde{b}_N(X_i)^\intercal \left( S(X_i) + H(X_i) ~\tilde{\delta}_N(X_i)\right) \right] \\
        &\overset{(1)}{=} \left| \tfrac{1}{2} E\left[  \left( \tilde{b}_N(X_i)^\intercal   H(X_i) \right) \left( \tilde{\delta}_N(X_i) - \tilde{\delta}_{\infty}(X_i)  \right) \right] \right|\\
        &\overset{(2)}{\le} \tfrac{1}{2} E\left[  \left| \left| \tilde{b}_N(X_i)^\intercal   H(X_i)    \right| \right|^2 \right]^{1/2}  E\left[  \left| \left| \tilde{\delta}_N(X_i) - \tilde{\delta}_{\infty}(X_i)    \right| \right|^2 \right]^{1/2} \overset{(3)}{=} O(1) \times O(N^{-\varphi})~,
    \end{align*}
    where (1) holds by Assumption~\ref{asm:information}(ii), (2) follows by the Cauchy-Schwarz inequality, and (3) by Assumptions~\ref{asm:EM}(iii), \ref{asm:SLE}(ii), and \ref{asm:information}(i).
\end{proof}
 
\section{Supporting Technical Results}\label{appendix:aux-results}
\subsection{Auxiliary results for Section \ref{sec:result1}}\label{appendix:aux-results1}

We use the following notation in Appendix \ref{appendix:main-results}. Let $e_j$ be the $j$-th column of the identity matrix $\mathbb{I}_d \in \mathbf{R}^{d \times d}$. Recall $\cup_{k=1}^K \mathcal{I}_k=[n]$. For every $k \in [K]$ define
    \begin{align*}
         a_k &=   n_k^{-1/2} \sum_{i \in \mathcal{I}_k} G^{-1} m_i~, \qquad  b_k  =  n_k^{-1/2} \sum_{i \in \mathcal{I}_k} \left( G^{-1}\psi^a_i - \mathbb{I}_d\right)~,    \\
        \hat{a}_k &=   n_k^{-1/2} \sum_{i \in \mathcal{I}_k} G^{-1} \widehat{m}_i~, \qquad \hat{b}_k  =  n_k^{-1/2} \sum_{i \in \mathcal{I}_k} \left( G^{-1} \hat{\psi}^a_i - \mathbb{I}_d\right)~. 
        \end{align*} 
For the Taylor-expansion terms, we write        
        \begin{align*}  
        D_\eta m_i[\hat{\eta}_i-\eta_i] &= \sum_{t=1}^d (\hat{\eta}_i - \eta_i)^\intercal \partial_\eta m_t(W_i,\theta_0,\eta_i) \cdot e_t~, \\
        D_\eta \psi_i^a[\hat{\eta}_i-\eta_i] &=  \sum_{s=1}^d \sum_{t=1}^d (\hat{\eta}_i - \eta_i)^\intercal \partial_\eta\psi_{t,s}^a(W_i,\theta_0,\eta_i) \cdot e_t ~e_s^\intercal~, \\
        D_\eta^2 \tilde{m}_i[\hat{\eta}_i-\eta_i,\hat{\eta}_i-\eta_i] &= \sum_{t=1}^d (\hat{\eta}_i - \eta_i)^\intercal \partial_\eta^2 m_t(W_i,\theta_0,\tilde{\eta}_{t,i}) (\hat{\eta}_i - \eta_i) \cdot e_t~, \\
        D_\eta^2 \psi_i^a[\hat{\eta}_i-\eta_i,\hat{\eta}_i-\eta_i] &=  \sum_{s=1}^d \sum_{t=1}^d (\hat{\eta}_i - \eta_i)^\intercal \partial_\eta^2\psi_{t,s}^a(W_i,\theta_0,\tilde{\eta}_{t,s,i}) (\hat{\eta}_i - \eta_i) \cdot e_t~ e_s^\intercal ~,
    \end{align*}
where $\tilde{\eta}_{t,i}$ and $\tilde{\eta}_{t,s,i}$ are in $\mathcal{E}$ and exist due to Taylor expansions with Lagrange remainder for each $t$ and $s$.
Using this notation, define  
\begin{align*}
        \mathcal{D}m_n &= \max_{1\le k \le K_n} \left | \left|    n_k^{-1/2} \sum_{i \in \mathcal{I}_k} D_\eta m_i[\hat{\eta}_i-\eta_i]   \right| \right| ~,\\
        \mathcal{D}\psi^a_n &= \max_{1\le k \le K_n} \left | \left|    n_k^{-1/2} \sum_{i \in \mathcal{I}_k} D_\eta \psi^a_i[\hat{\eta}_i-\eta_i]   \right| \right| ~.
\end{align*}

We also use the next matrix identity 
\begin{equation}\label{eq:identity-appendix} 
   (\mathbb{I}_d + \mathbb{M})^{-1} = \mathbb{I}_d - (\mathbb{I}_d + \mathbb{M})^{-1}\mathbb{M} 
\end{equation}
and matrix inequality for square matrices $\mathbb{M}$ and $\mathbb{M}_1$
\begin{equation}\label{eq:inequality-appendix} 
     \left| \left| (\mathbb{I}_d + \mathbb{M})^{-1} - (\mathbb{I}_d + \mathbb{M}_1)^{-1}  \right| \right|  \le \left| \left| (\mathbb{I}_d + \mathbb{M})^{-1}  \right| \right| \cdot \left| \left| \mathbb{M}  - \mathbb{M}_1  \right| \right| \cdot \left| \left|  (\mathbb{I}_d + \mathbb{M}_1)^{-1}  \right| \right| ~.
\end{equation}
 
The proofs of the following lemmas appear in Appendix \ref{proofs:appendix-result1}.

\begin{lemma}\label{lemma:bounds-L2CR} 
    Let Assumption \ref{asm:EM} hold and let $\gamma: \mathcal{X} \to \mathbf{R}^p$ be a given function. Then, there exists a constant $C>0$ independent of $\gamma$ such that
    \begin{enumerate}
         \item[(i)] $E[ (e_j^\intercal (D_\eta m_{i}[  \gamma_i]) )^2 ] \le C E \left[\left | \left| \gamma_i \right| \right|^2 \right]$  for $i \in \mathcal{I}_k$ and $\gamma_i = \gamma(X_i)$.

        \item[(ii)] $ \frac{1}{2}\left | \left| D_\eta^2\tilde{m}_i[ \gamma_i , \gamma_i ] \right| \right| \le C \left | \left|  \gamma_i \right| \right|^2$ for $i \in \mathcal{I}_k$ and $\gamma_i = \gamma(X_i)$.

        \item[(iii)] $  \frac{1}{2}\left | \left| D_\eta^2\tilde{\psi^a}_i[ \gamma_i , \gamma_i ] \right| \right| \le C \left | \left|  \gamma_i  \right| \right|^2 $ for $i \in \mathcal{I}_k$ and $\gamma_i = \gamma(X_i)$.
    \end{enumerate} 
\end{lemma}

\begin{lemma}\label{lemma:sqrt-Kn}
Let Assumptions \ref{asm:EM} and \ref{asm:L2rate} hold and let $K_n$ be such that $2 \le K_n \le n$ and $K_n = O(\sqrt{n})$. Then, 
\begin{enumerate}
    \item[(i)] $\mathcal{D}m_n = o_p(1)$ and  $\mathcal{D}\psi^a_n = o_p(1)$.
    
    \item[(ii)] $ \max_{1\le k \le K_n}  \left | \left| (\mathbb{I}_d + n_k^{-1/2} {b}_k)^{-1}  \right| \right|  = O_p(1)$ .
    
    \item[(iii)] $ \max_{1\le k \le K_n}  \left | \left| (\mathbb{I}_d + n_k^{-1/2} \hat{b}_k)^{-1}  \right| \right|  = O_p(1) $ .
\end{enumerate}
\end{lemma}

\begin{lemma}\label{lemma:Kn}
Let Assumptions \ref{asm:EM}(i)--(iii) and \ref{asm:L2rate} hold and let $K_n$ be such that $K_n \le n$. Then,
\begin{enumerate}
    \item[(i)]   $n^{-1/2} \sum_{i=1}^n \left | \left| \hat{\eta}_i-\eta_i \right| \right|^2 = o_p(1)$.
    
    \item[(ii)] $ n^{-1/2} K_n^{-1/2} \sum_{k=1}^{K_n}  \hat{b}_k-b_k  =  o_p\left( 1 \right) $.
     
   \item[(iii)]  $\left(  \mathbb{I}_d +  n_k^{-1/2} K_n^{-1}\sum_{k=1}^{K_n}  {b}_k \right)^{-1}$ and $\left(  \mathbb{I}_d +  n_k^{-1/2} K_n^{-1}\sum_{k=1}^{K_n}  \hat{b}_k \right)^{-1}$ are $O_p(1)$.
   
\end{enumerate} 
\end{lemma}

\begin{lemma}\label{lemma:variance-consistency}
Let Assumptions~\ref{asm:EM} and \ref{asm:L2rate} hold and let $K_n$ be such that $2 \le K_n \le n$. If $\bar{\theta}_n \overset{p}{\to} \theta_0$, then,
\begin{enumerate}
    \item[(i)] $\hat{G}_n \overset{p}{\to} G$.
    \item[(ii)]  $n^{-1}\sum_{i=1}^n m(W_i,\bar{\theta}_n,\hat{\eta}_i)\, m(W_i,\bar{\theta}_n,\hat{\eta}_i)^\intercal \overset{p}{\to} \Omega$.
\end{enumerate} 
\end{lemma}

\subsection{Auxiliary results for Section \ref{sec:result2}}\label{appendix:aux-results2}

Recall  $H(x) = G^{-1}E[ \partial_\eta^2 m_i \mid X_i=x]$, $\tilde{\delta}_N(x) = N^{1/2-\varphi} E[(G^{-1}m_j) \delta_N(W_j,x)]$, and  $\tilde{b}_N(x) = E[b_N(X_j,x)]$. To describe $ E[\mathcal{T}_{n,K_n}]$ and $Var[ \mathcal{T}_{n,K_n}]$, consider the next notation:  
    \begin{align*}
        \mathcal{B}_N  &= E\left[ \left( \delta_N(W_j,X_i) + \tilde{b}_N(X_i)\right)^\intercal H(X_i) \left( \delta_N(W_j,X_i) + \tilde{b}_N(X_i)\right) \right] ~,\\
        \mathcal{C}_{N} &=   E\left[   \tilde{b}_N(X_i)^\intercal \left( S(X_i) + H(X_i) ~\tilde{\delta}_N(X_i)\right) \right]~.
    \end{align*} 
The proofs of the following results appear in Appendix~\ref{proofs:appendix-result2}.
    
\begin{prop}\label{prop:so-bias-variance-dml2}
    Let Assumptions \ref{asm:EM}, \ref{asm:SLE}, and \ref{asm:leading-terms} hold and let $K_n$ be such that $2 \le K_n \le n$.   
    Then, both $\mathcal{B}_N$ and $\mathcal{C}_N$ are $O(1)$, and
    \begin{align}
        E[\mathcal{T}_{n,K_n}] &=  \tfrac{1}{2}\mathcal{B}_N \left( 1 + \frac{1}{K_n-1} \right)^{2\varphi} n^{1/2-2\varphi}  + \Lambda n^{-1/2} + o(n^{-1/2})~, \label{eq:bias-so-DML2-exact}\\
        %3
         Var[ \mathcal{T}_{n,K_n}]  &= \Sigma +  2 \mathcal{C}_{N}  \left( 1 + \frac{1}{K_n-1} \right)^{\varphi}  n^{-\varphi}  + O(n^{-2\varphi})~,\label{eq:variance-so-DML2}
    \end{align} 
    where  $\Sigma$ and $\Lambda$ are as in \eqref{eq:asymptotic-variance} and \eqref{eq:Lambda}, respectively.
\end{prop}

\begin{lemma}\label{lemma:moments-so-dml2}
    Let Assumptions \ref{asm:EM}, \ref{asm:SLE}, and \ref{asm:leading-terms} hold and let  $K_n$ be such that $2\le K_n \le n$. Then,
    \begin{enumerate}
        \item[(i)]  $E\left[ \mathcal{T}_{n,K_n}^{q} \right] = \tfrac{1}{2}\mathcal{B} \left( \tfrac{K_n}{K_n-1}\right)^{2\varphi} n^{1/2-2\varphi} + o\left( n^{1/2-2\varphi} \right) $ .
        
        \item[(ii)]  $Var[\mathcal{T}_{n,K_n}^{\ell}] = \left( \tfrac{1}{2}\mathcal{V}^{(\ell)} + E\left[ \left( \tilde{b}_N(X_i)^\intercal (G^{-1} \partial_\eta m_i) \right)^2 \right] \right)   \left( 1 +  \tfrac{1}{K_n-1}\right)^{2\varphi} n^{-2\varphi} +  o\left( n^{-2\varphi} \right)  $. 
    \end{enumerate}
\end{lemma}

 \begin{lemma}\label{lemma:approximation-dml2}
    Let Assumptions \ref{asm:EM}, \ref{asm:SLE}, \ref{asm:leading-terms}, and \ref{asm:AS-SO} hold and let  $K_n$ be such that $2 \le K_n \le n$. Then,
    \begin{enumerate}
        \item[(i)] $n^{-1/2} \sum_{i=1}^n G^{-1}\left( m(W_i,\theta_0,\hat{\eta}_i)- m(W_i,\theta_0,{\eta}_i) \right) = \mathcal{T}_{n,K_n}^{\ell} + \mathcal{T}_{n,K_n}^{q} + O_p\left( n^{1/2-3\varphi} \right) $.
        
        \item[(ii)]  $n^{-1} \sum_{i=1}^n G^{-1}\left( \psi^a(W_i,\hat{\eta}_i)- \psi^a(W_i,{\eta}_i) \right) = O_p\left( n^{-2\varphi} \right) $.

        \item[(iii)] $\left(  1 +  n_k^{-1/2} K_n^{-1}\sum_{k=1}^{K_n}  {b}_k \right)^{-1} = O(1)$. 

        \item[(iv)] $\left(  1 +  n_k^{-1/2} K_n^{-1}\sum_{k=1}^{K_n}  \hat{b}_k \right)^{-1} = 1+ O_p(n^{-1/2})$.
    \end{enumerate}
\end{lemma}
  
\section{Examples} \label{sec:appendix_examples}

\begin{example}[Average Treatment Effect]\label{example_ate}
Let $A \in \{0,1\}$ denote a binary treatment status, $Y(a)$ denote the potential outcome under treatment $a \in \{0,1\}$, $X$ denote a vector of covariates, and 
$Y = A Y(1) + (1-A) Y(0)$
denote the observed outcome. The available data is modeled by the vector $W = (Y,A,X)$. 
The parameter of interest is 
$\theta_0 = E[Y(1) - Y(0)]~,$
which is the expectation of the treatment effect when the treatment is mandated across the entire population, also known as the ATE. 
Under the selection-on-observables assumption, 
$(Y(1),Y(0)) \perp A \mid X,$
and the overlap condition, 
the ATE can be identified by a moment condition such as \eqref{eq:moment_for_theta} using a moment function like \eqref{eq:moment_funct} with $\psi^a(W,\eta) = 1$ and
\begin{align*}
        \psi^b(W,\eta) &= \eta_1 - \eta_2 + A(Y-\eta_1) \eta_3 - (1-A)(Y-\eta_2) \eta_4~,
\end{align*}
%$$
for $\eta \in \mathbf{R}^4$, and where the nuisance parameter $\eta_0(X)$ has four components:%
\begin{align*}
        \eta_{0}(X) = (E[Y \mid X, A=1],~ E[Y \mid X, A=0],~ (E[A \mid X])^{-1},~ (E[1 -A \mid X])^{-1})^\intercal.
\end{align*} 
This moment function corresponds to the augmented inverse propensity weighted (AIPW) estimator (\cite{robins1994estimation}, \cite{scharfstein1999adjusting}). It also appears as the efficient influence function for the ATE in  \cite{hahn1998role} and \cite{hirano2003efficient}.  
\end{example}

\begin{example}[Difference-in-Differences] \label{example_DID}
This example considers the average treatment effect on the treated in difference-in-differences research designs with two periods and panel data. Let $A \in \{0,1\}$ denote a binary treatment status on the post-treatment period, $Y_1(a)$ denote the potential outcome on the post-treatment period under treatment status $ a \in \{0,1\}$, $Y_0$ denote the outcome of interest in a pre-treatment period, $X$ denote a vector of covariates, and 
$Y_1 = A Y_1(1) + (1-A) Y_1(0)~$
denote the observed outcome in the post-treatment period. The available data is modeled by the vector $W = (Y_0, Y_1, A, X)$. The parameter of interest is 
$\theta_0 = E[Y_1(1) - Y_1(0) \mid A=1]~,$
which represents the treatment effect for the treated group in the post-treatment period, also known as ATT-DID.  \cite{sant2020doubly} used a conditional parallel trend assumption,
$ E[Y_1(0) - Y_0 \mid X, A=1 ] = E[Y_1(0) - Y_0 \mid X, A=0 ], $
and the treatment overlap condition, to identify the ATT-DID by a moment condition, such as \eqref{eq:moment_for_theta}, using a moment function like \eqref{eq:moment_funct} with $\psi^a(W,\eta) = A$ and
\begin{align*}
    \psi^b(W,\eta) &=  A( Y_1 - Y_0 - \eta_1) + (1-A) (1-\eta_2) (Y_1-Y_0 - \eta_1)~,
\end{align*} 
    for $\eta \in \mathbf{R}^2$, and where the nuisance parameter $\eta_0(X)$ has two components: 
    \begin{align*}
        \eta_{0}(X) &= (E[ Y_1 - Y_0 \mid X, A=0],~(E[1 - A \mid X])^{-1})^\intercal~.  
    \end{align*} 
This moment function is the efficient influence function for the ATT-DID under the conditions in
\cite{sant2020doubly}. 
\end{example}

\begin{example}[Local Average Treatment Effect]\label{example_LATE} 
Let $Z \in \{0,1\}$ denote a binary instrumental variable, $D(z)$ denote potential treatment decisions under the intervention $z \in \{0,1\}$, and assume the observed treatment decision is given by
$D = Z D(1) + (1-Z) D(0).$
Let $X$ denote a vector of covariates, $Y(d)$ denote the potential outcome under treatment decision $d \in \{0,1\}$, and $Y = D Y(1) + (1-D) Y(0)$ denote the observed outcome. The available data is modeled by the vector $W = (Y,Z,D,X)$. The parameter of interest is 
$\theta_0 = E[ Y(1) - Y(0) \mid D(1) > D(0)],$
which is the expected treatment effect for the sub-population that complies with the assigned treatment, also known as LATE \citep{imbens1994identification}.  \cite{frolich2007nonparametric} and \cite{singh2024double} used a selection-on-observables assumption, 
$(Y(1),Y(0),D(1),D(0)) \perp Z \mid X,$
monotonicity and overlap conditions, to identify the LATE by a moment condition, such as \eqref{eq:moment_for_theta}, using a moment function like \eqref{eq:moment_funct}:
    \begin{align*}
        \psi^a(W,\eta) &= \eta_3 - \eta_4 + Z(D-\eta_3)\eta_5 - (1-Z) (D-\eta_4) \eta_6~,\\
        \psi^b(W,\eta) &= \eta_1 - \eta_2 + Z(Y-\eta_1)\eta_5 - (1-Z)(Y-\eta_2)\eta_6~,
    \end{align*} 
    for $\eta \in \mathbf{R}^6$, and where the nuisance parameter $\eta_0(X)$ has six components:
    \begin{align*}
        \eta_{0}(X) = (E[Y \mid X, Z = 1],~ &E[Y \mid X, Z = 0],~ E[D \mid X, Z = 1]~, \\
        &E[D \mid X, Z = 0],~ (E[Z \mid X])^{-1},~ (E[1-Z \mid X])^{-1})^\intercal~.  
    \end{align*}
This moment function appears in \cite{frolich2007nonparametric} as the efficient influence function for the LATE. This moment function corresponds to the estimators proposed in \cite{tan2006regression}. 
\end{example} 
 
 \newpage
\section{Proofs of Auxiliary Results} 

\subsection{Proofs of Results in Appendix \ref{appendix:aux-results1}}\label{proofs:appendix-result1}

\begin{proof}[Proof of Lemma \ref{lemma:bounds-L2CR}]
    
Part (i) follows by recalling that $e_j^\top (D_\eta m_{i}[\gamma_i]) = \gamma_i^\top \partial_{\eta} m_{j} $ for $j \in [p]$, by the Law of Iterated Expectations, and  by Assumption~\ref{asm:EM}(iii). Parts (ii) follows by Assumption \ref{asm:EM}(iii) and the definition of $L_2$-operation norm and its standard properties.  Parts (iii) is similar to part (ii) but using Assumption~\ref{asm:EM}(iv) instead of Assumption~\ref{asm:EM}(iii).
\end{proof}

\begin{proof}[Proof of Lemma \ref{lemma:sqrt-Kn}]

\textit{Part (i):} We present the proof for $\mathcal{D}m_n = o_p(1)$, since the proof of $\mathcal{D}\psi_n^a = o_p(1)$ is analogous but using Assumption~\ref{asm:EM}(iv) instead of Assumption~\ref{asm:EM}(iii). By Markov's inequality and the union bound,
$ P\left( \mathcal{D}m_n > \varepsilon\right) \le \varepsilon^{-2} K_n E\left[\left | \left|  \mathcal{D}m_{k,n}   \right| \right|^2 \right], $
where $ \mathcal{D}m_{k,n}  = n_k^{-1/2} \sum_{i \in \mathcal{I}_k} D_\eta m_i[\hat{\eta}_i-\eta_i] $. Therefore, it is sufficient to show that $E\left[\left | \left|   \mathcal{D}m_{k,n}    \right| \right|^2 \right]$ is $o(n^{-1/2})$ since $K_n=O(n^{1/2})$. To prove this, we consider the following derivations:
\begin{align*}
    E\left[\left | \left|   \mathcal{D}m_{k,n}    \right| \right|^2 \right]  \overset{(1)}{=} n_k^{-1} \sum_{i \in \mathcal{I}_k} E\left[\left | \left| D_\eta m_i[\hat{\eta}_i-\eta_i]  \right| \right|^2 \right]  \overset{(2)}{\le} d C  E\left[\left | \left| \hat{\eta}_i-\eta_i \right| \right|^2 \right]  \overset{(3)}{=} o(n^{-1/2})~,
\end{align*}
where (1) holds by the Law of Iterated Expectations and Assumption~\ref{asm:EM}(iii), (2) holds by Lemma~\ref{lemma:bounds-L2CR}(i), and (3) holds by Assumption~\ref{asm:L2rate} and recalling that $N \asymp n$. 
 
\textit{Part (ii):} We first assume that $n_k^{-1/2} \max_{ 1 \le k \le K_n} b_k = o_p(1)$. We then define the event $E_n = \{ \|n_k^{-1/2} \max_{ 1 \le k \le K_n} b_k\| < 1/2\}$, which holds with probability approaching 1. We conclude by conditioning on $E_n$, where  $\max_{1\le k \le K_n}  \left | \left| (\mathbb{I}_d + n_k^{-1/2} {b}_k)^{-1}  \right| \right| \le (1-1/2)^{-1}$ holds due to standard properties of the $L_2$-operator norm (Neumann series bound).

To prove $n_k^{-1/2} \max_{ 1 \le k \le K_n} b_k = o_p(1)$, we consider the following derivations:
$$  P\left(n_k^{-1/2} \max_{ 1 \le k \le K_n} b_k> \varepsilon\right) \le \varepsilon^{-4} n^{-2} K_n^{3} E\left[ \|b_k\|^4  \right] \overset{(1)}{=} O(n^{-1/2})~,$$
where (1) holds by Assumption~\ref{asm:EM}(i)--(ii) and since $K_n = O(n^{1/2})$. 
  
\textit{Part (iii):} To use the same arguments as in the proof of part (ii), it is sufficient to show that $n_k^{-1/2} \max_{ 1 \le k \le K_n} \| \hat{b}_k - b_k\| = o_p(1)$. To prove this, we first write   
$$ \hat{b}_k - b_k  = n_k^{-1/2} \sum_{i \in \mathcal{I}_k} G^{-1} D_\eta \psi^a_i[\hat{\eta}_i-\eta_i] + \tfrac{1}{2}  G^{-1}D_\eta^2\tilde{\psi^a}_i[\hat{\eta}_i-\eta_i,\hat{\eta}_i-\eta_i]~. $$
We then consider the following derivations
\begin{align*}
    n_k^{-1/2} \max_{ 1 \le k \le K_n} \| \hat{b}_k - b_k\| &\overset{(1)}{\le} \|G^{-1}\| \left(  n_k^{-1/2} \mathcal{D}\psi_n^a + \tfrac{1}{2} C n_k^{-1} \sum_{i=1}^n \|\hat{\eta}_i-\eta_i\|^2 \right) \\
    &\overset{(2)}{=} O(1) \times \left(  o_p(1) + K_n n^{-1/2} o_p(1) \right)~\overset{(3)}{=} o_p(1)~,
\end{align*}
where (1) holds by definition of $\mathcal{D}\psi_n^a$ and Lemma~\ref{lemma:bounds-L2CR}(iii), (2) holds by Assumption~\ref{asm:EM}(i), part (i), and Lemma~\ref{lemma:Kn}(i), and (3) holds since $K_n = O(n^{1/2})$.
\end{proof}

\begin{proof}[Proof of Lemma~\ref{lemma:Kn}]

\textit{Part (i):} It holds by Assumption~\ref{asm:L2rate} and Markov's inequality.

\textit{Part (ii):} We can write
$  n^{-1/2} K_n^{-1/2} \sum_{k=1}^{K_n}  \hat{b}_k-b_k = G^{-1} \left( I_{1,n} + \tfrac{1}{2}I_{2,n}\right),$
where $I_{1,n}  = n^{-1} \sum_{i =1}^n  D_\eta \psi^a_i[\hat{\eta}_i-\eta_i] $ and $I_{2,n}  = n^{-1} \sum_{i =1}^n    D_\eta^2\tilde{\psi^a}_i[\hat{\eta}_i-\eta_i,\hat{\eta}_i-\eta_i]$. By Assumption~\ref{asm:EM}(i), it is sufficient to show that both $I_{1,n}$ and $I_{2,n}$ are $o_p(1)$.

To conclude $I_{1,n} = o_p(1)$, we prove $e_t^\intercal I_{1,n} e_s = o_p(1)$ for any $t,s \in [d]$, where $e_j$ is the $j$-th column of the identity matrix $\mathbb{I}_d \in \mathbf{R}^{d \times d}$. We consider the following derivations:
\begin{align*}
     E[(e_t^\intercal I_{1,n} e_s)^2] &\overset{(1)}{=}  E\left[ \left( n^{-1} \sum_{i =1}^n (\hat{\eta}_i-\eta_i)^\intercal \partial_\eta\psi_{t,s}^a(W_i,\theta_0,\eta_i)  \right)^2 \right] \\
     &\overset{(2)}{\le} K^{-1} \sum_{k=1}^K  E\left[ \left( n_k^{-1} \sum_{i \in \mathcal{I}_k}  (\hat{\eta}_i-\eta_i)^\intercal \partial_\eta\psi_{t,s}^a(W_i,\theta_0,\eta_i)  \right)^2 \right] \\
     &\overset{(3)}{\le}    E\left[ \left(    (\hat{\eta}_i-\eta_i)^\intercal \partial_\eta\psi_{t,s}^a(W_i,\theta_0,\eta_i)  \right)^2 \right] \overset{(4)}{=}  o(n^{-1/2})~,
\end{align*} 
where (1) holds by definition, (2) and (3) hold by Jensen's inequality, and (4) holds by the Law of Iterated Expectations  and Assumptions~\ref{asm:EM}(iii) and \ref{asm:L2rate}.

To conclude $I_{2,n} = o_p(1)$, we rely on Lemma~\ref{lemma:bounds-L2CR}(iii) and part (i) to conclude that $|I_{2,n}| \le C n^{-1} \sum_{i=1}^n \|\hat{\eta}_i-\eta_i\|^2 = o_p(1)$. 

\textit{Part (iii):} To prove $\left(  \mathbb{I}_d +  n_k^{-1/2} K_n^{-1}\sum_{k=1}^{K_n}  {b}_k \right)^{-1} = O_p(1)$, we write
$   n_k^{-1/2} K_n^{-1}\sum_{k=1}^{K_n}  {b}_k  =  n^{-1} \sum_{i = 1 }^n \left( G^{-1}\psi^a_i - \mathbb{I}_d\right),$
which is $O_p(n^{-1/2})$ due to the Central Limit Theorem and Assumption~\ref{asm:EM}(i)--(ii). We then define the event $E_n = \{ \| n_k^{-1/2} K_n^{-1}\sum_{k=1}^{K_n}  {b}_k\| < 1/2 \} $, which holds with probability approaching 1. We conclude by conditioning on $E_n$, where $\| \left(  \mathbb{I}_d +  n_k^{-1/2} K_n^{-1}\sum_{k=1}^{K_n}  {b}_k \right)^{-1} \| \le (1-1/2)^{-1}$ holds due to the properties of the $L_2$-operator norm (Neumann series bound). 

To prove $\left(  \mathbb{I}_d +  n_k^{-1/2} K_n^{-1}\sum_{k=1}^{K_n}  \hat{b}_k \right)^{-1} = O_p(1)$ by the same arguments used above, it is sufficient to show that $n_k^{-1/2} K_n^{-1}\sum_{k=1}^{K_n}  \hat{b}_k -b_k = o_p(1)$, which holds by part (ii). 
\end{proof}
  
\begin{proof}[Proof of Lemma \ref{lemma:variance-consistency}] Denote $ \bar{\Omega}_n = n^{-1}\sum_{i=1}^n m(W_i,\bar{\theta}_n,\hat{\eta}_i)  m(W_i,\bar{\theta}_n,\hat{\eta}_i)^\intercal$. %Recall that $\hat{m}_i = m(W_i,\theta_0,\hat{\eta}_i)$ and $\hat{\psi}^a_i= \psi^a(W_i,\hat{\eta}_i)$.

\textit{Part (i):} It holds by noting $\hat{G}_n - G =  n^{-1}\sum_{i=1}^n (\psi^a_i - G) + n^{-1/2} K_n^{-1/2} \sum_{k=1}^{K_n} (\hat{b}_k-b_k)$, and using Law of Large Numbers and Lemma~\ref{lemma:Kn}(ii). 

\textit{Part (ii):} We first write $m(W_i,\bar{\theta}_n,\hat{\eta}_i) =    m(W_i,\theta_0,\hat{\eta}_i) - \psi^a(W_i,\hat{\eta}_i)(\bar{\theta}_n-\theta_0)$. We then write $\bar{\Omega}_n = I_1 + I_2 + I_3$, where $I_1 = n^{-1}\sum_{i=1}^n \hat{m}_i \hat{m}_i^\intercal$, $I_2 = 2 n^{-1}\sum_{i=1}^n \hat{\psi}^a_i (\bar{\theta}_n - \theta_0) \hat{m}_i^\intercal$, and $I_3 = n^{-1}\sum_{i=1}^n \hat{\psi}^a_i (\bar{\theta}_n - \theta_0) (\bar{\theta}_n - \theta_0)^\intercal (\hat{\psi}^a_i)^\intercal$. Claims 1, 2, and 3 below imply that $I_1 = \Omega + o_p(1)$, $I_2 = o_p(1)$, and $I_3=o_p(1)$, which is sufficient to complete part (ii).

\noindent \textbf{Claim 1:} $I_1 = \Omega + o_p(1)$. It follows by showing that $n^{-1}\sum_{i=1}^n \hat{m}_i \hat{m}_i^\intercal - m_i m_i^\intercal = o_p(1)$. By Cauchy-Schwartz and $n^{-1}\sum_{i=1}^n \|m_i\|^2 = O_p(1)$ (Assumption~\ref{asm:EM}(ii)), it suffices to show $n^{-1}\sum_{i=1}^n \| \hat{m}_i-m_i\|^2 = o_p(1)$. By a Taylor expansion with Lagrange remainder and Lemma~\ref{lemma:bounds-L2CR}(i)--(ii), we conclude $\| \hat{m}_i-m_i\|^2 \le C\big(\|D_\eta m_i[\hat{\eta}_i-\eta_i]\|^2 + \|\hat{\eta}_i-\eta_i\|^4\big) $. By Lemma~\ref{lemma:bounds-L2CR}(i), $E[ \|D_\eta m_i[\hat{\eta}_i-\eta_i]\|^2] \le C E[\| \hat{\eta}_i-\eta_i\|^2]$, which implies $n^{-1} \sum_{i=1}^n \|D_\eta m_i[\hat{\eta}_i-\eta_i]\|^2 = o_p(1)$ due to Assumption~\ref{asm:L2rate}. Since $\mathcal{E}$ is bounded, it follows that $\|\hat{\eta}_i-\eta_i\|^4 \le \mathrm{diam}(\mathcal{E})^2 \|\hat{\eta}_i-\eta_i\|^2$, which implies $n^{-1} \sum_{i=1}^n \|\hat{\eta}_i-\eta_i\|^4 = o_p(1)$ due to Assumption~\ref{asm:L2rate}.

\noindent \textbf{Claim 2:}  $I_2 = o_p(1)$. As in the proof of claim 1, we have $n^{-1}\sum_{i=1}^n \| \hat{\psi}^a_i-\psi^a_i\|^2 = o_p(1)$. Therefore, $n^{-1}\sum_{i=1}^n  \| \hat{m}_i\|^2$ and $n^{-1}\sum_{i=1}^n  \| \hat{\psi}^a_i\|^2$ are both $O_p(1)$ using Assumption~\ref{asm:EM}(ii). Since $\bar{\theta}_n = \theta_0 + o_p(1)$, the triangle inequality and Cauchy-Schwartz completes the proof.  

\noindent \textbf{Claim 3:}  $I_3 = o_p(1)$. Similar as the proof of Claim 2; therefore, omitted.
\end{proof} 

\subsection{Proofs of Results in Appendix \ref{appendix:aux-results2}}\label{proofs:appendix-result2}

\textit{Notation:} Let $k(i) \in [K_n]$ be such that $i \in \mathcal{I}_{k}$ and $k=k(i)$. Define 
$$  \Delta_{k,i}^{(\delta)} = N^{-1/2 } \sum_{j \in \mathcal{J}} N^{-\varphi}\delta_{N}(W_j,x) \qquad \text{and} \qquad \Delta_{k,i}^{(b)} =  N^{-1} \sum_{j \in \mathcal{J}} N^{-\varphi} b_{N}(X_{j},x)~.$$

\begin{proof}[Proof of Proposition \ref{prop:so-bias-variance-dml2}]
    Recall that $\varphi \in (1/4,1/2)$ by Assumption~\ref{asm:SLE} and $N \asymp n$ since $N=(1-\tfrac{1}{K_n})n$. We divide the proof in three parts. 

    \textit{Part (i):} $\mathcal{B}_N = O(1)$ and $\mathcal{C}_N = O(1)$. Note that   $\mathcal{B}_N = O(1)$ follows by the definition of $\mathcal{B}_N$, the Law of Iterated Expectations, and  Assumption~\ref{asm:leading-terms}(ii). To prove $\mathcal{C}_N = O(1)$, we first write $\mathcal{C}_N = \mathcal{C}_{1,N} + \mathcal{C}_{2,N}$, where $\mathcal{C}_{1,N}  = E\left[ \tilde{b}_N(X_i)^\intercal S(X_i) \right]$ and $\mathcal{C}_{2,N}  =  E\left[ \tilde{b}_N(X_i)^\intercal (H(X_i) \tilde{\delta}_N(X_i)) \right]$.  
    Note that $\mathcal{C}_{1,N} = G^{-2}E\left[ m_i (\tilde{b}_N(X_i)^\intercal \partial_\eta m_i) \right]$ by Law of Iterated Expectations. Then, $\mathcal{C}_{1,N} = O(1)$ holds by Cauchy-Schwarz and Assumptions~\ref{asm:EM}(i)--(iii) and \ref{asm:SLE}(iii). Finally, $\mathcal{C}_{2,N} = O(1)$ holds by Assumptions~\ref{asm:EM}(i)--(iii) and \ref{asm:SLE}(iii)--(iv).
 
    \textit{Part (ii):} We now show that \eqref{eq:bias-so-DML2-exact} holds. Recall $\mathcal{T}_{n,K} = \mathcal{T}_{n}^{*}+\mathcal{T}_{n,K}^{\ell}+\mathcal{T}_{n,K}^{q}+\mathcal{T}_{n}^{*,(2)}$. By the Law of Iterated Expectations and Assumption~\ref{asm:EM}(iii), it follows that $E[\mathcal{T}_{n,K}^{\ell}]=0$. The moment condition \eqref{eq:moment_for_theta} and standard calculations imply $E[\mathcal{T}_{n}^{*}]=0$  and $E[\mathcal{T}_{n}^{*,(2)}] = \Lambda n^{-1/2}$. 
    
    It is sufficient to show that  $E\left[ \mathcal{T}_{n,K}^{q} \right] = \tfrac{1}{2}  \mathcal{B}_N N^{-2\varphi}n^{1/2} + o(N^{-1/2})$ holds since $N \asymp n$ and $N = (1-\tfrac{1}{K_n})n$. To prove this, we first recall the definition of $\Delta_{k,i}^{(b)}$ and  $\Delta_{k,i}^{(\delta)}$, and we then consider the following derivations
    \begin{align*}
       E\left[ \mathcal{T}_{n,K}^{q} \right] &= \tfrac{1}{2}  n^{-1/2 } \sum_{k = 1}^K \sum_{i \in \mathcal{I}_k} E\left[ \Delta_{k,i}^\intercal   \left(G^{-1} \partial_\eta^2 m_i \right)  \Delta_{k,i} \right] \\
       &=  \tfrac{1}{2}  n^{1/2 } E\left[ \Delta_{k,i}^\intercal   \left(G^{-1} \partial_\eta^2 m_i \right)  \Delta_{k,i} \right] \\
       &=   \tfrac{1}{2}  n^{1/2 }  \left(  I_{k,i}^{(\delta,\delta)} + 2I_{k,i}^{(\delta,b)} + I_{k,i}^{(b,b)} \right)~,
    \end{align*}
    where $ I_{k,i}^{(\iota_1,\iota_2)} = E\left[ (\Delta_{k,i}^{(\iota_1)})^\intercal   \left(G^{-1} \partial_\eta^2 m_i \right)  \Delta_{k,i}^{(\iota_2)} \right] $ for $\iota_1, \iota_2 \in \{\delta,b\}$.
    
    Claims 1, 2, and 3 below imply that $I_{k,i}^{(\delta,b)} = 0$  and $I_{k,i}^{(\delta,\delta)} + I_{k,i}^{(b,b)}=  N^{-2\varphi} \mathcal{B}_N + o(N^{-1}) $, which completes the proof, where we used $E[\delta_N(W_j,X_i)^\intercal H(X_i) \tilde{b}_N(X_i)] = 0$ for $i\neq j$ due to Assumption~\ref{asm:SLE}(ii).

    \noindent \textbf{Claim 1:} $I_{k,i}^{(\delta,\delta)} = N^{-2\varphi} E[\delta_N(W_j,X_i)^\intercal H(X_i) \delta_N(W_j,X_i)]$. It follows by the definition of $\Delta_{k,i}^{(\delta)}$, the Law of Iterated Expectations, and Assumption~\ref{asm:SLE}(ii).

    \noindent \textbf{Claim 2:} $I_{k,i}^{(\delta,b)} = 0$.  It follows by the definition of $\Delta_{k,i}^{(\delta)}$ and $\Delta_{k,i}^{(b)}$, the Law of Iterated Expectations, and Assumption~\ref{asm:SLE}(ii). 

    \noindent \textbf{Claim 3:} $I_{k,i}^{(b,b)} = N^{-2\varphi} E[\tilde{b}_N(X_i)^\intercal H(X_i) \tilde{b}_N(X_i)] + o(N^{-1}) $. The  definition of  $\Delta_{k,i}^{(b)}$ and the Law of Iterated Expectations imply
    \begin{align*}
         I_{k,i}^{(b,b)} &=   N^{-2\varphi} E\left[ \tilde{b}_N(X_i)^\intercal H(X_i) \tilde{b}_N(X_i) \right] \\
        &~~ +  N^{-1-2\varphi} E\left[ \left( b_N(X_j,X_i) - \tilde{b}_N(X_i)\right)^\intercal H(X_i)  \left( b_N(X_j,X_i) - \tilde{b}_N(X_i)\right) \right] ~ .
    \end{align*}
    Therefore, it is sufficient to show
    \begin{equation}\label{eq:aux1-proof-prop1}
        E\left[ \left( b_N(X_j,X_i) - \tilde{b}_N(X_i)\right)^\intercal H(X_i)  \left( b_N(X_j,X_i) - \tilde{b}_N(X_i)\right) \right] = O(N^{1-2\varphi})~
    \end{equation} 
    to complete the proof since $\varphi \in (1/4,1/2)$. 
    To prove this, we rely on the next observation:
    \begin{align*}
        \left | E\left[ \left( b_N(X_j,x) - \tilde{b}_N(x)\right)^\intercal H(x)  \left( b_N(X_j,x) - \tilde{b}_N(x)\right) \right] \right| \overset{(1)}{\lesssim} E[\|b_N(X_j,x)\|^2] \overset{(2)}{\lesssim}  N^{1-2\varphi} ~, 
    \end{align*}
    where (1) holds by Assumption~\ref{asm:EM}(i)--(iii) and variance decomposition, and (2) holds by Assumption~\ref{asm:SLE}(iii). Then, \eqref{eq:aux1-proof-prop1} follows by the Law of Iterated Expectations. 

    \textit{Part (iii):} We now show that \eqref{eq:variance-so-DML2} holds.  Recall $\mathcal{T}_{n,K} = \mathcal{T}_{n}^{*}+\mathcal{T}_{n,K}^{\ell}+\mathcal{T}_{n,K}^{q}+\mathcal{T}_{n}^{*,(2)}$. Standard derivations and Assumption~\ref{asm:EM}(i)--(ii) imply $Var[\mathcal{T}_{n}^{*}] = \Sigma$ and $Var[\mathcal{T}_{n}^{*,(2)}] = O(n^{-1})$. Lemma~\ref{lemma:moments-so-dml2}(ii) and Assumption~\ref{asm:leading-terms}(ii)--(iv) imply $Var[\mathcal{T}_{n,K}^{\ell}]$ and $Var[\mathcal{T}_{n,K}^{q}]$ are  $O(n^{-2\varphi})$. 
    
    By expanding the variance and applying Cauchy-Schwarz, we have
    $$ Var[ \mathcal{T}_{n,K_n}] = \Sigma +   2Cov[ \mathcal{T}_{n}^*,\mathcal{T}_{n,K_n}^{\ell} ]   + 2Cov[ \mathcal{T}_{n}^*, \mathcal{T}_{n,K_n}^{q}] + O(n^{-2\varphi})~,$$
    where we used $n^{-1} = o(n^{-2\varphi})$ and $n^{-1/2-\varphi}=o(n^{-2\varphi})$. Recall that $\mathcal{C}_{N} = \mathcal{C}_{1,N} + \mathcal{C}_{2,N}$. Therefore, it is sufficient to show the following holds: 
    \begin{align}
        Cov[ \mathcal{T}_{n}^*,\mathcal{T}_{n,K_n}^{\ell} ] &=  N^{-\varphi} \mathcal{C}_{1,N}, \label{eq:aux2-proof-prop1}\\
        Cov[ \mathcal{T}_{n}^*,\mathcal{T}_{n,K_n}^{q} ] &= N^{-\varphi} \mathcal{C}_{2,N} + O(N^{-2\varphi})~.\label{eq:aux3-proof-prop1}
    \end{align}

    To prove that \eqref{eq:aux2-proof-prop1} holds, we consider the following derivations:
    \begin{align*}
        Cov[ \mathcal{T}_{n}^*,\mathcal{T}_{n,K_n}^{\ell} ] %&= n^{-1/2} \sum_{i=1}^n E\left[ (G^{-1}m_i) \mathcal{T}_{n,K_n}^{\ell}  \right] \\
        &\overset{(1)}{=}  n^{1/2}  E\left[ (G^{-1}m_i) \mathcal{T}_{n,K_n}^{\ell}  \right] \\
        &\overset{(2)}{=}  \sum_{k=1}^K \sum_{i_1 \in \mathcal{I}_k} E\left[  (G^{-1}m_i) \Delta_{k,i_1}^\intercal (G^{-1} \partial_\eta m_{i_1})  \right] \\
        &\overset{(3)}{=}  \sum_{k=1}^K \sum_{i_1 \in \mathcal{I}_k} E\left[  (G^{-1}m_i) (\Delta_{k,i_1}^{(b)})^\intercal (G^{-1} \partial_\eta m_{i_1})  \right] \\
        &\overset{(4)}{=}   E\left[  (G^{-1}m_i) (\Delta_{k,i}^{(b)})^\intercal (G^{-1} \partial_\eta m_{i})  \right] \\
        &\overset{(5)}{=} N^{-\varphi}E\left[  (G^{-1}m_i) (b_N(X_j,X_i)^\intercal (G^{-1} \partial_\eta m_{i})  \right] \overset{(6)}{=}  N^{-\varphi} \mathcal{C}_{1,N}~,
    \end{align*}
    where (1) and (2) hold by the definition of $\mathcal{T}_{n}^*$ and $\mathcal{T}_{n,K_n}^{\ell}$, respectively, (3) holds by writing $ \Delta_{k,i_1} =  \Delta_{k,i_1}^{(\delta)} +  \Delta_{k,i_1}^{(b)}$ and because $E\left[  (G^{-1}m_i) (\Delta_{k,i_1}^{(\delta)})^\intercal (G^{-1} \partial_\eta m_{i_1})  \right] = 0$ due to the Law of Iterated Expectations and Assumptions~\ref{asm:EM}(iii) and \ref{asm:SLE}(ii), (4) holds by the Law of Iterated Expectations and Assumptions~\ref{asm:EM}(iii), (5) holds by the definition of  $\Delta_{k,i_1}^{(b)}$, and (6) holds by the Law of Iterated Expectations.

    To prove that \eqref{eq:aux3-proof-prop1} holds, we consider the following derivations:
     \begin{align*}
         Cov[ \mathcal{T}_{n}^*,\mathcal{T}_{n,K_n}^{q} ] %&= n^{-1/2} \sum_{i=1}^n E\left[ (G^{-1}m_i) \mathcal{T}_{n,K_n}^{q}  \right] \\
         &\overset{(1)}{=} n^{1/2} E\left[ (G^{-1}m_i) \mathcal{T}_{n,K_n}^{q}  \right] \\
         &\overset{(2)}{=} \tfrac{1}{2} \sum_{k=1}^K \sum_{i_1 \in \mathcal{I}_k} E\left[  (G^{-1}m_i) \Delta_{k,i_1}^\intercal (G^{-1} \partial_\eta^2 m_{i_1}) \Delta_{k,i_1} \right] \\
         &= \tfrac{1}{2} I_{q,i}^{(\delta,\delta)} +   I_{q,i}^{(\delta,b)}  + \tfrac{1}{2} I_{q,i}^{(b,b)} ~,
     \end{align*}
     where (1) and (2) holds by the definition of $\mathcal{T}_{n}^*$ and $\mathcal{T}_{n,K_n}^{q}$, respectively, and 
     $$I_{q,i}^{(\iota_1,\iota_2)} = \sum_{k=1}^K \sum_{i_1 \in \mathcal{I}_k} E\left[  (G^{-1}m_i) (\Delta_{k,i_1}^{(\iota_1)})^\intercal (G^{-1} \partial_\eta^2 m_{i_1}) \Delta_{k,i_1}^{(\iota_2)} \right]~, \quad \iota_1, \iota_2 \in \{\delta, b\}~. $$
     Claims 4, 5, and 6 below imply that $I_{q,i}^{(\delta,b)} =  N^{-\varphi} \mathcal{C}_{2,N} + O(N^{-2\varphi})$ and both $I_{q,i}^{(\delta,\delta)}$  and $I_{q,i}^{(b,b)}$  are $O(N^{-2\varphi})$, which completes the proof.

     \noindent \textbf{Claim 4:} $I_{q,i}^{(\delta,\delta)} = O(N^{-2\varphi})$. To prove this, we consider the following derivations
     \begin{align*}
          I_{q,i}^{(\delta,\delta)} &\overset{(1)}{=}  N^{-2\varphi-1} \sum_{k=1}^K \sum_{i_1 \in \mathcal{I}_k} \sum_{j_1, j_2 \notin \mathcal{I}_k} E\left[ (G^{-1}m_i) \delta_{N}(W_{j_1},X_{i_1})^\intercal (G^{-1} \partial_\eta^2 m_{i_1}) \delta_{N}(W_{j_2},X_{i_1}) \right] \\
          &\overset{(2)}{=}  N^{-2\varphi-1}  \sum_{j_1, j_2 \notin \mathcal{I}_{k(i)}} E\left[ (G^{-1}m_i) \delta_{N}(W_{j_1},X_{i})^\intercal (G^{-1} \partial_\eta^2 m_{i}) \delta_{N}(W_{j_2},X_{i}) \right] \\
          &~~ + N^{-2\varphi-1} \sum_{i_1 =1, i_1 \neq i}^n ~\sum_{j_1, j_2 \notin \mathcal{I}_{k(i_1)}} E\left[ (G^{-1}m_i) \delta_{N}(W_{j_1},X_{i_1})^\intercal (G^{-1} \partial_\eta^2 m_{i_1}) \delta_{N}(W_{j_2},X_{i_1}) \right]  \\
          &\overset{(3)}{=}   N^{-2\varphi} E\left[ G^{-1} m_i~ \delta_N(W_j,X_i)^\intercal (G^{-1} \partial_\eta^2 m_i)~ \delta_N(W_j,X_i) \right] \\
         &~~+  N^{-2\varphi} (\tfrac{n-1}{N}) E\left[ G^{-1} m_j \delta_N(W_j,X_i)^\intercal H(X_i) \delta_N(W_j,X_i) \right] \\
         &\overset{(4)}{=}  O(N^{-2\varphi})~,
     \end{align*}
     where (1) holds by the definition of $\Delta_{k,i_1}^{(\delta)}$, (2) holds by dividing the analysis into  two  cases ($i_1 = i$; $i_1 \neq i$) and defining $k(i) \in [K_n]$ such that $i \in \mathcal{I}_k$, (3) holds by the Law of Iterated Expectations and Assumptions~\ref{asm:SLE}(ii), and (4) holds by Cauchy-Schwarz, the Law of Iterated Expectations, and Assumption~\ref{asm:SLE}(ii) and (iv).
      
     \noindent \textbf{Claim 5:} $I_{q,i}^{(b,b)} = O(N^{-2\varphi})$. To prove this, we consider the following derivations\vspace{-0.8cm}
     \begin{align*}
         I_{q,i}^{(b,b)} &\overset{(1)}{=}  N^{-2\varphi-2} \sum_{k=1}^K \sum_{i_1 \in \mathcal{I}_k} \sum_{j_1, j_2 \notin \mathcal{I}_k} E\left[ (G^{-1}m_i) b_{N}(X_{j_1},X_{i_1})^\intercal (G^{-1} \partial_\eta^2 m_{i_1}) b_{N}(X_{j_2},X_{i_1}) \right] \\
         &\overset{(2)}{=}   N^{-2\varphi-2}   \sum_{j_1, j_2 \notin \mathcal{I}_{k(i)}} E\left[ (G^{-1}m_i) b_{N}(X_{j_1},X_{i})^\intercal (G^{-1} \partial_\eta^2 m_{i}) b_{N}(X_{j_2},X_{i}) \right] \\
         &~~+   N^{-2\varphi-2} \sum_{\substack{i_1 = 1 \\ k(i_1) \neq k(i)}}^n \sum_{ \substack{j_2 \notin \mathcal{I}_{k(i_1)} \\ j_2 \neq i}} E\left[ (G^{-1}m_i) b_{N}(X_{i},X_{i_1})^\intercal (G^{-1} \partial_\eta^2 m_{i_1}) b_{N}(X_{j_2},X_{i_1}) \right] \\
          &~~+  N^{-2\varphi-2} \sum_{\substack{i_1 = 1 \\ k(i_1) \neq k(i)} }^n \sum_{ \substack{ j_1 \notin \mathcal{I}_{k(i_1)} \\ j_1 \neq i}  } E\left[ (G^{-1}m_i) b_{N}(X_{j_1},X_{i_1})^\intercal (G^{-1} \partial_\eta^2 m_{i_1}) b_{N}(X_{i},X_{i_1}) \right]\\
           &~~+   N^{-2\varphi-2} \sum_{ \substack{i_1 = 1 \\ k(i_1) \neq k(i)} }^n   E\left[ (G^{-1}m_i) b_{N}(X_{i},X_{i_1})^\intercal (G^{-1} \partial_\eta^2 m_{i_1}) b_{N}(X_{i},X_{i_1}) \right]  \\
           &\overset{(3)}{=} N^{-2\varphi-1}  E\left[ (G^{-1}m_i) b_{N}(X_{j},X_{i})^\intercal (G^{-1} \partial_\eta^2 m_{i}) b_{N}(X_{j},X_{i}) \right] \\
           &~~+ N^{-2\varphi}(\tfrac{N-1}{N}) E\left[ (G^{-1}m_i) \tilde{b}_{N}(X_{i})^\intercal (G^{-1} \partial_\eta^2 m_{i}) \tilde{b}_{N}(X_{i}) \right] \\
           &~~+2 N^{-2\varphi } (\tfrac{n-2}{N})    E\left[ (G^{-1}m_i) b_{N}(X_{i},X_{i_1})^\intercal (G^{-1} \partial_\eta^2 m_{i_1}) \tilde{b}_{N}( X_{i_1}) \right] \\
           &~~+N^{-2\varphi-1}(\tfrac{n-1}{N}) E\left[ (G^{-1}m_i) b_{N}(X_{i},X_{i_1})^\intercal (G^{-1} \partial_\eta^2 m_{i_1}) b_{N}(X_{i},X_{i_1}) \right]  \\
           &\overset{(4}{=} O(N^{-2\varphi})~,
     \end{align*}
     where (1) holds by the definition of $\Delta_{k,i_1}^{(b)}$, (2) holds by dividing the analysis into four nontrivial cases ($i_1=i$; $i_1 \neq i$ and $j_1 = i \neq j_2$; $i_1 \neq i$ and $j_2 = i \neq j_1$; $i_1 \neq i$ and $j_1 = i = j_2$) and defining $k(i) \in [K_n]$ such that $i \in \mathcal{I}_k$, (3) holds by  the Law of Iterated Expectations, and (4) holds by Assumptions~\ref{asm:EM}(iii) and \ref{asm:SLE}(ii)--(iv).

      \noindent \textbf{Claim 6:} $I_{q,i}^{(\delta,b)} = \tfrac{1}{2} N^{-\varphi} \mathcal{C}_{2,N} + O(N^{-2\varphi})$. To prove this, we consider the derivations:\vspace{-0.1cm}
      \begin{align*}
          I_{q,i}^{(\delta,b)} &\overset{(1)}{=}  N^{-2\varphi-3/2} \sum_{k=1}^K \sum_{i_1 \in \mathcal{I}_k} \sum_{j_1, j_2 \notin \mathcal{I}_k} E\left[ (G^{-1}m_i) \delta_{N}(W_{j_1},X_{i_1})^\intercal (G^{-1} \partial_\eta^2 m_{i_1}) b_{N}(X_{j_2},X_{i_1}) \right] \\
          &\overset{(2)}{=}  N^{-2\varphi-3/2} \sum_{j_1, j_2 \notin \mathcal{I}_{k(i)}} E\left[ (G^{-1}m_i) \delta_{N}(W_{j_1},X_{i})^\intercal (G^{-1} \partial_\eta^2 m_{i}) b_{N}(X_{j_2},X_{i}) \right] \\
          &~~+   N^{-2\varphi-3/2} \sum_{\substack{i_1 = 1 \\ k(i_1) \neq k(i)}}^n \sum_{ \substack{j_2 \notin \mathcal{I}_{k(i_1)} \\ j_2 \neq i}} E\left[ (G^{-1}m_i) \delta_{N}(W_{i},X_{i_1})^\intercal (G^{-1} \partial_\eta^2 m_{i_1}) b_{N}(X_{j_2},X_{i_1}) \right] \\
          &~~+  N^{-2\varphi-3/2} \sum_{\substack{i_1 = 1 \\ k(i_1) \neq k(i)} }^n \sum_{ \substack{ j_1 \notin \mathcal{I}_{k(i_1)} \\ j_1 \neq i}  } E\left[ (G^{-1}m_i) \delta_{N}(W_{j_1},X_{i_1})^\intercal (G^{-1} \partial_\eta^2 m_{i_1}) b_{N}(X_{i},X_{i_1}) \right]\\
           &~~+   N^{-2\varphi-3/2} \sum_{ \substack{i_1 = 1 \\ k(i_1) \neq k(i)} }^n   E\left[ (G^{-1}m_i) \delta_{N}(W_{i},X_{i_1})^\intercal (G^{-1} \partial_\eta^2 m_{i_1}) b_{N}(X_{i},X_{i_1}) \right]  \\
           &\overset{(3)}{=} N^{-2\varphi+1/2}  (\tfrac{N-1}{N}) E\left[ (G^{-1}m_i) \delta_{N}(W_{i},X_{i_1})^\intercal (G^{-1} \partial_\eta^2 m_{i_1}) \tilde{b}_{N}(X_{i_1}) \right] \\
           &~~+  N^{-2\varphi-1/2} E\left[ (G^{-1}m_i) \delta_{N}(W_{i},X_{i_1})^\intercal (G^{-1} \partial_\eta^2 m_{i_1}) b_{N}(X_{i},X_{i_1}) \right] \\
           &\overset{(4)}{=}  N^{-\varphi} \mathcal{C}_{2,N} + O(N^{-\varphi-1}) +  O(N^{-3\varphi})~,
      \end{align*}
      where (1) holds by the definition of $\Delta_{k,i_1}^{\delta}$ and $\Delta_{k,i_1}^{b}$, (2) holds by dividing the analysis into four cases ($i_1=i$; $i_1 \neq i$ and $j_1 = i \neq j_2$; $i_1 \neq i$ and $j_2 = i \neq j_1$; $i_1 \neq i$ and $j_1 = i = j_2$) and the definition of $k(i)$, (3) holds because $E\left[ (G^{-1}m_i) \delta_{N}(W_{j_1},X_{i})^\intercal (G^{-1} \partial_\eta^2 m_{i}) b_{N}(X_{j_2},X_{i}) \right]$ and $E\left[ (G^{-1}m_i) \delta_{N}(W_{i},X_{i_1})^\intercal (G^{-1} \partial_\eta^2 m_{i_1}) b_{N}(X_{i},X_{i_1}) \right]$ are both zero by Assumption~\ref{asm:SLE}(ii) and the Law of Iterated Expectation, 
      and (4) holds by definition of $\mathcal{C}_{2,N}$, Law of Iterated Expectations, Cauchy-Schwarz, and Assumption~\ref{asm:SLE}(iii)--(iv). 
\end{proof}

\begin{proof}[Proof of Lemma~\ref{lemma:moments-so-dml2}] Recall that $\varphi \in (1/4,1/2)$ by Assumption \ref{asm:SLE} and $N \asymp n$. 

\textit{Part (i):} In the proof of Proposition~\ref{prop:so-bias-variance-dml2}, we show  
    $ E[\mathcal{T}_{n,K_n}^{q}] = \tfrac{1}{2}  \mathcal{B}_N N^{-2\varphi} n^{1/2} + o(N^{-1/2})~.$
    Hence, Assumption~\ref{asm:leading-terms}(ii) completes the proof since $\mathcal{B} = \mathcal{B}_N + o(1)$ . 

\textit{Part (ii)}: We use that $E[\mathcal{T}_{n,K_n}^{\ell}]=0$, which we show in the proof of Proposition~\ref{prop:so-bias-variance-dml2}, and $\Delta_{k,i} = \Delta_{k,i}^{(\delta)} +\Delta_{k,i}^{(b)}$ to write $ Var[\mathcal{T}_{n,K_n}^{\ell}] = I_1 + I_2 + 2I_3$, where
\begin{align*}
    I_1 &= E \left[ \left( n^{-1/2} \sum_{k=1}^{K_n}  \sum_{i \in \mathcal{I}_k} (\Delta_{k,i}^{(\delta)})^{\top} (G^{-1} \partial_\eta m_i)  \right)^2 \right] \\
    I_2 &= E \left[ \left( n^{-1/2} \sum_{k=1}^{K_n}  \sum_{i \in \mathcal{I}_k} (\Delta_{k,i}^{(b)})^{\top} (G^{-1} \partial_\eta m_i)  \right)^2 \right] \\
    I_3 &= E \left[ \left( n^{-1/2} \sum_{k=1}^{K_n}  \sum_{i \in \mathcal{I}_k} (\Delta_{k,i}^{(\delta)})^{\top} (G^{-1} \partial_\eta m_i)  \right)\left( n^{-1/2} \sum_{k=1}^{K_n}  \sum_{i \in \mathcal{I}_k} (\Delta_{k,i}^{(b)})^{\top} (G^{-1} \partial_\eta m_i)  \right) \right]~.
\end{align*}
Define  $\mathcal{V}_N = E\left[ \left( \delta_{N}(W_j,X_i)^\intercal (G^{-1} \partial_\eta m_i) +\delta_{N}(W_i,X_j)^\intercal (G^{-1} \partial_\eta m_j)  \right)^2 \right]$. Claims 1 and 2 below show that $ I_1 = \tfrac{1}{2}\mathcal{V}_N^{(\ell)} N^{-2\varphi}$ and $I_2 = N^{-2\varphi} E\left[ \left( \tilde{b}_N(X_i)^\intercal (G^{-1} \partial_\eta m_i) \right)^2 \right] + O(N^{-4\varphi})$, which is sufficient to prove Part~(ii) since Claim 3 below implies $I_3 = 0$.
 
\noindent \textbf{Claim 1:}  $ I_1 = \tfrac{1}{2}\mathcal{V}_N^{(\ell)} N^{-2\varphi}$. To prove this, we first define
$$h(W_j,W_i) = \delta_N(W_j,X_i)^\intercal (G^{-1} \partial_\eta m_i)  + \delta_N(W_i,X_j)^\intercal (G^{-1} \partial_\eta m_j)~,$$
which is a degenerate kernel by Assumptions~\ref{asm:EM}(iii) and \ref{asm:leading-terms}(ii), i.e., $E[h(W_j,W_i) \mid W_j]  = 0 $. Note that $ E\left[ h(W_j,W_i) ^2\right] = \mathcal{V}_N $. 
Define also $\mathfrak{D} = \{ \{i,j\}: 1 \le i \le n,~ j \notin \mathcal{I}_{k(i)} \}$ the set of all the pairs that are not in the same fold. The number of pairs in $\mathfrak{D}$ is $\tfrac{1}{2}n N $. We now rewrite $I_1$ as follows:
$$ I_1 =  E \left[ \left( n^{-1/2}  N^{-1/2-\varphi} \sum_{ \{i,j\} \in \mathfrak{D}}  h(W_j,W_i)  \right)^2 \right]~.$$
Finally, we use \citet[Theorem 2 in Chapter 4.3]{lee2019u} for incomplete U-statistics to conclude that $ I_1 = n^{-1} N^{-1-2\varphi} |\mathfrak{D}| E\left[ h(W_j,W_i) ^2\right] = \tfrac{1}{2} N^{-2\varphi} \mathcal{V}_N^{(\ell)}$.

\noindent \textbf{Claim 2:} $I_2 = N^{-2\varphi} E\left[ \left( \tilde{b}_N(X_i)^\intercal (G^{-1} \partial_\eta m_i) \right)^2 \right] + O(N^{-4\varphi})$. To prove this, we first note that for any $i_1 \neq i_2$:
$$E\left[  (\Delta_{k(i_1),i_1}^{(b)})^{\top} (G^{-1} \partial_\eta m_{i_1}) (\Delta_{k(i_2),i_2}^{(b)})^{\top} (G^{-1} \partial_\eta m_{i_2}) \right] = 0~,$$
which holds by the Law of Iterated Expectations and Assumption~\ref{asm:EM}(iii). We then use this result to write:
\begin{align*}
    I_2 &=  n^{-1} \sum_{k=1}^{K_n}  \sum_{i \in \mathcal{I}_k} E \left[ \left(  (\Delta_{k,i}^{(b)})^{\top} (G^{-1} \partial_\eta m_i)  \right)^2 \right]  = E \left[ \left(  (\Delta_{k,i}^{(b)})^{\top} (G^{-1} \partial_\eta m_i)  \right)^2 \right]~.
\end{align*}
Finally, we use consider the following derivations:
\begin{align*}
    I_2 &\overset{(1)}{=}N^{-2-2\varphi} \sum_{ \substack{j_1,j_2 \notin \mathcal{I}_k \\ j_1 \neq j_2 } } E \left[ \left(  b_N(X_{j_1},X_i)^{\top} (G^{-1} \partial_\eta m_i)  \right) \left(  b_N(X_{j_2},X_i)^{\top} (G^{-1} \partial_\eta m_i)  \right) \right] \\
    &~~ + N^{-2-2\varphi} \sum_{ j \notin \mathcal{I}_k } E \left[ \left(  b_N(X_{j},X_i)^{\top} (G^{-1} \partial_\eta m_i)  \right)^2  \right] \\
    &\overset{(2)}{=} N^{-2\varphi}(1-N^{-1}) E \left[ \left(  \tilde{b}_N(X_i)^{\top} (G^{-1} \partial_\eta m_i)  \right)^2  \right] + N^{-1-2\varphi}  E \left[ \left(  b_N(X_{j},X_i)^{\top} (G^{-1} \partial_\eta m_i)  \right)^2  \right] \\
    &\overset{(3)}{=}  N^{-2\varphi}E \left[ \left(  \tilde{b}_N(X_i)^{\top} (G^{-1} \partial_\eta m_i)  \right)^2  \right]  + O(N^{-4\varphi})~,
\end{align*}
where (1) holds by the definition of $\Delta_{k,i}^{(b)}$, (2) holds by the Law of Iterated Expectations and definition of $\tilde{b}_N$, and (3) holds by Assumption~\ref{asm:SLE}(iii) and since $\varphi <1/2$.
 
\noindent \textbf{Claim 3:} $I_3 = 0$. To prove this, we first note that for any $i_1 \neq i_2$:
$$ E \left[ \left(   (\Delta_{k(i_1),i_1}^{(\delta)})^{\top} (G^{-1} \partial_\eta m_{i_1})  \right)\left( (\Delta_{k(i_2),i_2}^{(b)})^{\top} (G^{-1} \partial_\eta m_{i_2})  \right) \right] = 0~,$$
which holds by the Law of Iterated Expectations and Assumption~\ref{asm:EM}(iii). We then use this result to write:
$I_3 = n^{-1} \sum_{k=1}^{K_n}  \sum_{i \in \mathcal{I}_k} E \left[ \left(  (\Delta_{k,i}^{(\delta)})^{\top} (G^{-1} \partial_\eta m_i)  \right)\left(  (\Delta_{k,i}^{(b)})^{\top} (G^{-1} \partial_\eta m_i)  \right) \right]$. 
 Finally, we conclude by noting that 
 $ E \left[ \left(  (\Delta_{k,i}^{(\delta)})^{\top} (G^{-1} \partial_\eta m_i)  \right)\left(  (\Delta_{k,i}^{(b)})^{\top} (G^{-1} \partial_\eta m_i)  \right) \right]=0$, 
 which holds by the Law of Iterated Expectations and Assumption~\ref{asm:SLE}(ii).
\end{proof}

\begin{proof}[Proof of Lemma~\ref{lemma:approximation-dml2}] Recall that $\hat{R}_{k}(x) = \hat{\eta}_{k}(x) - \eta_0(x) - \Delta_{k}(x)$, where $\Delta_{k}(x) = \Delta_{[n]\setminus \mathcal{I}_k}(x)$, $\varphi \in (1/4,1/2)$ by Assumption~\ref{asm:SLE}, and $N \asymp n$ since $N = (1-\tfrac{1}{K_n})n$. Assumption~\ref{asm:SLE}(i) implies $E[\|\hat{R}_{k}(X_i)\|^2]^{1/2} \lesssim N^{-2\varphi}$ for $i \in \mathcal{I}_k$.

\textit{Part (i):} Recall $m_i = m(W_i,\theta_0,\eta_i)$ and $\Delta_{k,i} = \Delta_k(X_i)$, where $\eta_i = \eta_0(X_i)$. Denote $\widehat{m}_i = m(W_i,\theta_0,\hat{\eta}_i)$ and ${m}_i^{(\Delta)} = m(W_i,\theta_0,{\eta}_i+\Delta_{k,i})$ for $i \in \mathcal{I}_k$. Using this notation, we write
\begin{align*}
    n^{-1/2} \sum_{i=1}^n G^{-1} (\widehat{m}_i - m_i) % &= n^{-1/2} \sum_{i=1}^n  G^{-1} (\widehat{m}_i - m_i^{(\Delta)}) + G^{-1} ({m}_i^{(\Delta)} - m_i) \\
    &= I_1 + I_2~,
\end{align*}
where $I_1  = n^{-1/2} \sum_{i=1}^n  G^{-1} (\widehat{m}_i - m_i^{(\Delta)})$ and $I_2 = n^{-1/2} \sum_{i=1}^n   G^{-1} ({m}_i^{(\Delta)} - m_i)$.  
Claims 1 and 2 below shows $I_1 = O_p(n^{1/2-3\varphi})$ and $I_2 = \mathcal{T}_{n,K_n}^{\ell} + \mathcal{T}_{n,K_n}^{q} +  O_p(n^{1/2-3\varphi})$, which completes the proof.

\noindent \textbf{Claim 1:} $I_1 = O_p(n^{1/2-3\varphi})$. We consider the following derivations: 
\begin{align*}
    I_1 &\overset{(1)}{=} n^{-1/2} \sum_{k=1}^{K_n} \sum_{i \in \mathcal{I}_k} G^{-1} \hat{R}_{k}(X_i)^\intercal \partial_\eta m(W_i,\theta_0,\eta_i+\Delta_{k,i}) \\
    &~~ + \tfrac{1}{2} n^{-1/2} \sum_{k=1}^{K_n} \sum_{i \in \mathcal{I}_k} G^{-1} \hat{R}_{k}(X_i)^\intercal \partial_\eta^2 m(W_i,\theta_0,\tilde{\eta}_i) \hat{R}_{k}(X_i) \\
    &\overset{(2)}{=} n^{-1/2} \sum_{k=1}^{K_n} \sum_{i \in \mathcal{I}_k} G^{-1} \hat{R}_{k}(X_i)^\intercal \partial_\eta m(W_i,\theta_0,\eta_i)  \\
    &~~ + n^{-1/2} \sum_{k=1}^{K_n} \sum_{i \in \mathcal{I}_k} G^{-1} \hat{R}_{k}(X_i)^\intercal \partial_\eta^2 m(W_i,\theta_0,\check{\eta}_i)  \Delta_{k,i} + O_p(n^{1/2-4\varphi})\\
    &\overset{(3)}{=} O_p(n^{-2\varphi}) + O_p(n^{1/2-3\varphi}) + O_p(n^{1/2-4\varphi})
\end{align*}
 where (1) holds by a Taylor expansion with Lagrange remainder term, (2) holds by the Mean Value Theorem, Assumption~\ref{asm:EM}(i)--(iii) and  $E[\| \hat{R}_{k}(X_i)\|^2]  \lesssim n^{-4\varphi}$, (3) holds by Assumption~\ref{asm:EM}(i)--(iii), Cauchy-Schwarz, $E[\| \hat{R}_{k}(X_i)\|^2]^{1/2}  \lesssim n^{-2\varphi}$, $E[\| \Delta_{k,i} \|^2]^{1/2}  \lesssim n^{-\varphi}$, and because 
 \begin{align*}
     E\left[n^{-1/2} \sum_{k=1}^{K_n} \sum_{i \in \mathcal{I}_k} G^{-1} \hat{R}_{k}(X_i)^\intercal \partial_\eta m(W_i,\theta_0,\eta_i) \right] &\overset{(4)}{=}  0~,\\
     E\left[ \left( n^{-1/2} \sum_{k=1}^{K_n} \sum_{i \in \mathcal{I}_k} G^{-1} \hat{R}_{k}(X_i)^\intercal \partial_\eta m(W_i,\theta_0,\eta_i) \right)^2 \right] &\overset{(5)}{=}  O(n^{-4\varphi})~,
 \end{align*}
 where (4) holds by the Law of Iterated Expectations and Assumption~\ref{asm:EM}(iii), and (5) holds by expanding the expression and Claims 1.1 and 1.2 below.
  
 Define
 \begin{align*}
     I_{1,1} &= E\left[ n^{-1} \sum_{k=1}^{K_n} \sum_{i \in \mathcal{I}_k} (G^{-1} \hat{R}_{k}(X_i)^\intercal \partial_\eta m_i)^2 \right] \\
      I_{1,2} &= E\left[ n^{-1} \sum_{ \substack{ k_1,k_2=1 \\ k_1 \neq k_2}  }^{K_n} \sum_{i_1 \in \mathcal{I}_{k_1}} \sum_{i_2 \in \mathcal{I}_{k_2}} (G^{-1} \hat{R}_{k_1}(X_{i_1})^\intercal \partial_\eta m_{i_1})(G^{-1} \hat{R}_{k_2}(X_{i_2})^\intercal \partial_\eta m_{i_2}) \right]
 \end{align*}
 \noindent \textbf{Claim 1.1} $I_{1,1} = O(n^{-4\varphi})$. This follows from the Law of Iterated Expectations and Assumptions~\ref{asm:EM}(iii) and \ref{asm:SLE}(i).

 \noindent \textbf{Claim 1.2} $I_{1,2} = O(n^{-4\varphi})$. This follows from the Law of Iterated Expectations, Cauchy-Schwarz, and Assumptions~\ref{asm:EM}(iii) and~\ref{asm:AS-SO}, as in the proof of Lemma~\ref{lemma:DML2_Kn}. 

\noindent \textbf{Claim 2:}  $I_2 = \mathcal{T}_{n,K_n}^{\ell} + \mathcal{T}_{n,K_n}^{q} +  O_p(n^{1/2-3\varphi})$. By a Taylor expansion with Lagrange remainder, we obtain
$$ G^{-1} (m_i^{(\Delta)} - m_i) = \Delta_{k,i}^\intercal (G^{-1}  \partial_\eta m_i) + \tfrac{1}{2} \Delta_{k,i}^\intercal ( G^{-1}  \partial_\eta^2 m_i) \Delta_{k,i} + \tilde{r}_{k,i}~,$$
where $i \in \mathcal{I}_k$. Using this decomposition, we have 
$$I_2 = \mathcal{T}_{n,K_n}^{\ell} + \mathcal{T}_{n,K_n}^{q} + n^{-1/2} \sum_{k=1}^{K_n} \sum_{i \in \mathcal{I}_k} \tilde{r}_{k,i}~.$$
Assumption~\ref{asm:EM}(i)--(iii) implies $|\tilde{r}_i| \lesssim \|\Delta_{k,i}\|^3$; therefore, it is sufficient to show that 
$$n^{-1/2} \sum_{k=1}^{K_n} \sum_{i \in \mathcal{I}_k} \| \Delta_{k,i} \|^3 = O_p(n^{1/2-3\varphi})~.$$
To prove this, we consider the following derivations:
\begin{align*}
    E\left[ n^{-1/2} \sum_{k=1}^{K_n} \sum_{i \in \mathcal{I}_k} \| \Delta_{k,i} \|^3 \right] &\overset{(1)}{=} n^{1/2} E\left[  \| \Delta_{k,i} \|^3 \right] \\
    &\overset{(2)}{\le} 4n^{1/2} E\left[  \| \Delta_{k,i}^{(\delta)} \|^3 \right] + 4 n^{1/2} E\left[   \| \Delta_{k,i}^{(b)} \|^3 \right] \\
    &\overset{(3)}{\le} 4n^{1/2} E\left[  \| \Delta_{k,i}^{(\delta)} \|^4 \right]^{3/4} + 4 n^{1/2} E\left[   \| \Delta_{k,i}^{(b)} \|^4 \right]^{3/4} \\
    &\overset{(4)}{=} O(n^{1/2-3\varphi})
\end{align*}
 where (1) holds since $\Delta_{k,i}$ are identically distributed, (2) hold by the Loève's inequality, (3) holds by Holder's inequality, and (4) holds by Lemma~\ref{appendix:lemma_C0}. 
\end{proof}

\begin{lemma}\label{appendix:lemma_C0} 
Let Assumption \ref{asm:SLE} holds. Then,  
\begin{enumerate}
    \item[(i)] $E\left[  \| \Delta_{k,i}^{(\delta)} \|^4 \right] = O(N^{-4\varphi}) $ 

    \item[(ii)] $E\left[  \| \Delta_{k,i}^{(b)} \|^4 \right] =O(N^{-4\varphi})$
\end{enumerate}
\end{lemma}
\begin{proof} It is sufficient to show the result for the case when $\delta_{N}$ and $b_{N}$ are real-valued functions since for any $x = (x_1,\ldots,x_p) \in \mathbf{R}^p$ it holds
$\| x \|^4 = \left(\sum_{\ell=1}^p x_\ell^2 \right)^2 \le p \sum_{\ell=1}^p |x_\ell|^4 $.
\textit{Part (i):} We consider the following derivations:
\begin{align*}
    E\left[  \| \Delta_{k,i}^{(\delta)} \|^4 \right] &\overset{(1)}{=} E\left[ E\left[ \left( N^{-1/2} \sum_{j \notin \mathcal{I}_k} N^{-\varphi} \delta_N(W_j,X_i) \right)^4 \mid X_i \right]  \right] \\
    &\overset{(2)}{=}  N^{-2-4\varphi} \left( E\left[ N  E\left[ \delta_N(W_j,X_i)^4 \mid X_i \right]     + 3 N(N-1)   E\left[ \delta_N(W_j,X_i)^2 \mid X_i \right]^2   \right]\right) \\
    &\overset{(3)}{=} O(N^{-6\varphi}) + O(N^{-4\varphi})~,
\end{align*}
where (1) holds by the Law of Iterated Expectations and definition of $\Delta_{k,i}^{(\delta)}$, (2) holds by expanding the fourth moment and using conditional independence given $X_i$, (3) holds by Assumption~\ref{asm:SLE}(ii).

\textit{Part (ii):} Recall $\tilde{b}_N(x) = E[b_N(X_j,x)]$. We consider the following derivations:
\begin{align*}
    E\left[  \| \Delta_{k,i}^{(b)} \|^4 \right] &\overset{(1)}{\le} 8 E\left[ \left( N^{-1} \sum_{ j \notin \mathcal{I}_k} N^{-\varphi}\left(b_N(X_j,X_i) - \tilde{b}_N(X_i)\right)  \right)^4 \right] + 8N^{-4\varphi} E\left[\tilde{b}_N(X_i)^4\right] \\
    &\overset{(2)}{=} 8 N^{-3}    E\left[ N^{-4\varphi}\left(b_N(X_j,X_i) - \tilde{b}_N(X_i)\right)^4  \right] \\
    &~~ + 24  N^{-3}  (N-1)   E\left[ E\left[ N^{-2\varphi} \left(b_N(X_j,X_i) - \tilde{b}_N(X_i)\right)^2  \mid X_i \right]^2 \right]  + O(N^{-4\varphi}) \\
    &\overset{(3)}{\lesssim} N^{-3-4\varphi}    E\left[ b_N(X_j,X_i)  ^4  \right] +  N^{-2-4\varphi} E\left[ E\left[  b_N(X_j,X_i)^2  \mid X_i \right]^2 \right]  + O(N^{-4\varphi})   \\
    &\overset{(4)}{=} O(N^{-6\varphi}) + O(N^{-4\varphi})~,
\end{align*}
where (1) hold by the Loève's inequality and definition of $\Delta_{k,i}^{(b)}$, (2) holds by Assumption~\ref{asm:SLE}(iii) and expanding the fourth moment and using conditional independence given $X_i$, (3) holds by  Loève's and Holder's inequality, and (4) holds by Assumption~\ref{asm:SLE}(iii). 
\end{proof}

\section{Cost of Using a Fixed Number of Folds for DML2}\label{sec:relative-efficiency-loss}

We quantify the cost of using DML2 with a fixed number of folds (e.g. 5 or 10) instead of the theoretical benchmark (DML2 with $n$ folds). Table \ref{tab:relative-efficiency-loss} reports the resulting relative losses of second-order bias and MSE across different choices of $K_n$ and values of $\varphi \in (1/4,1/2)$.   The table shows that the cost of using a small number of folds is especially large for bias when first-step estimators are more precise, that is, when $\varphi$ is larger.
  
To simplify the analysis, we use the leading-order terms of $E[\mathcal{T}_{n,K_n}]$ and $E[\mathcal{T}_{n,K_n}^2]$ in Theorem \ref{theorem:so-bias-mse-DML2} to measure the percentage loss from choosing $K_n$ folds instead of $n$ folds. We define the relative loss of the choice $K_n$ in terms of second-order bias by
\begin{equation*}
    \mathcal{RL}_{\text{bias}}(K_n) \equiv \left(  \tfrac{1 + \frac{1}{K_n-1}}{1 + \frac{1}{n-1}}   \right)^{2\varphi} - 1~,
\end{equation*}
and the relative loss of the choice $K_n$ in terms of second-order MSE by
\begin{equation*}
    \mathcal{RL}_{\text{MSE}}(K_n) \equiv   \tfrac{\Sigma + \mathcal{B}^2(1+\frac{1}{K_n-1})^{4\varphi} n^{1-4\varphi} }{ \Sigma + \mathcal{B}^2(1+\frac{1}{n-1})^{4\varphi} n^{1-4\varphi} } - 1~.
\end{equation*}
By construction, $\mathcal{RL}_{\text{bias}}(n) = \mathcal{RL}_{\text{MSE}}(n) = 0$ and  $\mathcal{RL}_{\text{bias}}(K_n) ,~\mathcal{RL}_{\text{MSE}}(K_n) \geq 0$ for all $K_n \le n$. The calculation of both relative losses requires information on $\varphi$ and $\mathcal{B}^2/\Sigma$, which often are unknown quantities.\footnote{The quantity $\mathcal{B}^2/\Sigma$ can take any positive value for some econometric models. For instance, the design in \citet[Section 6]{linton1995second} implies $\Sigma = 1$ and $\mathcal{B} = \tfrac{c_h^4}{25} \kappa_1 \log(\kappa_2+1) \log 3$ for any given $\kappa_1$ and $\kappa_2>0$.} One immediate solution is to obtain the worst relative losses across all possible  $\varphi \in (1/4,1/2)$ and $\mathcal{B}^2/\Sigma \in (0,\infty)$, which is equivalent to find upper bounds that depends only on $K_n$ and $n$, that is
$$ \mathcal{RL}_{\text{bias}}(K_n) \le \left(  \tfrac{1 + \frac{1}{K_n-1}}{1 + \frac{1}{n-1}}   \right)  - 1 \qquad \text{and} \qquad \mathcal{RL}_{\text{MSE}}(K_n)\le  \left(  \tfrac{1 + \frac{1}{K_n-1}}{1 + \frac{1}{n-1}}   \right)^2 -1~.$$
These upper bounds imply that the worst-case relative loss from using DML2 with $10$ folds is $11\%$ for second-order bias and $23\%$ for second-order MSE when $n \ge 1,000$.

\begin{table}[t]
\caption{Relative Loss for Different Numbers of Folds}
\label{tab:relative-efficiency-loss}
\begin{center}
\begin{tabular}{c c c c c c c}
\hline 
\multicolumn{7}{c}{Relative loss in terms of second-order bias} \\
\hline
\(K_n\) 
& \(\varphi=1/4\) 
& \(\varphi=2/7\) 
& \(\varphi=1/3\) 
& \(\varphi=2/5\) 
& \(\varphi=1/2\) 
& Worst RL \\
\hline
2  & 41\% & 49\% & 59\% & 74\% & 100\% & 100\% \\
5  & 12\% & 14\% & 16\% & 20\% & 25\%  & 25\%  \\
10 & 5\%  & 6\%  & 7\%  & 9\%  & 11\%  & 11\%  \\
20 & 3\%  & 3\%  & 3\%  & 4\%  & 5\%   & 5\%   \\
\hline
\multicolumn{7}{c}{Relative loss in terms of second-order MSE} \\
\hline
\(K_n\) 
& \(\varphi=1/4\) 
& \(\varphi=2/7\) 
& \(\varphi=1/3\) 
& \(\varphi=2/5\) 
& \(\varphi=1/2\) 
& Worst RL \\
\hline
2  & 99\% & 117\% & 133\% & 91\% & 10\% & 133\% \\
5  & 25\% & 28\%  & 30\%  & 19\% & 2\%  & 30\%  \\
10 & 11\% & 12\%  & 13\%  & 8\%  & 1\%  & 13\%  \\
20 & 5\%  & 6\%   & 6\%   & 4\%  & 0\%  & 6\%   \\
\hline 
\end{tabular}
\end{center}
\legend{ ``Worst RL'' denotes the worst relative loss over the considered range of $\varphi$. Sample size $n=3,000$ and $\mathcal{B}^2/\Sigma = 100$.}
\end{table}

Table \ref{tab:relative-efficiency-loss} complements our worst-case analysis. The table assumes $\mathcal{B}^2/\Sigma = 100$ and a sample size $n=3,000$, and it reports the relative losses for several values of $\varphi \in (1/4,1/2)$, including $\varphi = 1/4$ and $\varphi = 1/2$ for illustration purposes. Table \ref{tab:relative-efficiency-loss} shows that for sufficiently accurate first-step estimators (e.g., $\varphi>1/3$) may reduce the relative loss in terms of second-order MSE.

\section{Results for Local-Polynomial Estimators}\label{appendix:local-polynomials}

This appendix presents low-level conditions under which first-step estimators based on local-polynomial regressions verify the high-level conditions imposed in Section~\ref{sec:result2}.

\subsection{Conditions and $(\delta_N,b_N)-$functions}\label{app:lp-setup}

We consider a scalar nuisance function $\eta_0$ without loss of generality, since the argument can be applied component-wise and the resulting components can then be stacked. Let 
\begin{equation*} 
  \eta_0(x) = E[Y \mid X = x], \qquad x \in \mathcal{X} \subseteq \mathbf{R}^{d_x},
\end{equation*}
where $Y$ is a  scalar outcome. Let $\varepsilon_i = Y_i - \eta_0(X_i)$ be the regression error. By definition, $E[\varepsilon_i \mid X_i] = 0$. We now present conditions on the distribution of covariate $X$, the regression error $\varepsilon_i$, and smoothness of the nuisance function $\eta_0$:

\begin{con}[Covariate distribution]\label{con:lp-sampling}
$X$ has a continuously differentiable density $f_X$ with compact and convex support $\mathcal{X}\subseteq\mathbb{R}^{d_x}$ such that $f_X(x) \in [\underline f, \bar f]$ for all $x \in \mathcal{X}$.% $0<\underline f\leq f_X(x)\leq \bar f<\infty$ for all $x\in\mathcal{X}$.
\end{con}

\begin{con}[Regression error]\label{con:lp-error}
(i) For some $\nu > 0$ and $C < \infty$, $E[|\varepsilon_i|^{4+\nu} \mid X_i] \le C$ a.s., and (ii) for some $\underline \sigma, \bar \sigma \in (0,\infty)$, $\sigma^2(x) = E[\varepsilon_i^2 \mid X_i = x] \in [\underline \sigma, \bar \sigma]$ for all $x\in\mathcal{X}$.
\end{con}

\begin{con}[Smoothness and degree]\label{con:lp-smooth}
(i) $\eta_0$ has $s+1$ continuous derivatives, (ii) $s$ is even, and (iii)  $s > \max\{d_x/2,1\}$.
\end{con}

\subsubsection{Local-polynomial estimator with spectral trimming}

Let $p=s-1$ be a positive integer and $a = (a_1,\dots,a_{d_x}) \in \mathbf{N}^{d_x}$ denote a multi-index. We write $|a| = \sum_\ell a_\ell$, $u^a = \prod_\ell u_\ell^{a_\ell}$, and $a! = \prod_\ell a_\ell!$. Let $\mathcal{A}_p = \{a : |a| \le p\}$ be the set of all the multi-index with norm lower than or equal to $p$, $J_p = |\mathcal{A}_p| = \binom{d_x+p}{p}$, and we collect the basis in the vector
\begin{align*}
    r_p(u) = \left( \tfrac{u^a}{a!} \right)_{a \in \mathcal{A}_p} \in \mathbf{R}^{J_p},  
\end{align*}
where the first entry is the constant element ($a = 0$). Let $e_1 \in \mathbf{R}^{J_p}$ be the first column of~$\mathbb{I}_{J_p}$.

Recall that the first-step estimator $\hat{\eta}_k(\cdot)$ uses the sample $\{W_j = (Y_j,X_j) : j \notin \mathcal{I}_k\}$ with number of observations equal to $N = (1 - \tfrac{1}{K_n})n$. For a kernel $\mathcal{K} : \mathbf{R}^{d_x} \to \mathbf{R}$ and bandwidth $h_N$, define $\mathcal{K}_{h_N}(\cdot) = h_N^{-d_x} \mathcal{K}(\cdot / h_N)$ and $U_{i,x} = (X_i - x)/h_N$. 
Define  
\begin{align*}
     \mathcal{M}_h(x)  = E \left[ \mathcal{K}_{h}(X_i - x) r_p(U_{i,x}) r_p(U_{i,x})^\intercal \right]  = \int_{\mathcal{U}_h(x)} \mathcal{K}(u) r_p(u) r_p(u)^\intercal f_X(x + h u) du,
\end{align*} 
where $\mathcal{U}_h(x) = \{u : x + h u \in \mathcal{X}\}$. We next impose conditions on the kernel $\mathcal{K}$ and $\mathcal{M}_h$.

\begin{con}[Kernel and population design matrix]\label{con:lp-kernel}
(i) $\mathcal{K}$ is nonnegative, bounded, symmetric, and supported in $[-1,1]^{d_x}$. (ii) For some $\epsilon, c_Q, C_Q \in (0,\infty)$, $c_Q \le \lambda_{\min}\{\mathcal{M}_h(x)\} \le \lambda_{\max}\{\mathcal{M}_h(x)\} \le C_Q$ for all $x \in \mathcal{X}$ and all $h \in (0,\epsilon)$, and (iii) $\mathcal{M}_0(x) = \lim_{h \to 0} \mathcal{M}_h(x)$ exists for all $x \in \mathcal{X}$.
\end{con}

\begin{con}[Nondegenerate leading bias]\label{con:lp-bias} (i) The next limit exist and is finite,
\begin{equation}\label{eq:lp-leading-bias}
  B_p(x) = \lim_{h \to 0} \sum_{|a| = s} \frac{D^a \eta_0(x)}{a!}\,
    e_1^\intercal \mathcal{M}_h(x)^{-1} f_X(x)\int_{\mathcal{U}_h(x)} K(u) r_p(u) u^a du~, \forall x \in \mathcal{X}~,
\end{equation}
and (ii) $\inf_{x\in\mathcal{X}}|B_p(x)|>0$ and $\sup_{x\in\mathcal{X}}|B_p(x)|<\infty$.
\end{con}

To define the degree-$p$ local-polynomial estimator with spectral trimming for $\eta_0$, we need to introduce additional notation: $\hat{S}_k(x)  = \tfrac{1}{N}\!\sum_{j \notin \mathcal{I}_k}\! \mathcal{K}_{h_N}(X_j - x) r_p(U_{j,x}) Y_j$ and 
$$  \widetilde{\mathcal{M}}_k(x)  = \tfrac{1}{N}\!\sum_{j \notin \mathcal{I}_k}\! \mathcal{K}_{h_N}(X_j - x) r_p(U_{j,x}) r_p(U_{j,x})^\intercal~.$$ 
Define also the \emph{trimming} event $ \mathcal{G}_k(x) = \{\lambda_{\min} \{  \widetilde{\mathcal{M}}_k(x)\} \ge c_Q/2\}$. Recall $p=s-1$.

\begin{definition}[Local-polynomial estimator with spectral trimming]
    The estimator $\hat{\eta}_k$ of $\eta_0$ is defined as
    $$\hat{\eta}_k(x) = e_1^\intercal \widehat{\mathcal{M}}_k(x)^{-1} \hat{S}_k(x)~,$$
    where $\widehat{\mathcal{M}}_k(x) =   \widetilde{\mathcal{M}}_k(x)$ on the event $\mathcal{G}_k(x) $ and  $\widehat{\mathcal{M}}_k(x) =  c_Q \mathbb{I}_{J_p} + \widetilde{\mathcal{M}}_k(x)$ otherwise. 
\end{definition}

Note that $\hat{\eta}_k(x)$ coincides with the standard local polynomial estimator of degree $p$ whenever the trimming event $\mathcal{G}_k(x)$ holds, that is, when the sample Gram matrix $\widehat{\mathcal{M}}_k(x)$ is well conditioned. By construction, $\hat{\eta}_k(x)$ is well defined, since we replace the inverse of $\widetilde{\mathcal{M}}_k(x)$ with a uniformly bounded matrix on $\mathcal{G}_k(x)^c$. The next condition can be verified using Conditions~\ref{con:lp-sampling} and \ref{con:lp-kernel}, but it is presented as a condition due to limit constraints.

\begin{con}\label{con:lp-trimming}
    There is a constant $C_{\mathcal{M}}>0$ such that
\begin{align}
    \Big\{ E\big[ \| \widehat{\mathcal{M}}_k(X)^{-1} -\mathcal{M}_{h_N}(X)^{-1} \| ^4 \big] \Big\}^{1/4}
  &\le C_{\mathcal{M}}(N h_N^{d_x})^{-1/2}~,\label{eq:lp-inverse-fourth} \\ 
  \sup_{x \in \mathcal{X}}~\max_{\ell \notin \mathcal{I}_k}\, \Big\{ E\Big[ \big\| \widehat{\mathcal{M}}^{(\ell)}_k(x) - \widehat{\mathcal{M}}_k(x) \big\|^2 \Big] \Big\}^{1/2} &\le  C_{\mathcal{M}} N^{-1/2} \left( N h_N^{d_x}  \right)^{-1/2}~.\label{eq:lp-gram-stability-L2}
\end{align} 
\end{con} 
 
\subsubsection{The definition of the functions $\delta_N$ and $b_N$}

Define the population analogue of the sample equivalent kernel
\begin{equation}\label{eq:lp-equiv-kernel}
  \ell_N(z,x) = e_1^\intercal\mathcal{M}_{h_N}(x)^{-1} \mathcal{K}_{h_N}(z - x) r_p\!\left( \tfrac{z - x}{h_N} \right),
\end{equation}
which satisfies the  restrictions 
$E[\ell_N(X_i,x) ((X_i - x)/h_N)^a] = \mathbf{1}\{a = 0\} $ for $ |a| \le p.$

Let $Q_{p,x}(z)$ be the degree-$p$ Taylor polynomial of $\eta_0$ at $x$. Recall that $p=s-1$. Let $\rho_N(z,x) = \eta_0(z) - Q_{p,x}(z)$ be the Taylor residual. 

The next lemma presents a stochastic linear expansion for local-polynomial estimators with spectral trimming. Henceforth, we set the bandwidth $h_N = C_h N^{-1/(2s + d_x)}$ for some $C_h \in (0,\infty)$ and define $\varphi = \tfrac{s}{2s+d_x}$. Recall $N= (1-\tfrac{1}{K_n})n$ and $\varepsilon_i$ is the regression error.

\begin{lemma}\label{lem:lp-expansion}
Let Conditions~\ref{con:lp-sampling}--\ref{con:lp-kernel} and \ref{con:lp-trimming} hold. Then, for every $x \in \mathcal{X}$, we have
\begin{equation}\label{eq:lp-expansion}
  \hat{\eta}_k(x) - \eta_0(x)
  = \frac{1}{N}\sum_{j \notin \mathcal{I}_k} \ell_N(X_j,x) \varepsilon_j
  + \frac{1}{N}\sum_{j \notin \mathcal{I}_k} \ell_N(X_j,x) \rho_N(X_j,x)
  + \mathcal{R}_k(x),
\end{equation}
where  
\begin{align}
    \mathcal{R}_k(x) &= e_1^\intercal \{ \widehat{\mathcal{M}}_k(x)^{-1} -\mathcal{M}_{h_N}(x)^{-1} \}
  \{ \hat{S}_{\varepsilon,k}(x) + \hat{S}_{\rho,k}(x) \} \label{eq:lp-Rk}~, \\
  \hat{S}_{\varepsilon,k}(x) &= N^{-1}\sum_{j\notin\mathcal{I}_k}
K_{h_N}(X_j-x) r_p(U_{j,x}) \varepsilon_j~, \notag\\
  \hat{S}_{\rho,k}(x) &= N^{-1}\sum_{j\notin\mathcal{I}_k} K_{h_N}(X_j-x) r_p(U_{j,x}) \rho_N(X_j,x)~.\notag 
\end{align} 
Furthermore,  $ E[ |\mathcal{R}_k(X)|^2 ] ^{1/2} \le C N^{-2\varphi}$ for some $C>0$.
\end{lemma}

\begin{proof} Define $  \hat{\ell}_{k,N}(z,x) =  e_1^\intercal \widehat{\mathcal{M}}_k(x)^{-1} \mathcal{K}_{h_N}(z - x) r_p\!\left( \tfrac{z - x}{h_N} \right)$. We first consider the following derivations
\begin{align*}
    \hat{\eta}_k(x) &\overset{(1)}{=} N^{-1}\sum_{j\notin\mathcal{I}_k} \hat{\ell}_{k,N}(X_j,x) Y_j \overset{(2)}{=} N^{-1}\sum_{j\notin\mathcal{I}_k} \hat{\ell}_{k,N}(X_j,x) ( Q_{p,x}(X_j) + \rho_N(X_j,x) +
\varepsilon_j) \\
&\overset{(3)}{=} \eta_0(x) + N^{-1}\sum_{j\notin\mathcal{I}_k} \hat{\ell}_{k,N}(X_j,x) (\rho_N(X_j,x) + \varepsilon_j)~,
\end{align*}
where (1) holds by definition of  $ \hat{\ell}_{k,N}(X_j,x)$, (2) holds by definition of $\eta_0(X_j)$ and Taylor approximation with its remainder, and (3) holds by definition of $\hat{\ell}_{k,N}(z,x)$, which implies $N^{-1}\sum_{j\notin\mathcal{I}_k} \hat{\ell}_{k,N}(X_j,x) Q_{p,x}(X_j) = \eta_0(x)$. We then write
\begin{align*}
    N^{-1}\sum_{j\notin\mathcal{I}_k} \hat{\ell}_{k,N}(X_j,x) (\rho_N(X_j,x) + \varepsilon_j) = N^{-1}\sum_{j \notin \mathcal{I}_k} \ell_N(X_j,x) (\varepsilon_j + \rho_N(X_j,x) ) 
  + \mathcal{R}_k(x)~,
\end{align*}
where we used the definition of $\hat{S}_{\varepsilon,k}(x)$ and $\hat{S}_{\rho,k}(x)$. This completes the proof of \eqref{eq:lp-expansion}. 

Finally, $ E[ |\mathcal{R}_k(X)|^2 ] ^{1/2} \lesssim N^{-2\varphi}$ holds due to the following derivations:
\begin{align*}
    E[ |\mathcal{R}_k(X)|^2 ] % &= E\left[ |e_1^\intercal \{ \widehat{\mathcal{M}}_k(X)^{-1} -\mathcal{M}_{h_N}(X)^{-1} \}   \{ \hat{S}_{\varepsilon,k}(X) + \hat{S}_{\rho,k}(X) \} |^2 \right]^{1/2} \\
  &\overset{(1)}{\le} E\left[ \|  \widehat{\mathcal{M}}_k(X)^{-1} -\mathcal{M}_{h_N}(X)^{-1} 
\|^4 \right]^{1/4} E\left[ \|
   \hat{S}_{\varepsilon,k}(X) + \hat{S}_{\rho,k}(X)   \|^4 \right]^{1/4} \\
   &\overset{(2)}{\lesssim} (N h_N^{d_x})^{-1/2} \left( (N h_N^{d_x})^{-1/2} + h_N^{s} \right) \overset{(3)}{=} N^{-2\varphi}~,
\end{align*}
where (1) holds by Cauchy-Schwartz, (2) holds by Condition~\ref{con:lp-trimming}, the triangle inequality, and \eqref{eq:S_epsilon}  below, and (3) holds by definition of $\varphi$, which implies $(N h_N^{d_x})^{-1/2} \asymp h_N^s \asymp N^{-\varphi}$.
\begin{align}
    \{E[\|\hat{S}_{\varepsilon,k}(X)\|^4]\}^{1/4}  \lesssim   (N h_N^{d_x})^{-1/2}~, \qquad \{E[\|\hat{S}_{\rho,k}(X)\|^4]\}^{1/4}   \lesssim   h_N^s~. \label{eq:S_epsilon} 
\end{align}    
Note that \eqref{eq:S_epsilon} holds by similar arguments as in the proof of  Lemma \ref{appendix:lemma_C0} and Conditions~\ref{con:lp-sampling}--\ref{con:lp-kernel}, where we used that there is $C_{Q_{p}}>0$ such that $|\rho_N(z,x)| \le C_{Q_{p}} h_N^s$  when $\|z-x\| \le h_N$ due to Conditions~\ref{con:lp-sampling} and~\ref{con:lp-smooth}.
\end{proof}

\begin{definition}[($\delta_N$, $b_N$) for local-polynomial estimators $\hat{\eta}_k$ with bandwidth $h_N $] 
\begin{equation}\label{eq:lp-delta-b}
  \delta_N(W_j,x) = N^{\varphi - 1/2} \ell_N(X_j,x) \varepsilon_j,
  \qquad
  b_N(X_j,x) = N^{\varphi} \ell_N(X_j,x) \rho_N(X_j,x).
\end{equation}
\end{definition}
Using this notation, we can rewrite \eqref{eq:lp-expansion} as follows:
$  \hat{\eta}_k(x) - \eta_0(x)   = \Delta_k(x)   + \mathcal{R}_k(x),$
where $\Delta_k(x)$ is as in  Assumption~\ref{asm:SLE}. 
   
\begin{remark}\label{rem:delta-b}
    All our previous conditions and derivations were for $E[Y\mid X=x]$. We can use the delta-method to extend the derivation of $(\delta,b)$ for components of the nuisance function defined as $E[Y\mid X=x, A=1] = \tfrac{E[Y A \mid X=x ]}{E[A\mid X=x]}$ and $(E[A\mid X=x])^{-1}$. More concretely, suppose $\eta(x) = \Psi(\gamma(x))$ with $\gamma(x) = (\gamma_1(x),\ldots,\gamma_p(x))$ and $\gamma_t(x)$ verifying conditions of Lemma~\ref{lem:lp-expansion} with $(\delta_N^{(\gamma_t)},b_N^{(\gamma_t)})$; it follows that $(\delta_N^{(\eta)},b_N^{(\eta)})$ are defined as $\delta_N^{(\eta)}(W_j,x) = \sum_{t=1}^p \partial_{\gamma_t}\Psi(\gamma(x)) \delta_N^{(\gamma_t)}(W_j,x)$ and $b_N^{(\eta)}(X_j,x) = \sum_{t=1}^p \partial_{\gamma_t}\Psi(\gamma(x)) b_N^{(\gamma_t)}(X_j,x)$ when $\partial_{\gamma_t}\Psi(\gamma)$ and $\partial_{\gamma_t}^2 \Psi(\gamma)$ are uniformly bounded on the values $\gamma$ can take.
\end{remark}

\subsection{Verification of Assumptions in Section~\ref{sec:result2}}\label{app:lp-SLE}

We now verify the Assumptions in Section~\ref{sec:result2}. We assume that each component of the nuisance function $\eta_0=(\eta_{0,1},\ldots,\eta_{0,p})^\intercal$ satisfies Conditions~\ref{con:lp-sampling}--\ref{con:lp-trimming} using the same kernel, which implies the function $\ell_N$ in \eqref{eq:lp-equiv-kernel} is the same. For each component, the functions $\delta_N$ and $b_N$ are defined as in~\eqref{eq:lp-delta-b}; the scalar argument of Proposition~\ref{prop:lp-SLE} applies component-wise. Conditions~\ref{con:lp-sampling}--\ref{con:lp-bias} imply that there is a finite constant $C$ such that, for all $x,z\in\mathcal{X}$,
\begin{align}
     |\ell_N(z,x)| &\le C h_N^{-d_x}\mathbf{1}\{\|z-x\|_\infty \leq h_N\}~,\label{eq:lp-kernel-bound-proof}\\
     \sup_{x\in\mathcal{X}} E\left[|\ell_N(X_j,x)|^q\right] &\le C h_N^{-d_x(q-1)}~, \text{ for }q = 1,2,3,4,  \label{eq:lp-kernel-moment-proof} \\
      \|\rho_N(z,x)\| &\le C h_N^s \mathbf{1}\{\|z-x\|_\infty\leq h_N\} ~,\label{eq:lp-rho-bound-proof}\\
      E[\ell_N(X_j,x)\rho_N(X_j,x)] &= h_N^s B_p(x)+o(h_N^s)~, \text{uniformly in } x\in\mathcal{X}~. \label{eq:lp-leading-rho-proof}
\end{align}
 Throughout the proof, constants denoted by $C$ may change from line to line and do not depend on $N$, $x$, or $K_n$. We also use $\lesssim$ to ignore the use of the constant.

\begin{prop}\label{prop:lp-SLE}
Let Conditions~\ref{con:lp-sampling}--\ref{con:lp-bias} hold and let $\delta_N$ and $b_N$ be as in~\eqref{eq:lp-delta-b}. Then, Assumption~\ref{asm:SLE}(i)--(iii) holds with $\varphi = s/(2s + d_x) \in (1/4, 1/2)$. Furthermore, if the moment function $m$ is as in Appendix~\ref{sec:appendix_examples}, then Assumption~\ref{asm:SLE}(iv) holds.
\end{prop}

\begin{proof}  
Recall that $h_N=C_hN^{-1/(2s+d_x)}$ and $\varphi=s/(2s+d_x)$. Since Condition~\ref{con:lp-smooth} imposes $s>d_x/2$, we have $\varphi\in(1/4,1/2)$.  

\textit{Part (i).} It follows by Lemma~\ref{lem:lp-expansion} defining  $\hat{R}_k(x)$ as $\mathcal{R}_k(x)$.

We prove Assumption~\ref{asm:SLE}(ii)--(iii) in the scalar case (the $r$-component). The vector case follows component-wise and by equivalence of norms in finite dimensions.

\textit{Part (ii).} Since $E[\varepsilon_{j,r}\mid X_j]=0$ and $\delta_{N,r}(W_j,x) = N^{\varphi-1/2} \ell_N(X_j,x) \varepsilon_{j,r}$, it follows that $E[\delta_{N,r}(W_j,x)\mid X_j] = 0$. Next, algebra shows $E[\delta_{N,r}(W_j,x)^2] = N^{2\varphi-1}E[\ell_N(X_j,x)^2\sigma_r^2(X_j)]$; therefore,  by Condition~\ref{con:lp-error}(ii) and ~\eqref{eq:lp-kernel-moment-proof} with $q=2$ and the nondegeneracy of the equivalent kernel implied by Conditions~\ref{con:lp-sampling} and~\ref{con:lp-kernel}, it follows that $E[\delta_{N,r}(W_j,x)^2]\in(c,C)$ for constants $0<c<C<\infty$ that do not depend on $x$ or $N$. Similarly, using Condition~\ref{con:lp-error}(i) and~\eqref{eq:lp-kernel-moment-proof} with $q=4$, we conclude $E[\delta_{N,r}(W_j,x)^4] \le C N^{1-2\varphi}$.  

\textit{Part (iii).} By~\eqref{eq:lp-leading-rho-proof}, $E[b_{N,r}(X_j,x)]       =C_h^sB_{p,r}(x)+o(1)$. Condition~\ref{con:lp-bias} implies there are constants $c,C\in(0,\infty)$ such that $|E[b_N(X_j,x)]|\in(c,C)$ for all $x \in \mathcal{X}$ for all sufficiently large $N$. Finally, by~\eqref{eq:lp-kernel-moment-proof} and~\eqref{eq:lp-rho-bound-proof}, it follows that $ E[|b_N(X_j,x)|^{2q}] \leq C N^{2q\varphi}h_N^{2qs}E[|\ell_N(X_j,x)|^{2q}] \le C N^{(2q-1)(1-2\varphi)}$ for $q=1,2$. 

\textit{Part (iv).} Let $g_q(x) = E[m(W_j,\theta_0,\eta_j) \varepsilon_j^{q-1} \mid X_j = x]$ for $q=1,2,3$. By definition of $m$ in Appendix~\ref{sec:appendix_examples} and Assumption~\ref{asm:EM}, it follows that $\sup_{x \in \mathcal{X}} \|g_q(x)\|$ is finite. We use this and standard derivations to show: $ \left\|N^{1/2-\varphi}E[m(W_j,\theta_0,\eta_j)\delta_N(W_j,x)]\right\| \lesssim \int |\ell_N(z,x)|f_X(z)dz = O(1)$ due to \eqref{eq:lp-kernel-bound-proof}. Similarly, $\left\|E[m(W_j,\theta_0,\eta_j)b_N(X_j,x)]\right\| \lesssim N^{\varphi}h_N^s\int |\ell_N(z,x)|f_X(z)dz = O(1)$ and $E[\|m(W_j,\theta_0,\eta_j)\delta_N(W_j,x)\|^2] \lesssim N^{2\varphi-1}h_N^{-d_x} = O(1)$ by definition of $h_N$ and $\varphi$, and  due to \eqref{eq:lp-kernel-moment-proof}, respectively. \vspace{-0.2cm}
\end{proof}

\begin{prop}\label{prop:lp-leading}
Let Conditions~\ref{con:lp-sampling}--\ref{con:lp-bias} hold and let the moment function $m$ be as in Appendix~\ref{sec:appendix_examples}. If $P( \text{dist}(X,\partial \mathcal{X}) \le h_N) = O(h_N)$ (e.g., $X \sim \mathrm{Unif}[0,1]^{d_x}$), then Assumption~\ref{asm:leading-terms}(i),(ii), and (iv) hold, and $\mathcal{V}^{(\ell)}>0$ in  Assumption~\ref{asm:leading-terms}(iii). \vspace{-0.3cm} 
\end{prop}

\begin{proof} 
\textit{Part (i):} For the ATE, ATT-DID, and LATE moment functions in Appendix~\ref{sec:appendix_examples}, $m(W,\theta,\eta)$ is a polynomial in $\eta$ of degree at most two. Therefore, $    \partial_\eta^3 m(W_i,\theta_0,\eta)=0$ for all $\eta \in \mathcal{E}$; in particular, $\sup_{\eta\in\mathcal{E}}\|\partial_\eta^3m(W_i,\theta_0,\eta)\|_\infty=0.$ 

\textit{Part (ii):} 
 We calculate and present $\mathcal{V}^{(\ell)}$ and $\mathcal{B}$ for the ATE (Example \ref{example_ate}), the ATT-DID (Example \ref{example_DID}), and the LATE (Example \ref{example_LATE}). For brevity, we omit the intermediate algebra and report only the explicit expressions. We denote $B_p^{(g)}$ as in \eqref{eq:lp-leading-bias} replacing $\eta_0$ by $g$. Let $\kappa = \int_{[-1,1]^{d_x}}\left( e_1^{\intercal} M_0^{-1} r_p(u) K(u) \right)^2\, du$. Recall $h_N = C_h N^{-1/(2s+d_x)}$. 

\noindent \textit{ATE}. Let $p_A(x) = E[A \mid X=x]$, $\mu_a(x) = E[Y \mid X = x, A = a]$ and  $\sigma_a^2(x) = Var(Y \mid X = x, A = a)$ for $a=0,1$.  Note $G = 1$.  Then, $ \mathcal{V}^{(\ell)}  = 8 C_h^{-d_x} \kappa E\left[\tfrac{q_A(X)\, \sigma_1^2(X)}{p_A(X)^2}
+ \tfrac{p_A(X)\, \sigma_0^2(X)}{q_A(X)^2}\right]$ and $\mathcal{B} = 2 C_h^{2s}
E\left[B_p^{(p_A)}(X) \left(\tfrac{B_p^{(\mu_1)}(X)}{p_A(X)}
+ \tfrac{B_p^{(\mu_0)}(X)}{q_A(X)}\right)\right]$, where $q_A(x) = 1 - p_A(x)$. 
 
\noindent \textit{ATT-DID}. Let $p_A(x) = E[A \mid X=x]$, $\Delta Y = Y_1 - Y_0$, $\mu_0^{\Delta}(x) = E[\Delta Y \mid X = x, A = 0]$, and $\sigma_0^2(x) = Var(\Delta Y \mid X = x, A = 0)$. Note $G = E[A]$. Then, $\mathcal{V}^{(\ell)} = \tfrac{8\, C_h^{-d_x} \kappa}{E[A]^2}\;
E\left[\tfrac{p_A(X)\, \sigma_0^2(X)}{q_A(X)^2}\right]$ and $\mathcal{B} = \frac{2\, C_h^{2s}}{E[A]}\; E \left[\tfrac{B_p^{(\mu_0^{\Delta})}(X)\, B_p^{(p_A)}(X)} {q_A(X)}\right]$, where $q_A(x) = 1 - p_A(x)$.

\noindent \textit{LATE}. Let $ p_Z(x)=E[Z\mid X=x]$, $ \mu_{Yz}(x)=E[Y\mid X=x,Z=z]$, $\mu_{Dz}(x)=E[D\mid X=x,Z=z]$, $\varepsilon_{Yz}=Y- \mu_{Yz}(x)$, $ \varepsilon_{Dz}=D-\mu_{Dz}(x)$ and $ \xi_z=\varepsilon_{Yz}-\theta_0\varepsilon_{Dz}$ for $z \in \{0,1\}$. Note $G= E[\mu_{D1}(X)-\mu_{D0}(X)]$. Then, $\mathcal{V}^{(\ell)} = \tfrac{8\, C_h^{-d_x} \kappa}{G^2}  E \left[\tfrac{q_Z(X)\, V_1(X)}{p_Z(X)^2} + \tfrac{p_Z(X)\, V_0(X)}{q_Z(X)^2}\right]$ and 
$\mathcal{B} = \tfrac{2  C_h^{2s}}{G} E\left[B_p^{(p_Z)}(X) \left(\tfrac{B_p^{(\mu_{Y1} - \theta_0 \mu_{D1})}(X)}{p_Z(X)} + \tfrac{B_p^{(\mu_{Y0} - \theta_0 \mu_{D0})}(X)}{q_Z(X)}\right)\right]$, where $q_Z(x)=1-p_Z(x)$ and $ V_z(x)=E[\xi_z^2\mid X=x,Z=z]$. 

\textit{Part (iii): } $\mathcal{V}^{(\ell)}>0$ follows from the closed-form expression presented in part (ii). Note that $\mathcal{B} \neq 0$ requires additional information on the bias of each nuisance component. For instance, if each of them are positive, it follows that $\mathcal{B} \neq 0$.
 
\textit{Part (iv):} Let $\xi_N(W_j,x) = \ell_N(X_j,x)\varepsilon_j+\ell_N(X_j,x)\rho_N(X_j,x)$. Note that $G^{-1} \partial_\eta^2 m_i$ are bounded for $m$ as in Appendix \ref{sec:appendix_examples}, which together with \eqref{eq:lp-kernel-bound-proof} and \eqref{eq:lp-rho-bound-proof} imply $ \|E[\xi_N(W_j,x)]\|\lesssim h_N^{s}$ and $E[\|\xi_N(W_j,x)\|^{2q}] \lesssim h_N^{-(2q-1)d_x}$ for $q=1,2$. We now decompose $T^q_{n,K_n}$ as follows:\vspace{-0.25cm}
\begin{align*}
    \mathcal{T}^q_{n,K_n}     = n^{-1/2}N^{-2}
       \sum_{k=1}^{K_n}\sum_{i\in\mathcal I_k}  \sum_{j,\ell\notin\mathcal I_k}        H_i\xi_N(W_j,X_i)\xi_N(W_\ell,X_i)  = \tfrac{1}{2} D_n + \tfrac{1}{2} U_n,
\end{align*}
where \vspace{-0.5cm}
\begin{align*}
    D_n &= n^{-1/2}N^{-2} \sum_{k=1}^{K_n}\sum_{i\in\mathcal I_k} \sum_{j\notin\mathcal I_k}       \xi_N(W_j,X_i)^\intercal(G^{-1}  \partial_\eta^2 m_i)\xi_N(W_j,X_i),\\
    U_n &= n^{-1/2}N^{-2} \sum_{k=1}^{K_n}\sum_{i\in\mathcal I_k}       \sum_{\substack{j_1,j_2\notin\mathcal I_k\\ j_1\ne j_2}} \xi_N(W_{j_1},X_i)^\intercal (G^{-1}  \partial_\eta^2 m_i) \xi_N(W_{j_2},X_i).
\end{align*} 
Claims 1 and 2 below show that $Var(D_n) = o(n^{-2\varphi})$ and $Var(U_n) = O(n^{-2\varphi})$, which is sufficient to complete the proof due to Cauchy-Schwartz.

\noindent \textbf{Claim 1:} $Var(D_n) = O(n^{-2\varphi})$. Let $h(W_j,W_i) = \xi_N(W_j,X_i)^\intercal(G^{-1}  \partial_\eta^2 m_i)\xi_N(W_j,X_i)$. Consider the following derivations:\vspace{-0.4cm} 
\begin{align*}
     Var(D_n) &\overset{(1)}{\lesssim}   N^{-3} Var(h(W_j,W_i)) + N^{-2} Var(E[h(W_j,W_i)\mid W_i]) + N^{-2} Var(E[h(W_j,W_i) \mid W_j]) \\
     &\overset{(2)}{\lesssim} N^{-3} h_N^{-3d_x} + n^{-2} h_N^{-2d_x} \overset{(3)}{=} O(N^{-4\varphi}) + O(N^{-6\varphi}) \overset{(4)}{=} o(n^{-2\varphi})~,
\end{align*}
where (1) holds by a non-symmetric version of \citet[Theorem 2 in Chapter 4.3]{lee2019u} for incomplete U-statistics and using that $n \asymp N$, (2) holds by the properties of $\xi_N$ and using that variance of random variable is lower than its second moment, (3) holds by the definition of $h_N$ and $\varphi$, and (4) holds since $n \asymp N$ and $\varphi>0$. 

\noindent \textbf{Claim 2:} $Var(U_n) = O(n^{-2\varphi})$. Let $h(W_{j_1}, W_{j_2}, W_{i}) = \xi_N(W_{j_1},X_i)^\intercal (G^{-1}  \partial_\eta^2 m_i) \xi_N(W_{j_2},X_i)$, $I_{N,1} = \sum_{\iota \in \{j_1,j_2,i\} }Var(E[h(W_{j_1}, W_{j_2}, W_{i})\mid W_{\iota}])$, $I_{N,2} = Var(E[h(W_{j_1}, W_{j_2}, W_{i})\mid W_{j_1}, W_i]) + Var(E[h(W_{j_1}, W_{j_2}, W_{i})\mid W_{j_1}, W_{j_2}])$, and $I_{N,3} = Var(h(W_{j_1}, W_{j_2}, W_{i}))$. Consider the following derivations: 
\begin{align*}
     Var(U_n) &\overset{(1)}{\lesssim} n^{-1} N^{-4} \left( n^5 I_{N,1} + n^4 I_{N,2} + n^3 I_{N,3}\right)  \overset{(2)}{\lesssim} N^{-2\varphi} + N^{-1} h_N^{-d_x} + N^{-2} h_N^{-2d_x} \overset{(3)}{=} O(N^{-2\varphi})~,
\end{align*}
where (1) holds by a non-symmetric version of \citet[Theorem 2 in Chapter 4.3]{lee2019u} for incomplete U-statistics, using that $n \asymp N$, (2) holds by Claims 2.1, 2.2, and 2.3 below, and (3) holds by the definition of $h_N$ and $\varphi$.

\noindent \textbf{Claim 2.1:} $I_{N,1} = O(N^{-2\varphi})$. It follows by noting that $\| E[h(W_{j_1}, W_{j_2}, W_{i})  \mid W_i] \| \lesssim N^{-2\varphi}$ and $\| E[h(W_{j_1}, W_{j_2}, W_{i})  \mid W_{j_1}] \| \lesssim \| \varepsilon_{j_1} \| N^{-\varphi} + N^{-2\varphi}$ by the properties of $\xi_N$ and $\ell_N$, and using that the variance of a random variable is lower than its second moment. 

\noindent \textbf{Claim 2.2:} $I_{N,2} = O(h_N^{-d_x})$. It follows by noting that $\| E[h(W_{j_1}, W_{j_2}, W_{i})  \mid W_{j_1}, W_{i}] \| \lesssim \| \varepsilon_{j_1} \| N^{-\varphi} + N^{-2\varphi}$ and $\| E[h(W_{j_1}, W_{j_2}, W_{i})  \mid W_{j_1}, W_{j_2}] \| \lesssim h_N^{-2d_x}I\{ \|X_{j_1}-X_{j_2}\| \le 2h_N\}\left(\|\varepsilon_{j_1}\|^2+\|\varepsilon_{j_2}\|^2 + N^{-2\varphi} \right)$ by the properties of $\xi_N$ and $\ell_N$, and using that the variance of a random variable is lower than its second moment.

\noindent \textbf{Claim 2.3:} $I_{N,3} = O(h_N^{-2d_x})$. Algebra and the Law of Iterated Expectations imply $E[I_{N,3} \mid X_i] \lesssim E[ \| \xi_N(W_{j_1},X_i)\|^2 \mid X_i] E[ \| \xi_N(W_{j_2},X_i)\|^2 \mid X_i]$. We conclude by using that $E[ \| \xi_N(W_{j_1},X_i)\|^2 \mid X_i]  \lesssim h_N^{-d_X}$.
\end{proof}

\begin{prop}\label{prop:lp-AS}
Let Conditions~\ref{con:lp-sampling}--\ref{con:lp-kernel} and \ref{con:lp-trimming} hold. Then, Assumption~\ref{asm:AS-SO} holds. \vspace{-0.25cm}
\end{prop}
 \begin{proof}
 As in the proof of Lemma~\ref{appendix:lemma_C0}, it is sufficient to prove the claim for one coordinate. By Lemma~\ref{lem:lp-expansion} that $\mathcal{R}_k(x)
    =e_1^\intercal\mathcal A_k(x)\mathcal S_k(x)$, where  
    $\mathcal A_k(x) =\widehat{\mathcal M}_k(x)^{-1}-\mathcal M_{h_N}(x)^{-1}$ and $ \mathcal S_k(x)=\hat S_{\varepsilon,k}(x)+\hat S_{\rho,k}(x)$. 
 Let $\mathcal{R}_k^\ell(x)$ be the corresponding object after replacing $W_\ell$ by an independent copy $\widetilde{W}_\ell$, where $\ell \notin \mathcal{I}_k$. Let $\mathcal{A}_k^\ell(x)$ and
$\mathcal{S}_k^\ell(x)$ denote the corresponding replaced versions.  Then, we write $\mathcal R_k(x)-\mathcal R_k^\ell(x) = I_1(x) + I_2(x)$, where $I_1(x) = e_1^\intercal\{\mathcal A_k(x)-\mathcal A_k^\ell(x)\}\mathcal S_k(x)$ and $I_2(x) = e_1^\intercal\mathcal A_k^\ell(x)
       \{\mathcal S_k(x)-\mathcal S_k^\ell(x)\}$. 
Claims 1 and 2 below show that $E \left[ \|I_1(X)\|^2    \right]^{1/2}$ and $ E \left[  \| I_2(X)\|^2 \right]^{1/2}$ are both $O(N^{-1/2-2\varphi})$, which is sufficient to complete the proof due to Minkowski inequality.

\noindent \textbf{Claim 1:} $E \left[ \|I_1(X)\|^2    \right]^{1/2} = O( N^{-1/2-2\varphi})$. We first write $I_1(x) = I_{1,1}(x) + I_{1,2}(x)$, where $I_{1,1}(x) =  e_1^\intercal\{ \mathcal A_k(x)-\mathcal A_k^\ell(x)\} \mathcal S_{k,-\ell}(x)$ and $I_{1,2}(x) = e_1^\intercal\{\mathcal A_k(x)-\mathcal A_k^\ell(x)\}  s_\ell(x)$, here $s_\ell(x) = N^{-1}\mathcal K_{h_N}(X_\ell-x)r_p(U_{\ell,x}) \{\varepsilon_\ell+\rho_N(X_\ell,x)\}$ and $S_{k,-\ell}(x) = S_{k}(x) - s_\ell(x)$. We then consider the following derivations:
\begin{align*}
    E \left[ \|I_1(X)\|^2    \right]^{1/2} %&\overset{(1)}{\le}  E \left[ \|I_{1,1}(X)\|^2    \right]^{1/2} + E \left[ \|I_{1,2}(X)\|^2    \right]^{1/2} \\
    &\overset{(1)}{\le} E \left[ \| \widehat{\mathcal M}_k(X)^{-1}-\widehat{\mathcal M}_k^{\ell}(X)^{-1}\|^2   \| \mathcal S_{k,-\ell}(X)\|^2   \right]^{1/2} \\
    &~~+  E \left[  \| \widehat{\mathcal M}_k(X)^{-1}-\widehat{\mathcal M}_k^{\ell}(X)^{-1}\|^2   \|  s_{\ell}(X)\|^2   \right]^{1/2} \\
    &\overset{(2)}{\lesssim } E \left[ \| \widehat{\mathcal M}_k(X)-\widehat{\mathcal M}_k^{\ell}(X)\|^2   \| \mathcal S_{k,-\ell}(X)\|^2   \right]^{1/2} \\
    &~~+  E \left[  \| \widehat{\mathcal M}_k(X)-\widehat{\mathcal M}_k^{\ell}(X)\|^2   \|  s_{\ell}(X)\|^2   \right]^{1/2} \\
    &\overset{(3)}{\lesssim}  E \left[ E \left[\| \widehat{\mathcal M}_k(X)-\widehat{\mathcal M}_k^{\ell}(X)\|^2 \mid X \right] E \left[  \| \mathcal S_{k,-\ell}(X)\|^2 \mid X\right] \right]^{1/2} \\
    &~~+  E \left[   \| \widehat{\mathcal M}_k(X)-\widehat{\mathcal M}_k^{\ell}(X)\|^2 N^{-2}h_N^{-2d_x}(  \varepsilon_\ell^2 + h^{2s})   \right]^{1/2} \\
    &\overset{(4)}{\lesssim}   N^{-1/2} (N h_N^{d_x})^{-1/2} \left \{  \left( (N h_N^{d_x})^{-1/2} + h_N^{s}  \right) +   \left(N^{-1} h_N^{-d_x} \right) \right\} \overset{(5)}{\asymp} N^{-1/2-2\varphi}   
\end{align*}
where (1) holds by Minkowski inequality and definition of $\mathcal{A}_k(x)$ and $\mathcal{A}_k^{\ell}(x)$, (2) holds by \eqref{eq:inequality-appendix} and since $\|\widehat{\mathcal M}_k(x)^{-1}\|$ is uniformly bounded by construction, (3) holds by the Law of Iterated Expectations and definition of $s_\ell(x)$, (4) holds by Condition~\ref{con:lp-trimming}, and (5) holds by definition of $h_N$ and $\varphi$.

\noindent \textbf{Claim 2:} $E \left[ \|I_2(X)\|^2    \right] = O( N^{-1-4\varphi})$. The proof is analogs; therefore omitted. 
 \end{proof}

\begin{prop}\label{prop:lp-information}
Let Conditions~\ref{con:lp-sampling}--\ref{con:lp-kernel} hold. If the moment function $m$ is as in Appendix~\ref{sec:appendix_examples} and $x \mapsto E[m_j \varepsilon_{r,j} \mid X_j = x]$ has $s+1$ continuous derivatives. Then, Assumption~\ref{asm:information} holds. 
\end{prop}

\begin{proof} Recall $\mathcal U_{h}(x) = \{ u: x + hu \in \mathcal{X} \}$. Define $\tilde{\delta}_{\infty,r}(x)=G^{-1}E[m_i \varepsilon_{r,i}\mid X_i=x]$.

\textit{Part (i):} By definition $\delta_{N,r}(W_j,x)$ in \eqref{eq:lp-delta-b} and $ \tilde\delta_{N,r}(x)$, we first write
\begin{align}
   \tilde\delta_{N,r}(x) =e_1^\intercal\mathcal M_{h_N}(x)^{-1}   \int_{\mathcal U_{h_N}(x)}       \mathcal K(u)r_p(u)\tilde{\delta}_{\infty,r}(x+h_Nu)f_X(x+h_Nu)du.            \label{eq:tildedelta-smoother-audited}
\end{align}
We then use $\int \ell_N(z,x)f_X(z)dz=1$ and the Dominated Convergence Theorem to conclude that $\tilde{\delta}_{N,r}(x) \to \tilde{\delta}_{\infty,r}(x)$ for all $x \in \mathcal{X}$. Note that $\sup_{x \in \mathcal{X}}|\tilde{\delta}_N(x)| < C$ for some $C>0$ and any $N$ since $\sup_{x \in \mathcal{X}} \int |\ell_N(z,x)|f_X(z)dz$ is bounded due to \eqref{eq:lp-kernel-bound-proof}. Therefore, we can use the Dominated Convergence Theorem and that $\tilde{\delta}_{\infty,r}(x)$ has $s+1$ continuous derivatives to conclude that $  E\left[|\tilde{\delta}_{N,r}(X)-\tilde{\delta}_{\infty,r}(X)|^2\right]^{1/2} =O(h_N^s)$ which is equal to $O(N^{-\varphi})$ by definition of $h_N$ and $\varphi$. 

\textit{Part (ii):} We calculate and present $\tilde{\delta}_{\infty}(x)=(\tilde{\delta}_{\infty,1}(x),\ldots,\tilde{\delta}_{\infty,p}(x))$, $S(x)$, and $H(x)$ for the ATE (Example \ref{example_ate}), the ATT-DID (Example \ref{example_DID}), and the LATE (Example \ref{example_LATE}). For brevity, we omit the intermediate algebra and report only the explicit expressions. We use the notation introduced in the proof of Proposition~\ref{prop:lp-leading}. 
   
\noindent\textit{ATE:}  Algebra shows  $H(x)\tilde\delta_\infty(x) = -S(x)$, where
\begin{equation*}
     \tilde\delta_\infty(x)
    =
    \begin{pmatrix}
        \sigma_1^2(x)/p_A(x)\\
        -\sigma_0^2(x)/q_A(x)\\
        0\\
        0
    \end{pmatrix}, \quad
    S(x)
    =
    \begin{pmatrix}
        0\\
        0\\
        \sigma_1^2(x)\\
        \sigma_0^2(x)
    \end{pmatrix}, \quad
    H(x)
    =
    \begin{pmatrix}
        0 & 0 & -p_A(x) & 0\\
        0 & 0 & 0 & q_A(x)\\
        -p_A(x) & 0 & 0 & 0\\
        0 & q_A(x) & 0 & 0
    \end{pmatrix}.
\end{equation*} 

\noindent\textit{ATT-DID.} Let $\tau(x) = \eta_{0,1}(x)-\eta_{0,2}(x)-\theta_0$. Algebra shows  $H(x)\tilde\delta_\infty(x) = -S(x)$, where   
\begin{equation*}
       \tilde\delta_\infty(x)
    =
    \frac1G
    \begin{pmatrix}
        -p_A(x)\sigma_0^2(x)/q_A(x)\\
        p_A(x)\tau(x)/q_A(x)
    \end{pmatrix}, \quad 
    S(x)
    =
    \frac1{G^2}
    \begin{pmatrix}
        -p_A(x)\tau(x)\\
        p_A(x)\sigma_0^2(x)
    \end{pmatrix}, \quad 
    H(x)
    =
    \frac1G
    \begin{pmatrix}
        0 & q_A(x)\\
        q_A(x) & 0
    \end{pmatrix}.
\end{equation*}

\noindent\textit{LATE:} Let $C_{Yz}(x)=E[\xi_z\varepsilon_{Yz}\mid X=x,Z=z]$, and $C_{Dz}(x)=E[\xi_z\varepsilon_{Dz}\mid X=x,Z=z]$. Algebra shows $H(x)\tilde\delta_\infty(x) = -S(x)$, where 
\begin{equation*}
    \tilde\delta_\infty(x)
    =
    \frac1G
    \begin{pmatrix}
        C_{Y1}(x)/p_Z(x)\\
        -C_{Y0}(x)/q_Z(x)\\
        C_{D1}(x)/p_Z(x)\\
        -C_{D0}(x)/q_Z(x)\\
        0\\
        0
    \end{pmatrix}, \quad 
    S(x)
    =
    \frac1{G^2}
    \begin{pmatrix}
        0\\
        0\\
        0\\
        0\\
        V_1(x)\\
        V_0(x)
    \end{pmatrix},
\end{equation*}

and
\begin{equation*}
    H(x)
    =
    \frac1G
    \begin{pmatrix}
        0 & 0 & 0 & 0 & -p_Z(x) & 0\\
        0 & 0 & 0 & 0 & 0 & q_Z(x)\\
        0 & 0 & 0 & 0 & \theta_0p_Z(x) & 0\\
        0 & 0 & 0 & 0 & 0 & -\theta_0q_Z(x)\\
        -p_Z(x) & 0 & \theta_0p_Z(x) & 0 & 0 & 0\\
        0 & q_Z(x) & 0 & -\theta_0q_Z(x) & 0 & 0
    \end{pmatrix},
\end{equation*}
and using that $C_{Yz}(x)-\theta_0C_{Dz}(x)  =  E[\xi_z(\varepsilon_{Yz}-\theta_0\varepsilon_{Dz})\mid X=x,Z=z]  = V_z(x)$.
\end{proof}
 
 \end{appendix}
\end{document}